\documentclass[aps,prb,superscriptaddress,reprint
]{revtex4-2}

\usepackage{amsmath,amssymb,graphicx,bm,xspace,multirow}
\usepackage[caption=false]{subfig}
\usepackage[linktocpage=true]{hyperref}

\usepackage{etoolbox}

\usepackage{array}
\newcolumntype{L}[1]{>{\raggedright\let\newline\\\arraybackslash\hspace{0pt}}m{#1}}
\newcolumntype{R}[1]{>{\raggedleft\let\newline\\\arraybackslash\hspace{0pt}}m{#1}}

\newcommand{\sx}{_{\mathrm{x}}}
\newcommand{\suc}{_{\mathrm{c}}}
\newcommand{\sxc}{_{\mathrm{xc}}}
\newcommand{\nup}{n_{\uparrow}}
\newcommand{\ndn}{n_{\downarrow}}
\newcommand{\sus}{_{\mathrm{s}}}
\newcommand{\rrscan}{r$^2$SCAN\xspace}

\newcommand{\br}{\ensuremath{\bm{r}}}
\newcommand{\bR}{\ensuremath{\bm{R}}}
\newcommand{\bG}{\ensuremath{\bm{G}}}
\newcommand{\ef}{\ensuremath{ \varepsilon_{\mathrm{F}} }}
\newcommand{\texc}{\ensuremath{\widetilde E_{\mathrm{xc}}}\xspace}
\newcommand{\tu}{\tau_\text{unif}}
\newcommand{\tw}{\tau_\text{W}}

\newcommand{\kf}{k_{\mathrm{F}}}
\newcommand{\rs}{r_{\mathrm{s}}}
\newcommand{\apc}{\Delta^\text{PC}}

\newcommand{\ba}{\overline{\alpha}}

\newcommand{\n}{\nabla}
\newcommand{\nn}{\nabla^2}
\newcommand{\gn}{\n n}
\newcommand{\lan}{\nn n}

\newcommand{\dd}[2]{\frac{\partial #1}{\partial #2}}

\begin{document}

\title{Laplacian-level meta-generalized gradient approximation for solid and liquid metals}
\author{Aaron D. Kaplan}
\email{kaplan@temple.edu}
\affiliation{Department of Physics, Temple University, Philadelphia, PA 19122}
\author{John P. Perdew}
\email{perdew@temple.edu}
\affiliation{Department of Physics, Temple University, Philadelphia, PA 19122}
\affiliation{Department of Chemistry, Temple University, Philadelphia, PA 19122}

\date{\today}

\begin{abstract}
  We derive and motivate a Laplacian-level, orbital-free meta-generalized-gradient approximation (LL-MGGA) for the exchange-correlation energy, targeting accurate ground-state properties of $sp$ and $sd$ metallic condensed matter, in which the density functional for the exchange-correlation energy is only weakly nonlocal due to perfect long-range screening.
  Our model for the orbital-free kinetic energy density restores the fourth-order gradient expansion for exchange to the \rrscan meta-GGA [Furness \textit{et al.}, J. Phys. Chem. Lett. \textbf{11}, 8208 (2020)], yielding a LL-MGGA we call OFR2.
  OFR2 matches the accuracy of SCAN for prediction of common lattice constants and improves the equilibrium properties of alkali metals, transition metals, and intermetallics that were degraded relative to the PBE GGA values by both SCAN and \rrscan.
  We compare OFR2 to the \rrscan{}-L LL-MGGA [D. Mejia-Rodriguez and S.B. Trickey, Phys. Rev. B \textbf{102}, 121109 (2020)] and show that OFR2 tends to outperform \rrscan{}-L for the equilibrium properties of solids, but \rrscan{}-L much better describes the atomization energies of molecules than OFR2 does.
  For best accuracy in molecules and non-metallic condensed matter, we continue to recommend SCAN and \rrscan{}.
  Numerical performance is discussed in detail, and our work provides an outlook to machine learning.
\end{abstract}

\maketitle

\section{Introduction}

Practical Kohn-Sham density functional theory (DFT) \cite{kohn1965} seeks an accurate and computationally efficient description of the ground state energy $E[\nup,\ndn]$ and spin-densities $(\nup,\ndn)$ of any many-electron system.
This requires a density functional approximation (DFA) for the exchange-correlation energy $E\sxc$.
First-principles DFAs are derived from purely theoretical considerations, whereas empirical DFAs are fitted to data (especially for bonded systems).
Semi-empirical DFAs borrow from both approaches.
Empirical DFAs often cannot extrapolate well to systems unlike those used to parameterize them \cite{medvedev2017}.
Recent machine-learned, semi-empirical DFAs \cite{dick2021,kirkpatrick2021} which incorporate a greater number of exact constraints have overcome some of the limitations inherent to empiricism.
A semi-empirical, ``human-learned'' non-local DFA using a small number of parameters has been shown to rival highly-parametrized empirical DFAs' descriptions of thermochemical reactions \cite{becke2022}, supporting this analysis.
However, we will primarily discuss first-principles DFAs.

The most widely-known first-principles DFAs at the time of writing are the local spin density approximation (LSDA), and the Perdew-Burke-Ernzerhof generalized gradient approximation (PBE GGA or PBE) \cite{perdew1996}.
Both DFAs satisfy subsets of all known behaviors of the exact $E\sxc$: the $E\sxc$ of a uniform electron gas, spin-scaling of $E\sx$ \cite{oliver1979}, the behaviors of $E\sx$ and $E\suc$ under uniform scaling of the position vector \br \cite{levy1985,levy1991,levy1993}, among others.

LSDA and the gradient expansion approximation (GEA) \cite{kohn1965,svendsen1996,ma1968,wang1991} were the first two DFAs to be proposed (simultaneously).
The LSDA gives the exact $E\sxc$ of a uniform electron gas, and is the zeroth-order approximation to the $E\sxc$ of a slowly-varying electron gas.
The GEA of a given order describes the exact response of a uniform electron gas to a static, long-wavelength perturbation \cite{svendsen1995} (a slowly-varying electron gas).
While LSDA generally provides an accurate starting point for describing simple systems, the ungeneralized GEA offers no systematic correction to the LSDA \cite{langreth1979,langreth1980,springer1996}.

To quantify ``slowly-varying,'' we define a few dimensionless variables (in Hartree atomic units, $e^2 = m_e = \hbar = 1$, unless otherwise specified).
The appropriate length scale for the exchange energy is the Fermi wavevector
\begin{equation}
  \kf(n) = \left[3\pi^2 n(\br)\right]^{1/3}.
\end{equation}
Then let
\begin{equation}
  p(n,|\nabla n|) = \left[ \frac{|\nabla n(\br)|}{2 \kf(\br) n(\br)} \right]^2
\end{equation}
be a squared dimensionless gradient of the density, and
\begin{equation}
  q(n,\nabla^2 n) = \frac{\nabla^2 n(\br)}{4 [\kf(\br)]^2 n(\br)}
\end{equation}
be a dimensionless Laplacian of the density on this length scale.
For a uniform density, $p = q = 0$.
Let the positive definite kinetic energy density be
\begin{equation}
  \tau_\sigma = \frac{1}{2} \sum_i f_{i\sigma} |\nabla \phi_{i\sigma}(\br)|^2,
\end{equation}
with integer occupancies $f_{i\sigma}=0,\, 1$.
We also define a dimensionless kinetic energy variable
\begin{equation}
  \alpha(n,|\gn|, \tau) = \frac{\tau(\br)-\tw(n,|\gn|)}{\tu(n)},
  \label{eq:alpha}
\end{equation}
which depends upon the Weizs\"{a}cker kinetic energy density
\begin{equation}
  \tw(n,|\gn|) = \frac{|\nabla n(\br)|^2}{8 n(\br)}, \label{eq:tau_w}
\end{equation}
and the uniform electron gas, or Thomas-Fermi, non-interacting kinetic energy density
\begin{equation}
  \tu(n) = \frac{3}{10}\kf^2(n)n(\br). \label{eq:tau_unif}
\end{equation}
$\alpha=1$ for a uniform density.
Thus, a density is considered slowly-varying when
\begin{equation}
  p \ll 1 ~~\text{and}~~ |q| \ll 1 ~~\text{and}~~ |1-\alpha| \ll 1.
  \label{eq:svl_def}
\end{equation}

Approximating $\alpha$ using $p$ and $q$ will be the primary topic of this work; thus we discuss a few rigorous properties of $\alpha$.
$\alpha\to 0$ when $\tau$ approaches its lower bound, $\tw$ \cite{hoffmann1977}.
$\alpha=0$ uniquely identifies single-orbital densities where $\tau = \tw$ exactly.
A single-orbital (or ``iso-orbital'') density has only one occupied spatial orbital, such as a fully spin-polarized one electron density, or a spin-unpolarized two-electron density.
Density variables such as $\alpha$ that uniquely recognize single-orbital regions are often called iso-orbital indicators.
For a slowly-varying density, $\tau$ has a known gradient expansion like the GEA \cite{brack1976}.
These known limits are important, as they permit $\tau$-meta-GGAs (T-MGGAs) to be essentially exact for typical one- and two-electron densities and slowly-varying ones \cite{sun2015}.
Here, ``typical'' refers to compact, un-noded \cite{shahi2019} one-electron densities.
Such a balanced description between finite and extended systems is not possible when using only $p$ and $q$, as we shall demonstrate.

A meta-GGA that depends on $\alpha$ of Eq. \ref{eq:alpha} can mistakenly identify intershell regions in atoms as slowly-varying \cite{perdew2003}.
The same behavior will be demonstrated for a Laplacian-level meta-GGA (LL-MGGA).
To make an indicator like $\alpha$ that better distinguishes between finite and extended systems, one must consider the first and second derivatives of $\tau$, $\nabla \tau$ and $\nabla^2 \tau$ respectively, in addition to those of $n$ \cite{perdew2003}.
DFAs with all those ingredients are not currently available and are challenging to construct or use.

Most common LL-MGGAs are ``de-orbitalizations'' of T-MGGAs.
These orbital-free meta-GGAs replace the analytic expression for $\tau$ with an approximate form $\widetilde{\tau}_{\sigma}(n_{\sigma},|\nabla n_{\sigma}|,\nabla^2 n_{\sigma})$ that may be constrained to recover exact constraints.

The most popular correlation GGA in the quantum chemistry community, due to Lee, Yang, and Parr (LYP) \cite{lee1988}, was originally cast as an empirical LL-MGGA.
Miehlich \textit{et al.} \cite{miehlich1989} demonstrated that an integration by parts, such as that used in Appendix \ref{app:r2ge4}, could eliminate the density-Laplacian in favor of the density-gradient, yielding a conventional GGA.
This latter GGA form is generally called LYP, and the Laplacian-dependent variant is not commonly used.
Other authors \cite{proynov1994,filatov1998} have built upon LYP to derive Laplacian-dependent exchange and correlation DFAs.

Similarly, the exchange density matrix expansion (DME) of Negele and Vautherin \cite{negele1972}, originally derived in the context of nuclear Hartree-Fock theory, leads \cite{tao2003a} to an exchange energy density
\begin{equation}
  \frac{e\sx^\text{DME}(n,p,q,\alpha)}{e\sx^\text{LDA}(n)} = 1 + \frac{35}{27}(q - p) + \frac{7}{9}(1-\alpha),
\end{equation}
with $e\sx^\text{LDA} = -3 \kf n/(4\pi)$ the local density approximation (LDA) for exchange.
The DME was generalized and the $q$-dependence removed to construct the Van Voorhis-Scuseria (VS98) \cite{vanvoorhis1998} and the M06-L \cite{zhao2006} empirical meta-GGAs.
More recently, a similar $q$-independent generalization of the DME was used to construct the Tao-Mo meta-GGA \cite{tao2016}.

As will be discussed further, no single level of approximation (GGA, meta-GGA, etc.) in practical DFT can describe all systems with the same level of accuracy.
This has been demonstrated empirically, for example, in the derivations of the PBEsol \cite{perdew2008} and PBEmol \cite{delcampo2012} GGAs.
PBE, PBEsol, and PBEmol all use the same Becke 1986 \cite{becke1986} form for the exchange enhancement factor
\begin{equation}
  F\sx(p) \equiv \frac{e\sx(n,p)}{e\sx^\text{LDA}(n)} = 1 + \kappa - \frac{\kappa}{1 + \mu p/\kappa}
  \label{eq:b86_fx}
\end{equation}
and PBE-like correlation energy per electron (see Eqs. 7 and 8 of Ref. \cite{perdew1996}).
In all three variants, $\kappa=0.804$ to enforce an exact constraint \cite{perdew1996}.
The PBE GGA, which sets $\mu = 0.21951$, does not recover the correct second-order GEA coefficient for exchange (10/81), but does so for correlation ($\beta \approx 0.066725$, as in Eq. 4 of Ref. \cite{perdew1996}).
This choice is understood to improve PBE's description of atomic and molecular properties at the expense of those of solids \cite{perdew2008,cancio2018}.
By contrast, PBEsol \cite{perdew2008}, which sets $\mu = 10/81$ and $\beta=0.046$, recovers the second-order GEA coefficient for exchange, but not correlation.
PBEsol tends to describe solids well, at the expense of atoms and molecules.
PBEmol improves slightly \cite{delcampo2012} upon PBE's description of molecules by setting $\mu = 0.27583$ to recover the hydrogen atom exchange energy (and $\beta =0.08384$ to satisfy the same linear response constraint as PBE), thereby defining another GGA extreme.
PBE is a ``middle-path'' GGA, describing finite and extended densities with reasonable accuracy, but is not competitive with either extreme (PBEmol and PBEsol, respectively) in either category.

Similar but less severe limitations also appear at the meta-GGA level.
For example, the strongly constrained and appropriately normed (SCAN) \cite{sun2015} and regularized-restored SCAN (\rrscan) \cite{furness2020} T-MGGAs have achieved remarkable successes, not only for molecules, but also for semiconducting and insulating solids and liquids \cite{sun2016,yang2016,zhang2017,chen2017,shahi2018,zhang2018,dasgupta2021}, including strongly-correlated ones \cite{kitchaev2016,gautam2018,furness2018,zhang2020}.
But these T-MGGAs tend to predict unit cell magnetic moments that are somewhat too large compared to GGA predictions and experiment \cite{ekholm2018,fu2018,mejia2019a}.
SCAN also tends to predict longer lattice constants and smaller cohesive energies in alkali metals than PBE \cite{kovacs2019}, thereby providing a less correct description of simple metals.
Curiously, Ref. \cite{jana2018} found that SCAN predicts formation of a monovacancy in Pt to be energetically favorable.

PBE also describes the formation energies $\Delta E_\text{f}$ of many intermetallic alloys, such as HfOs, ScPt, and VPt$_2$, more accurately than SCAN \cite{isaacs2018}, although the PBE formation energies are substantially too large for these solids.
Kingsbury \textit{et al.} \cite{kingsbury2022} demonstrated that \rrscan makes modest improvements in $\Delta H_\text{f}$ of these three solids, and generally improves SCAN's description of formation enthalpies for all solids tested.
The random phase approximation (RPA, which depends upon the occupied and unoccupied orbitals) predicts slightly more accurate formation energies for HfOs and ScPt than SCAN \cite{nepal2020}.
For the convenience of the reader, we have compiled the results of Refs. \cite{isaacs2018} and \cite{kingsbury2022} in Sec. \ref{sec:intermet}.

A GGA is more nonlocal than the LSDA, because the existence of a derivative is conditioned upon the continuity of a function in the immediate neighborhood of a point $\br$.
Likewise, both variants of meta-GGAs are more nonlocal than GGAs, as these include higher-order derivatives of the density or Kohn-Sham orbitals.
However, because the Kohn-Sham orbitals are highly-nonlocal, implicit functionals of the density, a T-MGGA is more non-local than an LL-MGGA.
The exchange-correlation energy functional of a semi-local (SL) DFA (LSDA, GGA, or meta-GGA) can be written as
\begin{equation}
  E\sxc^\text{SL}[\nup,\ndn] = \int e\sxc(\nup,\ndn,...;\br) d^3 r,
\end{equation}
where the exchange-correlation energy density $e\sxc(\br)$ depends explicitly only on local variables: $n_\sigma(\br)$, $\nabla n_\sigma(\br)$, $\nabla^2 n_\sigma(\br)$, $\tau_\sigma(\br)$, etc.
A hybrid functional, which includes some fraction of single-determinant exchange in its energy density $e\sxc$
\begin{align}
  e\sxc^\text{hybrid}(\br) &= (1-a) e\sxc^\text{SL}(\br) + e\suc^\text{SL}(\br) \label{eq:hybrid_framework} \\
  & -\frac{a}{2} \sum_{\sigma} \int \frac{\left|\rho_1(\br\sigma,\br'\sigma)\right|^2}{|\br - \br'|} d\br' \nonumber,
\end{align}
is a non-local functional of the Kohn-Sham orbitals $\phi_{i\sigma}(\br)$ through the reduced one-body density matrix
\begin{equation}
  \rho_1(\br\sigma,\br'\sigma') = \delta_{\sigma,\sigma'} \sum_i \phi^*_{i\sigma}(\br)\phi_{i\sigma}(\br')
    \theta(\ef - \varepsilon_{i\sigma}). \label{eq:rho1}
\end{equation}
$\delta_{ij}=1$ if $i=j$ and 0 if $i\neq j$ is the Kronecker delta, and $\theta(x<0)=0$, $\theta(x>0)=1$ is the step function.
Single-determinant exchange using Eq. \ref{eq:rho1} delivers the exact exchange energy ($a=1$ in Eq. \ref{eq:hybrid_framework}).

Itinerant electron magnetism appears to be best described by more local DFAs.
As shown elsewhere \cite{ekholm2018,fu2018,mejia2019a} and here, LSDA, non-empirical GGAs, and LL-MGGAs tend to better predict transition metal magnetic properties than do T-MGGAs.
Global hybrids, which use a constant parameter $a$ in Eq. \ref{eq:hybrid_framework}, are much more nonlocal and thus even less accurate than meta-GGAs for transition metal magnetism \cite{fu2019}.
Range-separated hybrids, generalizations of global hybrids that separate the short- and long-range components of the Coulomb interaction, also tend to predict markedly worse equilibrium properties (e.g., lattice constants and bulk moduli) for structurally simple metals than they do for similarly simple insulators \cite{paier2006}.
To the best of our knowledge, no study of extended systems using local hybrids, which use a function $a(\br)$ in Eq. \ref{eq:hybrid_framework} (and may also be range-separated), has been undertaken.
As meta-GGAs and global hybrids are more non-local, it stands to reason that the exchange-correlation holes of elemental transition metals may be surprisingly local, with the gradient terms of GGAs and LL-MGGAs offering meaningful corrections to LSDA.

Why does the exact density functional for the exchange-correlation energy display a weaker nonlocality in metallic solids than in molecules and non-metallic solids?
A clue is provided by the exact expression \cite{langreth1975,gunnarsson1976}
\begin{equation}
  E\sxc = \frac{1}{2}\int d^3 r \, n(\br) \int d^3 r' \, \frac{n\sxc(\br',\br)}{|\br'-\br|},
\end{equation}
where $n\sxc(\br',\br)$ is the density at $\br'$ of the coupling-constant-averaged exchange-correlation hole around an electron at $\br$.
Starting from the exact exchange hole, correlation makes the exchange-correlation hole more negative at $\br'=\br$, with a faster decay to zero as $|\br'-\br| \to \infty$.
At long range, the exchange hole density in a solid is screened (divided) by a dielectric constant which is finite in non-metals but infinite in metals.
In the uniform electron gas \cite{gori2002}, for example, the exact exchange hole density (averaged over oscillations) at long range decays as $|\br'-\br|^{-4}$, while the exact exchange-correlation hole density (averaged over oscillations) decays much faster as $|\br' - \br|^{-8}$.
As the exact exchange-correlation hole becomes deeper and more localized around its electron, the exact exchange-correlation energy functional becomes less non-local in the electron density.
For example \cite{skone2014}, the optimum fraction $a$ of exact exchange in a global hybrid functional is the inverse of a long-wavelength dielectric constant, and vanishes for a metal.
Thus, highly nonlocal information (e.g., the fundamental energy gap, the dielectric constant, or the descriptors of Ref. \cite{perdew2003}) is required to determine the level of nonlocality needed in an approximate density functional.

The search for a computationally efficient DFA that is highly accurate for nearly all systems of interest has not yet found an unequivocal choice.
It has, however, shown that inclusion of exact constraints is perhaps the single most powerful aspect of DFA design \cite{furness2022}.
In this work, we derive an orbital-free LL-MGGA and determine its accuracy for a diverse set of common solid-state systems.
Section \ref{sec:of_mgga} reviews extant LL-MGGAs and motivates the new model derived in Sec. \ref{sec:rpp_deriv}.
Section \ref{sec:tests} applies this model to real solids: their structural properties in Sec. \ref{sec:latt_cons}; itinerant electron magnetism in Sec. \ref{sec:ferro}; bandgaps of insulators in Sec. \ref{sec:insul}; formation of a monovancancy in Pt in Sec. \ref{sec:defect}; intermetallic formation enthalpies in Sec. \ref{sec:intermet}; and alkali metals in Sec. \ref{sec:alkalis}.
Section \ref{sec:ae6} presents a test of molecular atomization energies.
A discussion of machine learning applications to LL-MGGAs is given in Sec. \ref{sec:HL}.

\section{Orbital-free meta-GGAs \label{sec:of_mgga}}

Orbital-free variants of T-MGGAs may be the most common LL-MGGAs to date.
Finding a suitable replacement for $\tau$ in terms of the density and its spatial derivatives alone permits, in principle, highly-accurate and computationally-efficient calculations within standard Kohn-Sham theory.
Early attempts, such as that of Perdew and Constantin \cite{perdew2007}, proposed de-orbitalized meta-GGAs but provided no self-consistent tests.
Later works \cite{smiga2015,smiga2017} in the context of subsystem DFT successfully proposed semi-local, orbital-free approximations of $\tau$ for use in calculating the meta-GGA embedding potential.
However, as noted in Ref. \cite{smiga2017}, a semi-local model of $\tau$ in subsystem-DFT only needs to accurately capture non-additive interactions between independent subsystems, which primarily involve the valence electrons.
More recently, Mej\'ia-Rodr\'iguez and Trickey \cite{mejia2017,mejia2018} have pioneered a general-purpose, \textit{self-consistent} ``de-orbitalization'' procedure to replace the analytic $\tau$ with an approximate expression.
Their work is the inspiration for ours.

This construction has two primary benefits: a more localized exchange-correlation hole, and potential for greater numerical efficiency \cite{mejia2020}.
We posit that the more localized exchange-correlation holes of metals, including ``atypical metals'', are unexpectedly local, a suggestion made long ago \cite{perdew1997}.
Thus meta-GGAs like SCAN and \rrscan tend to make their holes too non-local, and more insulator-like.
Indeed, Ref. \cite{mejia2020} demonstrates that orbital-free versions of SCAN and \rrscan predict smaller magnetic moments in ferromagnets (when evaluated at the same geometry), and that the orbital-free variants tend to predict more accurate lattice constants of simple metals.
However, the orbital-free variants worsen the cohesive energies of simple metals, presumably because these energy differences involve atoms as well as metallic solids.

Mej\'ia-Rodr\'iguez and Trickey have shown \cite{mejia2020} that an orbital-free version of \rrscan, called \rrscan{}-L, has a computational cost similar to PBE in solids, but is less accurate than \rrscan for describing their equilibrium properties.
We construct a similarly-efficient LL-MGGA that accurately describes solids (particularly metals) by restoring the gradient expansion to an orbital-free \rrscan.

The Perdew-Constantin (PC) \cite{perdew2007} model approximates $\tau$ using an enhancement factor similar to that of semi-local exchange energies,
\begin{align}
  \widetilde{\tau}(n,p,q) &= \tu(n)F^{\text{PC}}\sus(p,q).
\end{align}
We use the ``s'' subscript to indicate a single-electron property, i.e., $F\sus$ is used to approximate the non-interacting kinetic energy density of a spin-unpolarized system.
Such a description is useful because the kinetic energy and exchange energy share the same spin-scaling relationship \cite{oliver1979}
\begin{equation}
  T\sus[\nup, \ndn] =\frac{1}{2}\left(T\sus[2\nup] + T\sus[2\ndn] \right).
\end{equation}
For sufficiently slowly-varying densities,
\begin{equation}
  \lim_{\substack{p\ll 1 \\ |q| \ll 1}}F^{\text{PC}}\sus(p,q) \to
  F_{\text{SVL}} = 1 + \frac{5}{27}p + \frac{20}{9} q + \Delta + \mathcal{O}(|\nabla n|^6),
\end{equation}
where $\Delta$ stands for generalized fourth-order gradient expansion terms.
Because it employs only the variables $p$ and $q$, the Perdew-Constantin model recovers only the second-order gradient expansion of $\tau$ and (via integration by parts) the fourth-order gradient expansion of $T\sus$.

For iso-orbital regions,
\begin{equation}
  F^{\text{PC}}\sus(p,q) \to F_{\text{W}} = 5p/3 = \tw/\tau_{\text{unif}}.
\end{equation}
To approximately recover the iso-orbital limit of $\tau$, the PC model interpolates between these limits
\begin{align}
  F^{\text{PC}}\sus(p,q) &= F_\text{W} + \apc f_{ab}(\apc) \\
  \apc &= F_{\text{SVL}} - F_{\text{W}}. \label{eq:pc_alpha}
\end{align}
From Eq. (\ref{eq:alpha}), $\apc f_{ab}(\apc)$ approximates $\alpha$.
The PC interpolation function is a smooth, non-analytic two-parameter function
\begin{align}
  f_{ab}(z) &= \left\{ \begin{array}{lr} 0, & z \leq 0 \\
  \left[\frac{1 + g_{1a}(z)}{g_{2a}(z) + g_{1a}(z)} \right]^b, & 0 < z < a \\
  1, & z \geq a
  \end{array} \right. \\
  g_{1a}(z) &= \exp\left(\frac{a}{a-z} \right) \\
  g_{2a}(z) &= \exp\left(\frac{a}{z} \right).
\end{align}
The parameters $a=0.5389$ and $b=3$ were determined \cite{perdew2007} by fitting to the kinetic energies of neutral atoms, ions, and jellium clusters; we will discuss the lattermost system further in this work.
The PC model assumes that $\apc\leq 0$ indicates an iso-orbital density, and that $\apc\geq a$ indicates a sufficiently slowly-varying density.
For a uniform density, $\apc = 1$.
Thus, $a < 1$ is needed to recover both the uniform density limit of $\tau$ and its low-order gradient expansion for weakly-inhomogeneous densities.

If $ a<1$, as in the Perdew-Constantin work \cite{perdew2007}, then
\begin{align}
  f_{ab}(\apc) &\to 1 - 40 p/27 + 20q/9 + \Delta + \mathcal{O}(|\gn|^6),
\end{align}
because
\begin{equation}
  \frac{d^k f_{ab}}{d(\apc)^k}\bigg|_{\apc=1} = 0
\end{equation}
for all $k \in \mathbb{N}^+$.
However, if $a > 1$, as in the Mej\'ia-Rodr\'iguez and Trickey re-parameterization (MRT or PCopt) \cite{mejia2017} of the PC functional, then $f_{ab}$ no longer has a correct Taylor series about $\apc=1$,
\begin{align}
  f_{ab}(\apc) =& f_{ab}(1) + f'_{ab}(1)(\apc-1) \\
  & + \mathcal{O}[(\apc-1)^2]. \nonumber
\end{align}
The MRT parameters are $a=1.784720$ and $b=0.258304$; then the coefficients in the Taylor series of $f_{ab}(\apc)$ are
\begin{align}
  f_{ab}(1) &= \left\{\frac{1 + g_{1a}(1)}{ g_{2a}(1) + g_{1a}(1)} \right\}^b
  \approx 0.906485 \\
  f'_{ab}(1) &= b \left\{\frac{1 + g_{1a}(1)}{ g_{2a}(1) + g_{1a}(1)} \right\}^{b-1} \nonumber\\
  & \times \left\{ \frac{ g_{1a}'(1)[g_{2a}(1)-1] - g_{2a}'(1)[1 + g_{1a}(1)] }{
   [g_{1a}(1) + g_{2a}(1)]^2 }\right\} \nonumber \\
  & \approx 0.353363.
\end{align}
For reference,
\begin{align}
  g'_{1a}(z) &= \frac{a}{(a-z)^2}g_{1a}(z) \\
  g'_{2a}(z) &= -\frac{a}{z^2}g_{2a}(z).
\end{align}
Note that $\apc-1 = \mathcal{O}(|\nabla n|^2)$, and $(\apc-1)^2 = \mathcal{O}(|\nabla n|^4)$ to lowest order.
As $f'_{ab}(1)\neq0$ in the MRT model, the gradient expansion of the MRT $\tau$ no longer agrees with the known expansion, including the LSDA (uniform density) term,
\begin{align}
  \tau^\text{MRT}(n,p,q) &= \left[0.906485 + 1.143167 p \right. \nonumber \\
  &\left. + 0.785250q + \mathcal{O}(|\nabla n|^4) \right]\tu(n). \label{eq:tau_mrt_ge2}
\end{align}
Compare this to the exact expansion \cite{brack1976}
\begin{align}
  \tau^\text{GEA}(n,p,q) &= \left[1 + 0.185185p \right. \nonumber \\
  &\left. + 2.222222q + \mathcal{O}(|\nabla n|^4) \right]\tu(n).
\end{align}
The incorrect zeroth-order term in $\tau^\text{MRT}$ was identified in Ref. \cite{mejia2017}, but its relevance to the gradient expansion of $\tau$ was not.
Replacing the exact $\tau$ in SCAN or \rrscan{} by $\tau^\text{MRT}$ yields SCAN-L \cite{mejia2017} or \rrscan{}-L \cite{mejia2020}.

It has been shown, by the \rrscan authors and by many others \cite{santra2019,mejia2019,bartok2019a,yamamoto2020,kaplan2020} that the uniform density limit is critical for describing solid-state properties, molecular atomization energies, and molecular formation enthalpies.
The gradient expansion is expected to be particularly relevant to metals.
The present work parallels the restoration of the uniform density and gradient expansion constraints to the rSCAN T-MGGA \cite{bartok2019} by \rrscan \cite{furness2020}.

The loss of the correct uniform density and gradient expansion constraints reduces the accuracy of an orbital-free meta-GGA when applied to jellium prototypes of solids.
Table \ref{tab:deorb_jellium_norms} compares the XC surface formation energies calculated for the planar jellium surface and clusters from two $\tau$ meta-GGAs, SCAN \cite{sun2015} and r$^2$SCAN \cite{furness2020}, with their deorbitalized counterparts SCAN-L \cite{mejia2017,mejia2018} and
r$^2$SCAN-L \cite{mejia2020}.
It is clear that SCAN and r$^2$SCAN provide reasonably accurate descriptions of the jellium surface formation energies, while their deorbitalized counterparts do not.

\begin{table*}
  \begin{ruledtabular}
    \begin{center}
      \begin{tabular}{l|rr|rr|rr|rr}
        & \multicolumn{2}{c}{SCAN} & \multicolumn{2}{c}{SCAN-L} & \multicolumn{2}{c}{\rrscan} & \multicolumn{2}{c}{\rrscan{}-L} \\ \hline
        & Surface & Cluster & Surface & Cluster & Surface & Cluster & Surface & Cluster \\
        $\rs = 2$ & 3448 & 3424 & 3173 & 3072 & 3288 & 3299 & 3245 & 2863 \\
        $\rs = 3$ & 789 & 791 & 709 & 689 & 753 & 761 & 740 & 646  \\
        $\rs = 4$ & 274 & 277 & 242 & 235 & 262 & 266 & 257 & 223 \\
        $\rs = 5$ & 120 & 123 & 104 & 102 & 115 & 118 & 113 & 98 \\
        MAPE & 2.51 & 3.35 & 8.39 & 10.96 & 2.79 & 2.62 & 3.60 & 15.97 \\
      \end{tabular}
    \end{center}
    \caption{Jellium surface formation energies $\sigma\sxc$ in erg/cm$^2$ computed for two meta-GGAs, SCAN \cite{sun2015} and r$^2$SCAN \cite{furness2020}, and their de-orbitalized counterparts SCAN-L \cite{mejia2017,mejia2018} and r$^2$SCAN-L \cite{mejia2020}. Surface formation energies are calculated from LSDA reference densities for both the planar surface and the liquid drop model applied to spherical jellium clusters.
    The mean absolute percentage errors (MAPEs) are computed with respect to RPA+ values \cite{yan2000,almeida2002}, as motivated in the text.
    As 1 hartree $\approx 27.211386$ eV \cite{unit_data}, 1 erg/cm$^{2} \approx 0.0624151$ meV/\AA{}$^2$.
    \label{tab:deorb_jellium_norms}}
  \end{ruledtabular}
\end{table*}

\section{New model of the kinetic energy density \label{sec:rpp_deriv}}

We now sketch the derivation of a simplified Laplacian-level model of $\tau$, which is reasonably smooth and numerically stable.
Previous works attempting to construct an exchange enhancement factor with the density Laplacian demonstrated \cite{cancio2012} that the exchange-correlation potential
\begin{equation}
  v\sxc(\br) = \frac{\partial e\sxc}{\partial n} - \n \cdot \left(\frac{\partial e\sxc}{\partial \gn} \right) + \nn \left(\frac{\partial e\sxc}{\partial \lan}
  \right) \label{eq:vxc}
\end{equation}
is easily destabilized when the ``curvature'' term, rightmost in Eq. (\ref{eq:vxc}), is not well-constrained.
Note that $e\sxc$ is the exchange-correlation energy density, the integrand of the exchange-correlation energy functional.
It is not possible to eliminate all oscillations induced by this term into the Kohn-Sham potential, but these can be mitigated.

The Perdew-Constantin expression for the kinetic energy density enhancement factor $F\sus$ interpolates between the rigorous lower bound
\begin{equation}
  F_\mathrm{W} = \frac{5}{3}p \leq F\sus
\end{equation}
and a regulated fourth-order gradient expansion for $\tau$, whose asymptotic limit is $1 + 5p/3$.
The ``asymptotic limit'' is defined by $p,\, |q| \to \infty$ and typified by, e.g., a density tail.
Here, we will interpolate between the iso-orbital or von Weizs\"acker limit and the slowly-varying or second-order gradient expansion limit.
Other choices are more suitable for atoms \cite{cancio2017,sala2015}, but solid and liquid metals are the targets of our work.

A set of ``appropriate norms'' (see Sec. \ref{sec:app_norms}) could provide information about how best to extrapolate beyond these two limits, in line with the construction of SCAN and \rrscan.
However, an interpolation between these two limits suffices for an accurate description of solids.
Section \ref{sec:HL} presents a less numerically-stable model for $\tau$ that extrapolates beyond these limits by fitting to appropriate norms.

To recover the second-order gradient expansion for the exchange and correlation energies in
\rrscan, and the fourth-order gradient expansion for the exchange energy in SCAN, an approximate $\widetilde{\tau}$ must recover the second-order gradient expansion of $\tau$.
Therefore, we aim to recover only the second-order gradient expansion of $\tau$, and not the fourth-order gradient expansion of $T\sus$.
However, as shown in App. \ref{app:r2ge4}, we restore the fourth-order gradient expansion for the exchange energy to \rrscan by constraining the fourth-order terms in $\widetilde{\tau}$.

\begin{figure}
  \begin{center}
    \includegraphics[width=\columnwidth]{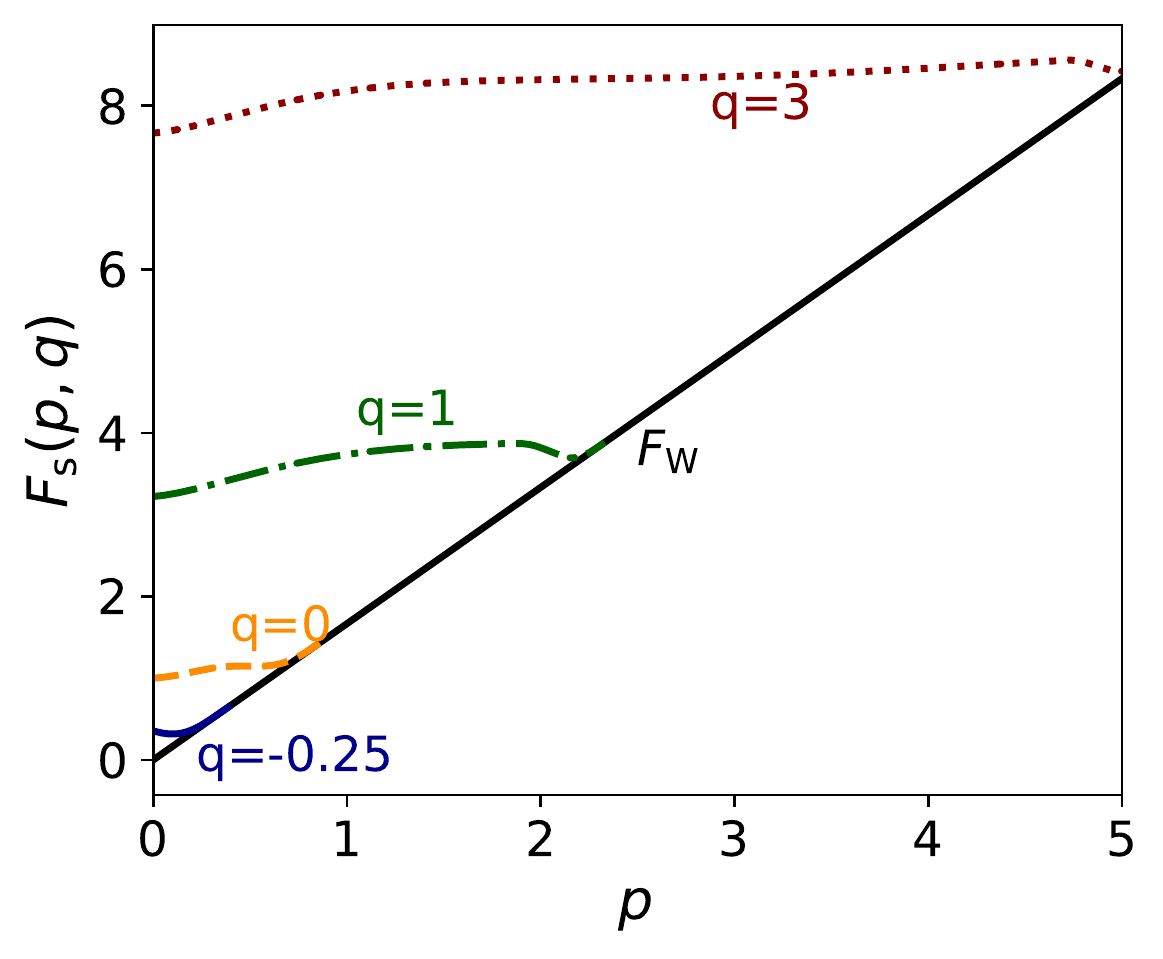}
  \end{center}
  \caption{The RPP kinetic energy density enhancement factor of
  Eq. (\ref{eq:fs}) compared to the Weizs\"acker lower bound
  $F_\mathrm{W} = 5p/3 $.
  For $q \lesssim -0.25$, $F\sus^\text{RPP}(p,q)\approx F_\text{W}(p)$.
   \label{fig:fs}}
\end{figure}

From Eq. (\ref{eq:alpha}),
\begin{equation}
  \alpha(\br) = F\sus - \frac{5}{3}p.
\end{equation}
$0 \leq \alpha < \infty $ is positive semi-definite, therefore we make a model of $\alpha$ with the same range as the true variable
\begin{align}
  \widetilde{\alpha}^\text{RPP}(x) &= \left\{ \begin{array}{lr} 0 , & x < 0 \\
  x^4(A + B x + C x^2 + D x^3), & 0 \leq x \leq x_0 \\
  x, & x > x_0
  \end{array}\right. \label{eq:alphas} \\
  x(p,q) &= 1 - \frac{40}{27}p + \frac{20}{9} q + c_3 p^2 e^{-|c_3|p} \label{eq:x_rpp1}\\
  & + x_4(p,q) \exp\left[-\left(\frac{p}{c_1} \right)^2
    -\left(\frac{q}{c_2} \right)^2 \right] \nonumber \\
  x_4(p,q) &= b_{qq}q^2 + b_{pq}p q + (b_{pp}-c_3)p^2 \\
  F\sus^\text{RPP}(p,q) &= \frac{5}{3}p + \widetilde{\alpha}^\text{RPP}(x(p,q))\label{eq:fs}
\end{align}
We call this model RPP for ``\rrscan piecewise-polynomial''.
Here, $A, \, B, \, C$ are determined by requiring that $\widetilde{\alpha}(x)$ is continuous up to its third derivative in $x$ at $x = x_0 $,
\begin{align}
  A &= 20/x_0^3 \\
  B &= -45/x_0^4 \\
  C &= 36/x_0^5 \\
  D &= -10/x_0^6.
\end{align}
$0 < x_0 < 1$, $c_1,c_2,$ and $c_3$ are model parameters determined by minimizing the residuum errors of a set of appropriate norms, described below.
Their optimal values are
\begin{align}
  x_0 &= 0.819411 \label{eq:x_0_fit} \\
  c_1 &= 0.201352 \\
  c_2 &= 0.185020 \\
  c_3 &= 1.53804  \label{eq:c_3_fit}
\end{align}
By construction, $\widetilde{\alpha}(x)$ is a $C^3$ function for all $x$.
While we model $\alpha$ as $\widetilde{\alpha}^\text{RPP}$, the actual quantity used to deorbitalize a meta-GGA is
\begin{equation}
  \tau^\text{RPP}(n,p,q) = \tu(n)F\sus^\text{RPP}(p,q),
\end{equation}
with $F\sus^\text{RPP}$ given by Eq. \ref{eq:fs}.
When $\tau^\text{RPP}$ is used to deorbitalize a T-MGGA, the resultant XC potential will be continuous.
$b_{qq}\approx 1.801019, b_{pq}\approx -1.850497,$ and $b_{pp}\approx 0.974002$ enforce the fourth-order gradient expansion for the exchange energy (GEX4); exact expressions are given in App. \ref{app:r2ge4}.
The Perdew-Constantin expression is a ``smooth non-analytic function,'' a $C^\infty$ function that has Taylor series with zero radius of convergence about at least one point ($z = 0 ,\, a$ in the Perdew-Constantin model).
The current model has a Taylor series of nonzero convergence radius about $x=0,x_0$.
Figure \ref{fig:fs} plots the enhancement factor over a range of $p$ typical for atoms and molecules (where the energetically important regions have $0 \leq p \leq 9$).

$\tau^\text{RPP}$ is intended for use in the \rrscan meta-GGA.
The numerical stability and general accuracy of \rrscan make it a good candidate for this kind of work, as noted in Ref. \cite{mejia2020}.
As \rrscan{} is still a relatively new meta-GGA, we briefly review its construction here.
The interested reader is encouraged to review Refs. \cite{furness2020,furness2022} for a more detailed presentation.
SCAN, while broadly accurate, tends to need dense numerical grids when performing self-consistent calculations \cite{yang2016}.

The rSCAN meta-GGA of Bart\'ok and Yates \cite{bartok2019} attempted to remedy this issue by replacing the iso-orbital indicator used in SCAN, $\alpha$, with a regularized indicator that tends to zero in density tails (where $\alpha$ diverges \cite{furness2019}), and by replacing the switching functions in SCAN, Eq. 9 of Ref. \cite{sun2015}, with a less-oscillatory function.
These modifications, while effective in improving the numerical performance of SCAN, broke exact constraints underpinning the construction of SCAN \cite{furness2022}.
The ablation of these constraints in rSCAN resulted in marked increases in computed atomization energy errors \cite{mejia2019}, for example.

The \rrscan{} meta-GGA \cite{furness2020} was constructed to maintain the numerical efficiency of rSCAN, but with accuracy comparable to SCAN.
This was accomplished by using an iso-orbital indicator,
\begin{equation}
  \ba = \frac{\tau - \tw}{\tu + \eta \, \tw}= \alpha\left[1 + \frac{5}{3}\eta p \right]^{-1}, \label{eq:alpha_to_bar_alpha}
\end{equation}
where $\eta=0.001$.
$\ba$ decays to zero in $s$-like density tails.
Furthermore, the slowly-varying limit (see Eq. \ref{eq:svl_def}) of rSCAN was modified to ensure recovery of the second-order gradient expansion constraints \cite{furness2022}.

The fourth-order terms in $x(p,q)$ restore the GEX4 terms to \rrscan.
The damped $x_4(p,q)$ term is modeled after the r$^4$SCAN meta-GGA \cite{furness2022}.
This meta-GGA restores the GEX4 to \rrscan using the exact $\tau$, at the price of some numerical stability and general accuracy.
We noticed in our testing that the gradient expansion terms need exponential cutoffs, like those used in r$^4$SCAN.
This is primarily due to the $b_{qq}q^2$ and $b_{pq}pq$ terms, which introduce numerical instabilities if they are not strongly regulated.
However, the $c_3 p^2$ term provides more meaningful corrections at large $p$.
For this reason, the damped $c_3p^2$ term has a much longer tail than $x_4(p,q)$.
We refer to the new orbital-free \rrscan, in which the exact $\tau$ is replaced by
\begin{equation}
  \tau^\text{RPP}(n,p,q) = \tu(n)[\widetilde{\alpha}^\text{RPP}(p,q) + 5p/3],
  \label{eq:tau_rpp}
\end{equation}
as ``OFR2,'' for orbital-free regularized-restored SCAN.
Equivalently, one could replace the exact $\alpha$ in the rightmost equality of Eq. \ref{eq:alpha_to_bar_alpha} with $\widetilde{\alpha}^\text{RPP}$; we make this distinction because \rrscan{} depends on $\ba$ instead of $\alpha$.
Of course, the cluster of \rrscan exact constraints associated with the iso-orbital limit $\tau =\tw$ can be satisfied only approximately by OFR2.

\begin{figure*}
  \begin{center}
    \subfloat[\label{fig:js_ked_a}]{
      \includegraphics[width=0.48\textwidth]{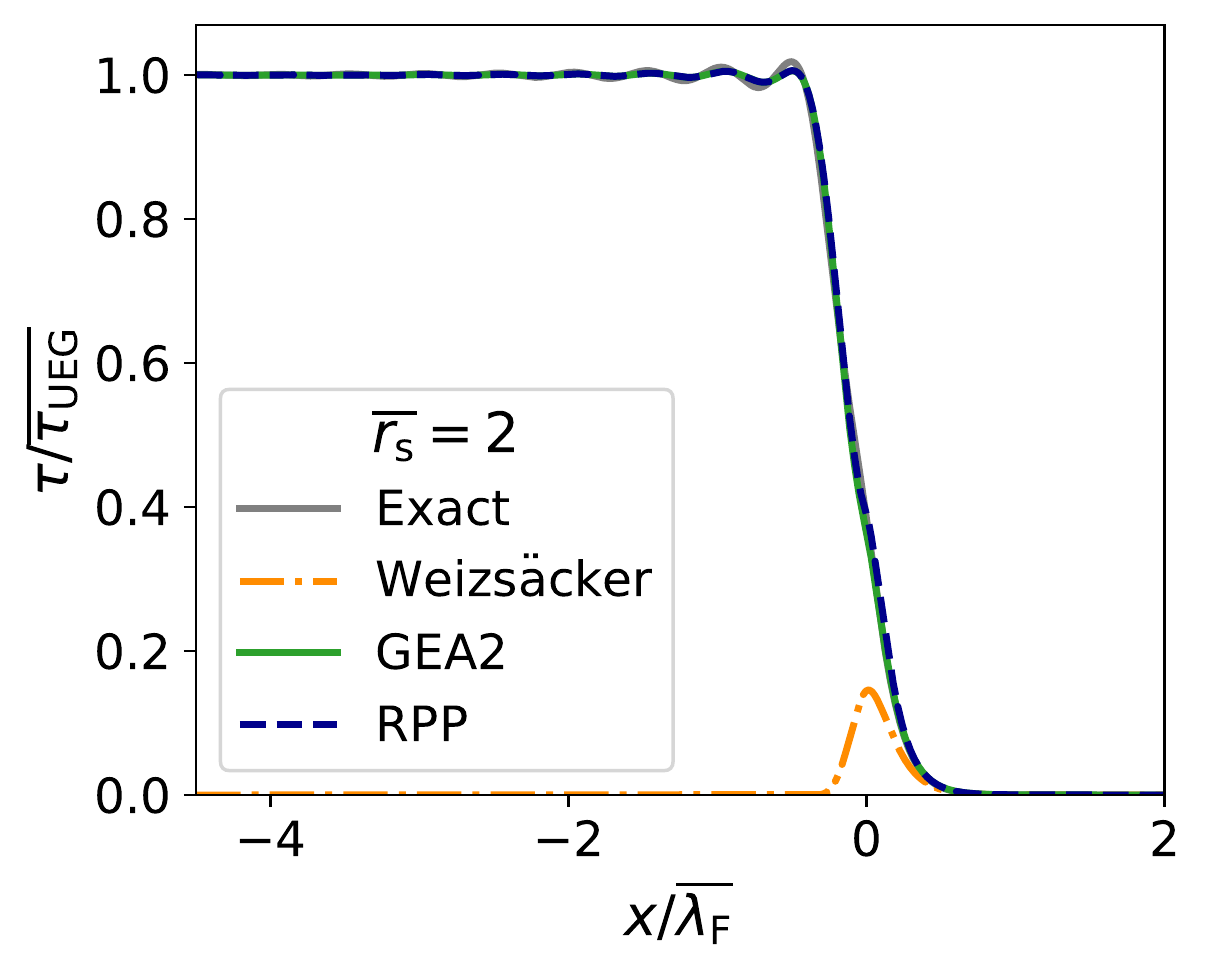}
    }
    \hfill
    \subfloat[\label{fig:js_ked_b}]{
      \includegraphics[width=0.48\textwidth]{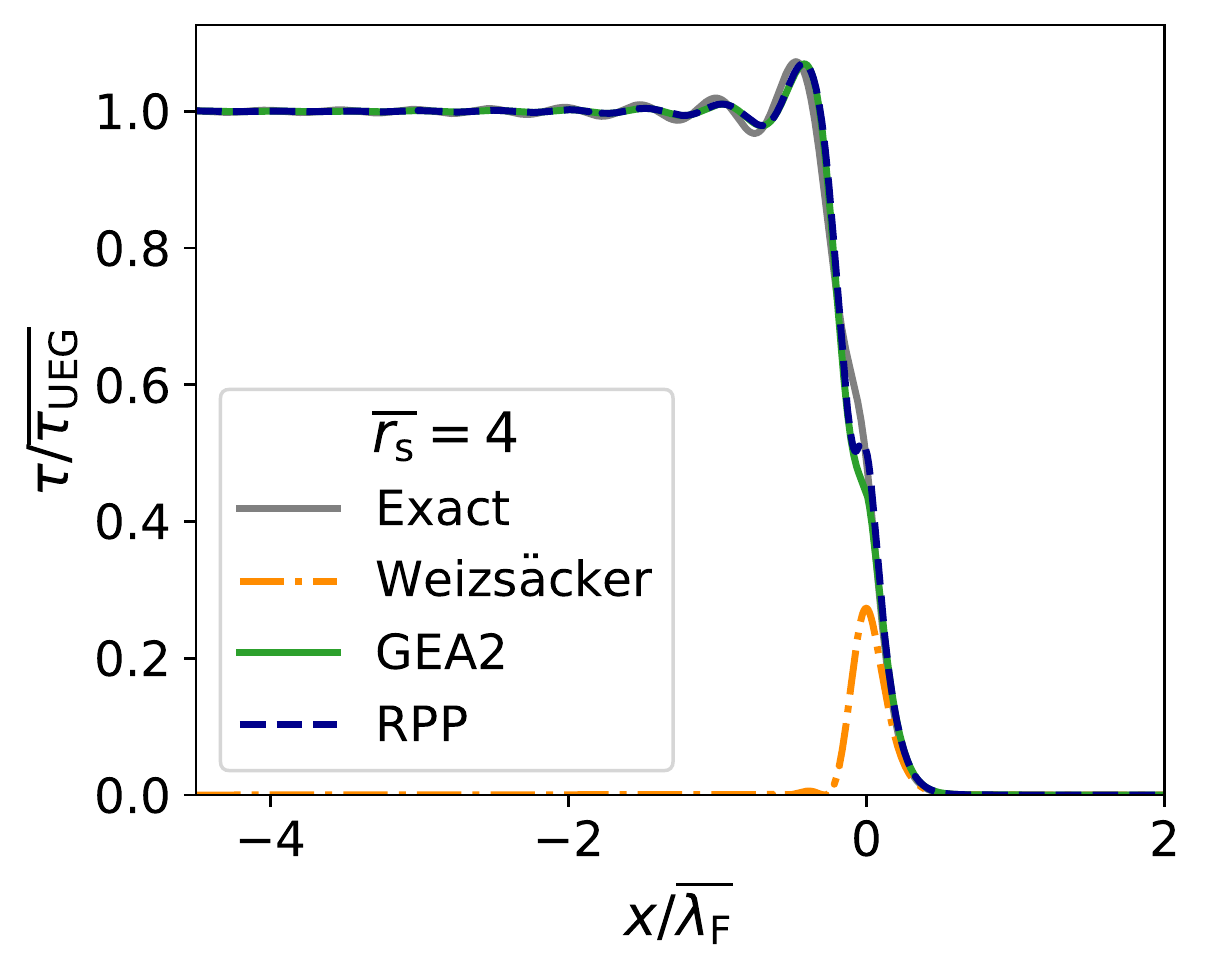}
    }
  \end{center}
  \caption{Plot of the exact $\tau$ (solid gray), second-order gradient expansion (GEA2, solid green), the RPP model (dashed blue), and Weizs\"acker (dash-dot orange) kinetic energy density for a jellium surface of bulk density parameter $\overline{\rs}=2,$ (left) and 4 (right).
  For a given density parameter $\overline{\rs}$, $ \overline{\tau_\text{UEG}} = (27/80)[3/(2\pi)]^{1/3}[\overline{\rs}]^{-5}$ and $\overline{\lambda_\text{F}} = 2(2\pi/3)^{2/3}\overline{\rs}$.
  The uniform positive background fills the half-space $x < 0$.
   \label{fig:js_ked}}
\end{figure*}

The second-order gradient expansion for $\tau$ is unexpectedly accurate in approximating the true $\tau$ in solids.
Figure \ref{fig:js_ked} plots the exact kinetic energy density of the jellium surface, second-order gradient expansion for $\tau$, the OFR2 model derived here (after fitting, described below), and the Weizs\"acker kinetic energy density for a bulk density parameter $\overline{\rs}=2,\, 4$.
We see that OFR2 reasonably approximates $\tau$ in the jellium surface (even in its density tail), despite predicting oscillations of too small magnitude and incorrect phase.

It is also worth noting that SCAN, \rrscan, and the orbital free variants SCAN-L, \rrscan{}-L, and OFR2 are among the first meta-GGAs to respect the conjectured tight bound on the exchange energy of a spin-unpolarized density \cite{perdew2014},
\begin{equation}
  E\sx[n] \geq 1.174 E\sx^\text{LDA}[n]
\end{equation}
where $n$ is an arbitrary density.
GGAs like PBE and PBEsol \cite{perdew2008} respect a more conservative bound \cite{lieb1981,perdew1991}
\begin{equation}
  E\sx[n] \geq 1.804 E\sx^\text{LDA}[n].
\end{equation}

\subsection{Appropriate norms \label{sec:app_norms}}

Reference \cite{sun2015} described the process of selecting systems which a DFA tier can describe exactly or with high accuracy.
This idea had been used previously in, e.g., the Tao-Perdew-Staroverov-Scuseria (TPSS) meta-GGA \cite{tao2003}, which was constrained to yield the exact exchange and correlation energies of the hydrogen atom when applied to its exact density.
Such auxiliary conditions, which may be satisfied by fitting to reference densities, are necessary in the absence of a sufficient number of known conditions on the exact exchange-correlation energy functional (exact constraints).

We distinguish first-principles DFAs, which build in all possible exact constraints prior to determining free parameters with appropriate norms, from empirical functionals.
Empirical functionals need not build in exact constraints first, however when the fit is done only with appropriate norms (e.g., rare gas atoms at the GGA level), they often emerge naturally \cite{elliott2009,kaplan2020}.
Semi-empirical functionals, like the Becke 1988 exchange GGA (B88) \cite{becke1988}, build in some constraints prior to determining free parameters by fitting to data sets.

At the LSDA level, the only appropriate norm available is the uniform electron gas, for which ``The LSDA'' \cite{kohn1965,perdew1992} is exact (as opposed to empirical LSDAs \cite{jones2015}).
The GGA level can add density-gradient expansions, or the lowest-order large-$Z$ coefficients \cite{elliott2009,kaplan2020} and the exchange-correlation energies of closed-shell atoms.

LL-MGGAs cannot uniquely identify one-electron and many-electron regions as T-MGGAs can.
Some appropriate norms used to parameterize SCAN \cite{sun2015} (the compressed Ar dimer; the hydrogen and helium atoms) are not appropriate norms for an LL-MGGA, whereas others (the noble gas atoms and jellium surface formation energies) are still applicable.

Thus we select the surface formation energies of planar jellium surfaces \cite{lang1970,monnier1978}, with $\rs$ values typical of metals ($\rs =2,3, 4,$ and 5), and spherical jellium clusters \cite{almeida2002} (with typical magic numbers $N=2,8,18,20,34,40,58,92,$ and 106) as LL-MGGA appropriate norms.
From the spherical jellium clusters, we extract surface formation energies $\sigma\sxc(\rs)$ and surface curvature energies $\gamma\sxc(\rs)$ via the liquid drop model \cite{fiolhais1992}
\begin{align}
  \frac{E\sxc}{N} = & \varepsilon\sxc^\text{UEG}(\rs) + 4\pi \rs^2 \sigma\sxc(\rs) N^{-1/3} \nonumber  \\
  & + 2\pi \rs \gamma\sxc(\rs) N^{-2/3}. \label{eq:liquid_drop}
\end{align}
The surface formation energies extracted from the jellium clusters will, in general, differ from those extracted from the planar surface, although the $N\to\infty$ limit of a spherical cluster is a planar surface.
Density functionals that are more sensitive to the shell structure of small-$N$ clusters, e.g., SCAN, predict less accurate $\sigma\sxc(\rs)$ values extracted from the clusters than the surfaces.
Moreover, to limit the effects of shell-structure oscillations, we always fit the difference $(E^{\text{approx}}\sxc - E^{\text{LSDA}}\sxc)/N$, as described in Ref. \citenum{almeida2002}.

\begin{figure}
  \includegraphics[width=\columnwidth]{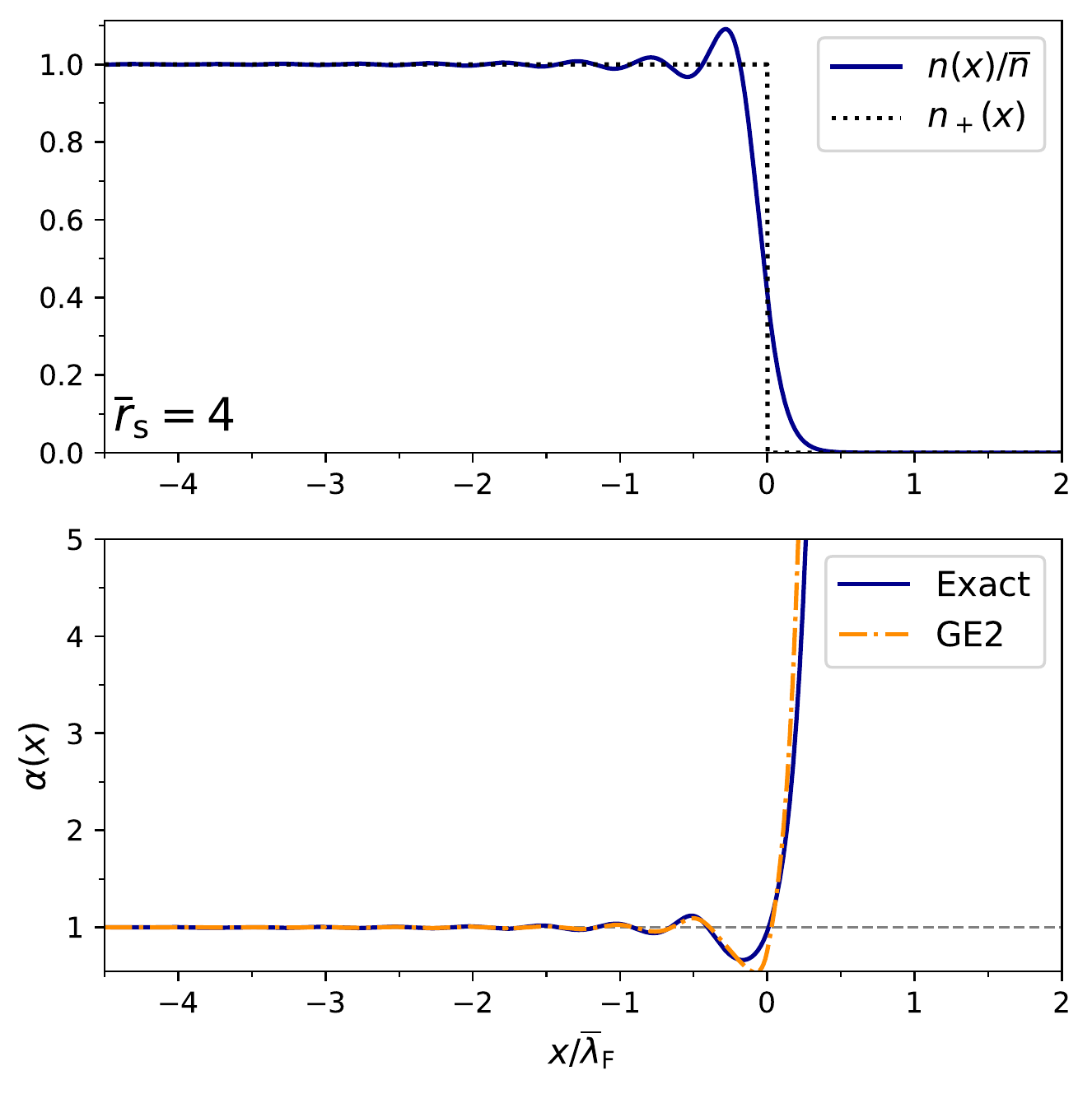}
  \caption{Upper: plot of the self-consistent LDA planar jellium surface density (blue, solid), scaled by the density of the corresponding bulk jellium $\overline{n}=3/(4\pi \overline{\rs}^3)$.
  Also shown is the neutralizing positive background (gray, dotted), which terminates at $x=0$.
  Lower: plot of the self-consistent LDA $\alpha=(\tau - \tw)/\tu$ (blue, solid) and the second-order gradient expansion (GE2) approximation for $\alpha_\text{GE2} = 1 - 40p/27 + 20q/9$ (orange, dot-dashed).
  Positions are scaled by the bulk Fermi wavevector $\overline{\lambda}_\mathrm{F} = 2\pi [4/(9\pi)]^{1/3}\overline{\rs}$, both plots are for $\overline{\rs}=4$.
  }
  \label{fig:js_rs_4}
\end{figure}

\begin{figure}
  \includegraphics[width=\columnwidth]{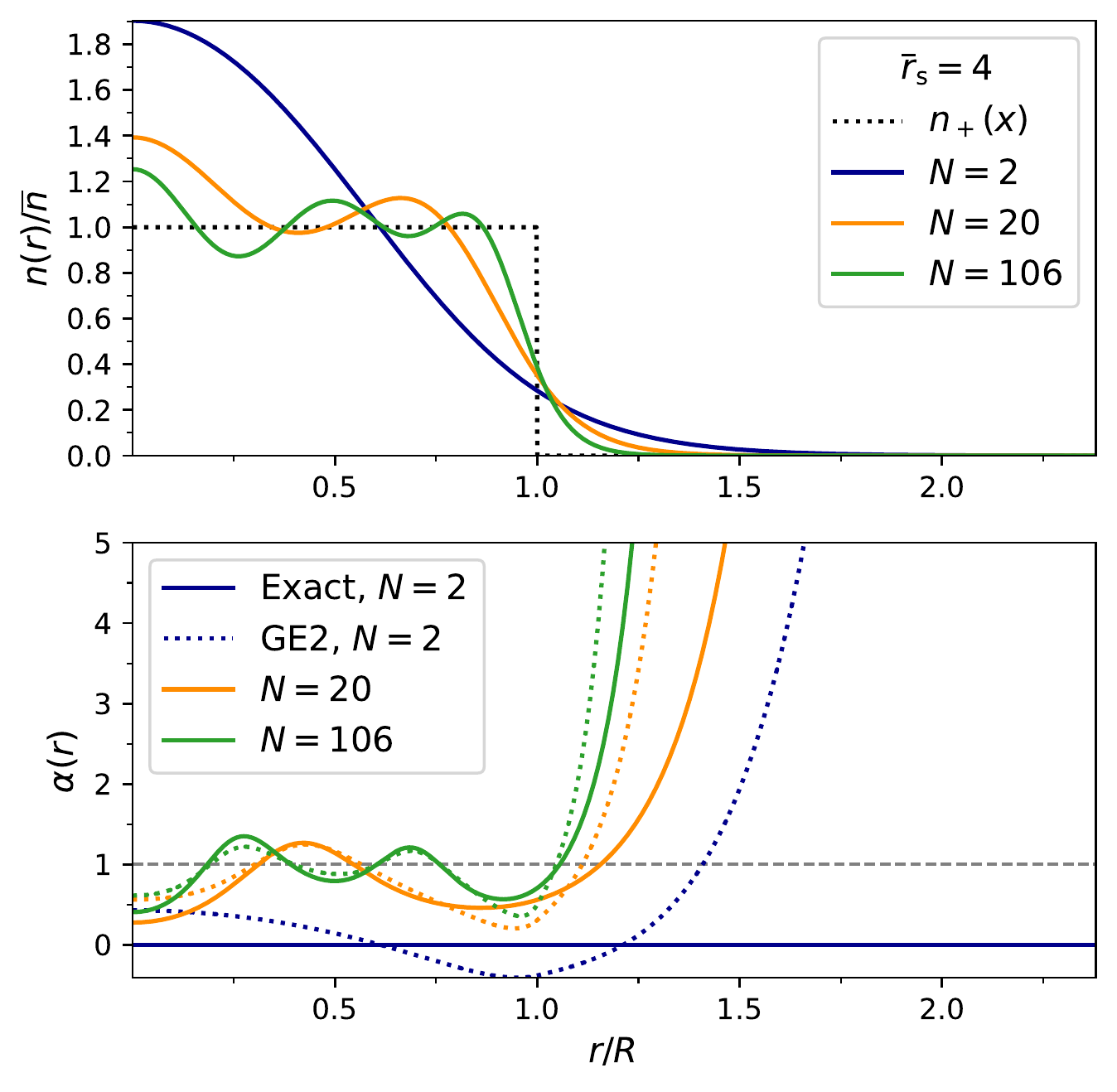}
  \caption{Upper: plot of the self-consistent LDA jellium cluster density (blue, solid), scaled by the density of the corresponding bulk jellium $\overline{n}=3/(4\pi \overline{\rs}^3)$, for a few values of $N=2$ (blue), 20 (orange), and 106 (green).
  Also shown is the neutralizing positive background (gray, dotted), which terminates at $r = R = \overline{\rs}N^{1/3}$.
  Lower: plot of the self-consistent LDA $\alpha=(\tau - \tw)/\tu$ (solid curves) and the second-order gradient expansion (GE2) approximation for $\alpha_\text{GE2} = 1 - 40p/27 + 20q/9$ (dotted curves).
  Both plots are for $\overline{\rs}=4$ bohr, as in Fig. \ref{fig:js_rs_4}.
  The GE2 only becomes relatively accurate as $N > 100$.
  }
  \label{fig:jc_rs_4}
\end{figure}

Plots of the self-consistent LDA planar jellium surface and jellium cluster densities for bulk background density-parameter $\overline{\rs}=4$ bohr can be found in Figs. \ref{fig:js_rs_4} and \ref{fig:jc_rs_4}, respectively.
These figures also plot the iso-orbital indicator $\alpha$ computed self-consistently with the LDA, and computed with the second-order gradient expansion (GE2) approximation for $\alpha$,
\begin{equation}
  \alpha_\text{GE2} = 1 - \frac{40}{27}p + \frac{20}{9}q.
  \label{eq:alpha_ge2}
\end{equation}
In these figures, $p$ and $q$ are computed from self-consistent LDA quantities.
When the GE2 is a reasonable approximation to $\alpha$, as for the planar surface in Fig. \ref{fig:js_rs_4}, a system can be considered slowly-varying, provided that $p$ and $|q|$ are both small (which we confirmed, but did not plot for reasons of clarity).

The jellium cluster densities for finite $N$ much more closely resemble the densities of atoms (see Fig. \ref{fig:cr_atom} in Sec. \ref{sec:tests}) than the planar jellium surface.
Indeed, the GE2 approximation for $\alpha$ only becomes reasonable for $N > 100$.
For $N=2$, where the exact $\alpha=0$ (iso-orbital), the GE2 is wildly off the mark, unphysically making $\alpha < 0$ near the cluster's surface.
Thus the jellium cluster densities are more characteristic of finite systems than the planar jellium surface, helping to balance the performance of OFR2.

The exchange-correlation energies of the noble gas atoms Ne, Ar, Kr, and Xe were also used as appropriate norms.
In these rare-gas atoms, and especially in their large-$Z$ limit, the exact exchange-correlation hole is reasonably short-ranged.
These atoms are needed to help RPP/OFR2 deal with nearly-iso-orbital regions like those near nuclei.
Furthermore, any error of the functional in the low-density tails of these atoms will be energetically negligible.
A Python library was written to generate self-consistent reference LSDA densities for the jellium appropriate norms, and to generate Roothaan-Hartree-Fock atomic densities \cite{bunge1993}.
The library is made available as a public code repository \cite{code_repo}.

To determine the model parameters, the objective function
\begin{equation}
  \delta = \sqrt{ \text{MAPE}^2_{\text{RGA}} + \text{MAPE}^2_{\text{JS}}
  +\text{MAPE}^2_{\text{JC}}}
\end{equation}
where ``RGA'' stands for the exchange-correlation energy of the rare-gas atoms Ne, Ar, Kr, and Xe; ``JS'' (``JC'') stands for the jellium surface (cluster) $\sigma\sxc$.
MAPE is the mean absolute percentage error.
For the planar jellium surfaces, $\rs \in \{2,3,4,5\}$ were used; for the jellium clusters, $\rs \in \{ 2,3,3.5,4,5\}$ were used.
The minimization was done in two steps: a Nelder-Mead simplex search, followed by a tiered grid search to (potentially) refine the parameters.
The fitting routine stopped when the change in the lowest $\delta$ over a few iterations stagnated.

\begin{figure}
  \begin{center}
    \includegraphics[width=\columnwidth]{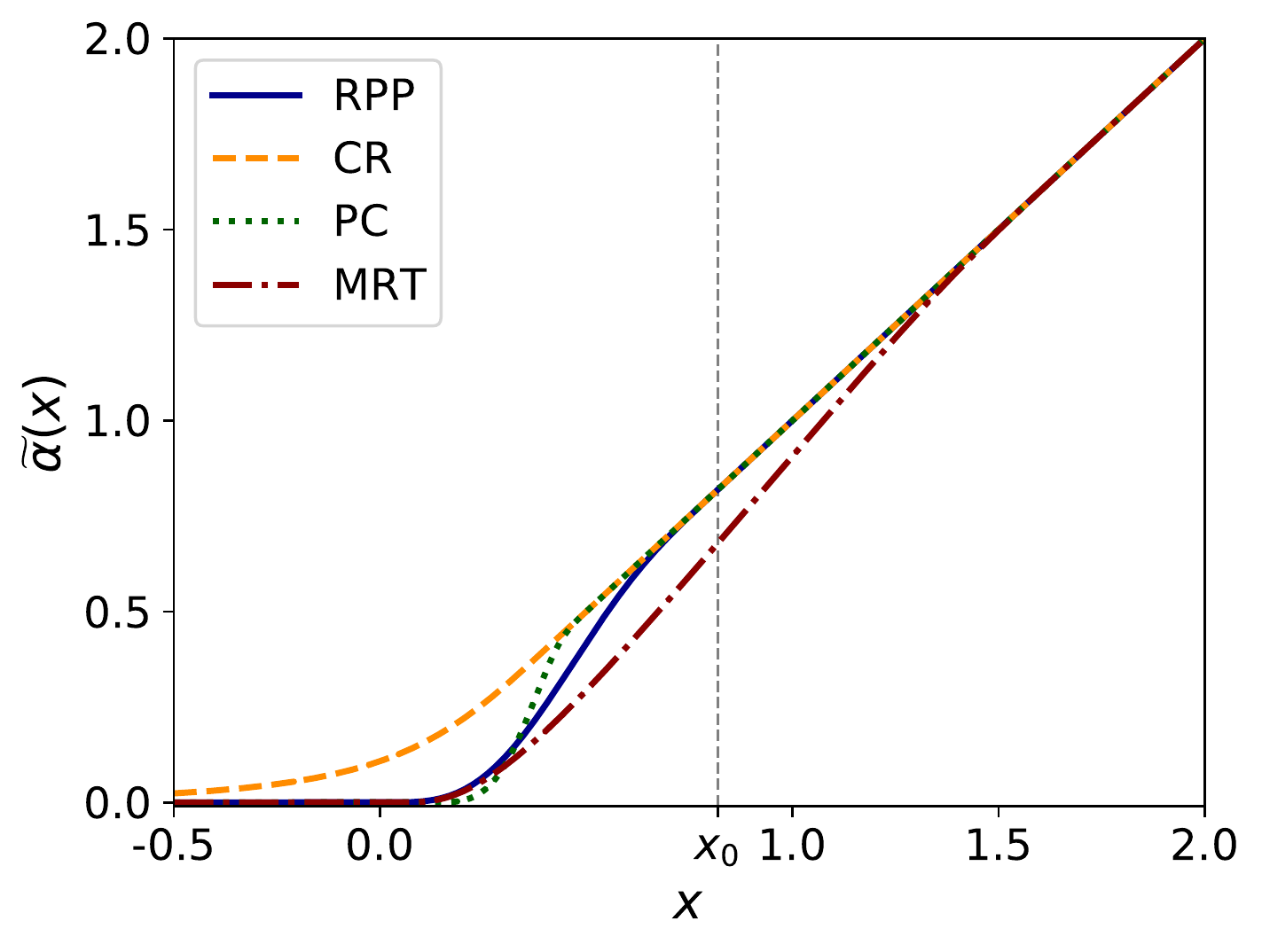}
  \end{center}
  \caption{Plot of the RPP model $\widetilde{\alpha}(x)$ of Eq. (\ref{eq:alphas}) as a function of an arbitrary measure of inhomogeneity $x$, which tends to one for a uniform density.
  The Perdew and Constantin (PC) \cite{perdew2007}, Cancio and Redd (CR) \cite{cancio2017}, and Mej\'ia-Rodr\'iguez and Trickey (MRT) \cite{mejia2017} models of $\widetilde{\alpha}(x)$ are also displayed.
   \label{fig:alphas}}
\end{figure}

A plot of the $\widetilde{\alpha}(x)$ function, compared with similar models \cite{perdew2007,cancio2017,mejia2017}, is given in Fig. \ref{fig:alphas}.
While the PC, MRT, and RPP models do not share a common inhomogeneity measure $x$, they assume that $x=1$ indicates a uniform density, $x\to \infty$ a density-tail, and $x \to -\infty$ a core.
Thus we can compare them using an arbitrary inhomogeneity measure $x$.
The Cancio-Redd model
\begin{align}
  \widetilde{\alpha}^\text{CR}(z^\text{CR}) &= 1
    + z^\text{CR}\{1 - \exp[-1/|z|^a]\}^{1/a} \Theta(-z^\text{CR}) \nonumber\\
  & + z^\text{CR} \Theta(z^\text{CR}) \\
  z^\text{CR} &= -\frac{40}{27}p + \frac{20}{9}q \\
  \Theta(z) &= \left\{\begin{array}{rr} 1 & z \geq 0 \\ 0 & z < 0 \end{array} \right.
\end{align}
with $a = 4$, tends to its uniform density limit when its inhomogeneity measure  $z^\text{CR}$ tends to zero, unlike the PC, MRT, and RPP models.
Thus we plot $\widetilde{\alpha}^\text{CR}$ as a function of $x\equiv z^\text{CR} + 1$, where $x \to 1$ indicates a uniform density.
The RPP model recovers the fourth-order gradient expansion for \textit{exchange} when combined with \rrscan.
The RPP, PC, and CR models all recover the second-order gradient expansion for $\tau$ by construction, whereas the MRT model does not.
This is seen in Fig. \ref{fig:alphas} by noting that $\widetilde{\alpha}(x\approx 1)\approx x$.

\begin{ruledtabular}
  \begin{table}[t]
    \begin{center}
      \begin{tabular}{p{2cm}p{1.2cm}p{1.2cm}p{1cm}}
        Atomic Norm & Reference (hartree) & OFR2 (hartree) & Percent error \\ \hline
        Ne & -12.499 & -12.229 & -2.16\% \\
        Ar & -30.913 & -30.326 & -1.90\% \\
        Kr & -95.740 & -94.308 & -1.50\% \\
        Xe & -182.202 & -179.837 & -1.30\% \\
        & & MAPE & 1.71\% \\ \hline
        Jellium surface $\rs$ (bohr) & Reference (erg/cm$^{2}$) & OFR2 (erg/cm$^{2}$) & Percent error \\ \hline
        2 & 3413 & 3336 & -2.25\% \\
        3 & 781 & 764 & -2.16\% \\
        4 & 268 & 265 & -1.19\% \\
        5 & 113 & 116 & 2.25\% \\
        & & MAPE & 1.96\% \\ \hline
        Jellium cluster $\rs$ (bohr) & Reference (erg/cm$^{2}$) & OFR2 (erg/cm$^{2}$) & Percent error \\ \hline
        2 & 3413 & 3363 & -1.47\% \\
        3 & 781 & 769 & -1.57\% \\
        3.25 & 582 & 578 & -0.84\% \\
        4 & 268 & 265 & -1.05\% \\
        5 & 113 & 116 & 2.98\% \\
        & & MAPE & 1.58\% \\
      \end{tabular}
    \end{center}
    \caption{Performance of the new orbital-free \rrscan (OFR2) for the appropriate norms.
    The reference atomic exchange-correlation energies are taken from Refs. \cite{santra2019,burke2016}, respectively.
    Reference jellium surface exchange-correlation formation energies are taken from the RPA+ values of Ref. \cite{yan2000}, and when needed, the fit to RPA+ data of Ref. \cite{almeida2002}.}
    \label{tab:appnorms}
  \end{table}
\end{ruledtabular}

Table \ref{tab:appnorms} shows the appropriate norms errors used to determine $x_0$, $c_1$, $c_2$, and $c_3$ (Eqs. \ref{eq:x_0_fit}--\ref{eq:c_3_fit}).
We use the RPA+ \cite{yan2000}, and the fit from Ref. \cite{almeida2002} as needed, as reference values for $\sigma\sxc$.
The RPA alone accounts for 100\% of exact exchange and the long-range part of correlation in a metal like the jellium surface.
The RPA+ makes a GGA-level correction to the RPA correlation energy at short range.
Thus the values of $\sigma\sxc$ found with the RPA+ are comparable to higher-level methods like the Singwi-Tosi-Land-Sj\"olander self-consistent spectral function method \cite{constantin2008}, or careful quantum Monte Carlo (QMC) calculations of finite jellium surfaces \cite{wood2007}.
Reference atomic exchange energies are taken from Ref. \cite{santra2019}, and correlation energies from Ref. \cite{burke2016}.

\section{Performance for real systems \label{sec:tests}}

OFR2 is constructed to accurately describe metallic densities.
While this is a niche goal, T-MGGAs adequately describe non-metallic densities, but exhibit too much non-locality for simple metallic solids.
This deficit can be rectified by an LL-MGGA like OFR2.

\begin{figure*}
  \centering
  \includegraphics[width=0.8\textwidth]{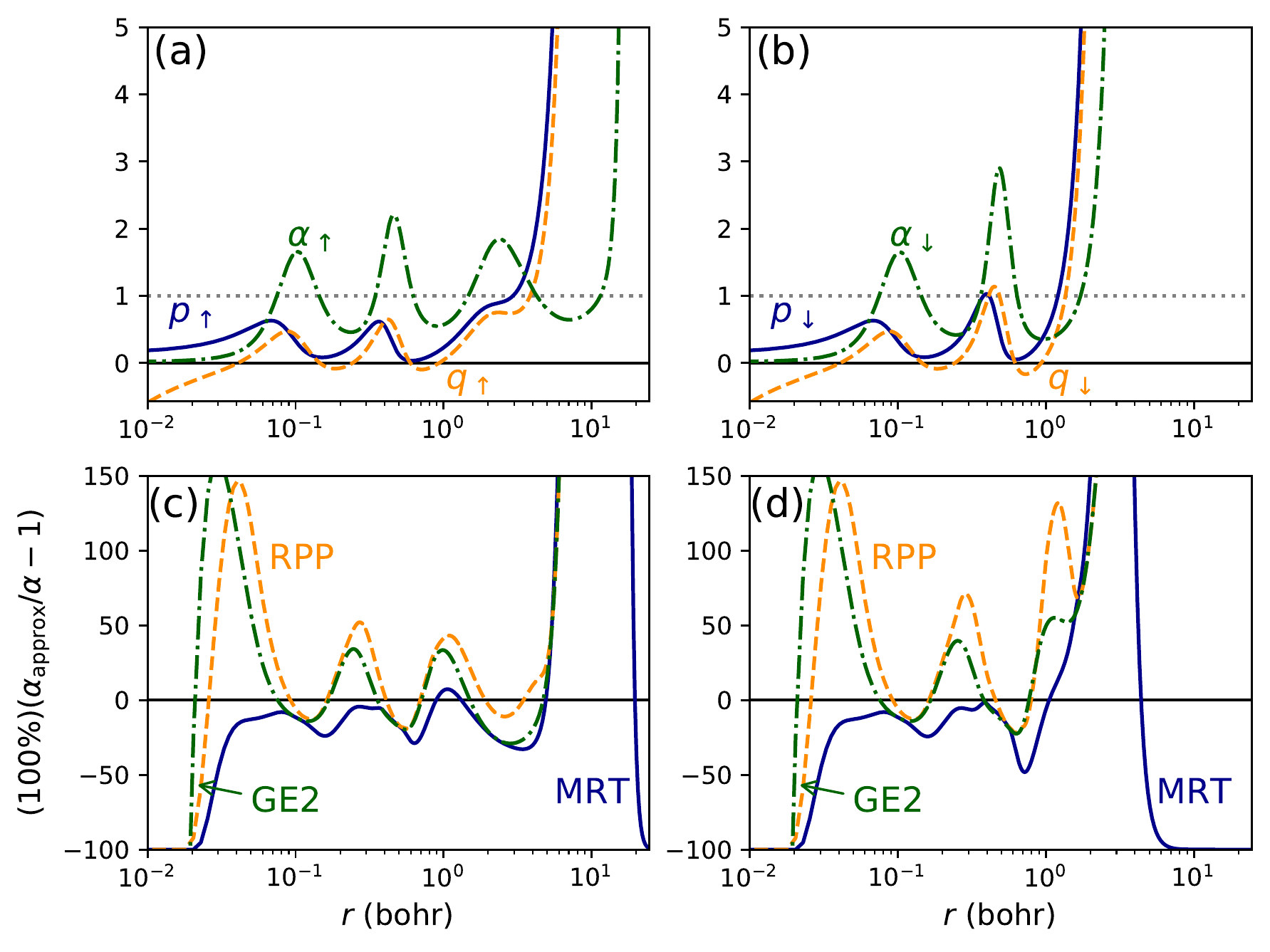}
  \caption{
  Upper: squared dimensionless density gradient $p$ (blue, solid), dimensionless Laplacian $q$ (orange, dashed), and iso-orbital indicator $\alpha$ (green, dot-dashed) in the Cr atom for the (a) up-spin ($\uparrow$) density, and (b) down-spin ($\downarrow$) density.
  The density, its derivatives, and kinetic energy density are spherically averaged after sampling 200 azimuthal points on a Gauss-Legendre grid, using Roothaan-Hartree-Fock Slater-type orbitals from Ref. \cite{koga1999}.
  Lower: the percent error, $100\left(\frac{\alpha_\mathrm{approx}}{\alpha}-1\right)$, made by the model of $\alpha$ from Ref. \cite{mejia2017} (MRT; blue, solid) and the present model, RPP (orange, dashed), for the (c) up-spin density and (d) down-spin density.
  Also shown is the second order gradient expansion, GE2 (green, dot-dashed).
  When $p\ll 1$, $|q|\ll 1$, and $|1-\alpha|\ll 1$, the density can be considered slowly-varying, and a semilocal model of $\tau$ can be approximately accurate.
  \label{fig:cr_atom}
  }
\end{figure*}

Panels (a) and (b) of Fig. \ref{fig:cr_atom} plot $p$, $q$, and $\alpha$ in the Cr atom for the up- and down-spin densities, respectively.
Note the similarity of $p$ and $q$ outside the 1$s$ shell of the atom.
In the region $0.07 \lesssim r \lesssim 2$ bohr, both $p$ and $|q|$ are less than one, and there are numerous points where $\alpha = 1$.
The density in this region would thus be characterized as approximately slowly-varying or metallic by a T-MGGA.
We define the spin-dependent variables as
\begin{align}
  p_\sigma &= p(2n_\sigma) = 2^{-2/3} \frac{|\nabla n_\sigma|^2}{4(3\pi^2)^{2/3}n_\sigma^{8/3}} \\
  q_\sigma &= q(2n_\sigma) = 2^{-2/3} \frac{\nabla^2 n_\sigma}{4(3\pi^2)^{2/3}n_\sigma^{5/3}} \\
  \alpha_\sigma &= \alpha(2n_\sigma,2\tau_\sigma)
    = 2^{-2/3} \frac{\tau_\sigma - |\nabla n_\sigma|^2/(8n_\sigma)}
    {3(3\pi^2)^{2/3} n_\sigma^{5/3} /10},
\end{align}
i.e., the density variables as seen by the exchange energy using its spin-scaling relation \cite{oliver1979}.

Panels (c) and (d) of Fig. \ref{fig:cr_atom} plot the errors made in approximating $\alpha$ with the MRT model \cite{mejia2017} and the RPP model, Eq. \ref{eq:fs}.
Because $p$ and $|q|$ are small, the second-order gradient expansion (GE2),
\begin{equation}
  \tau_\sigma = \left(1 + \frac{20}{9}q_\sigma + \frac{5}{27} p_\sigma \right)
    \tu(n_\sigma)
\end{equation}
is a reasonable approximation to $\tau$ in the region $0.07 \lesssim r \lesssim 2$ bohr only.
RPP closely follows the GE2 curve in this region.
These semi-local models of $\alpha$ better describe this region than the 1$s$ shell region, where they make $\alpha$ vanish too abruptly, or the density tail, where they make $\alpha$ diverge too quickly.
For the Cr atom, the MRT model better approximates $\alpha_\sigma$ than the RPP model of this work, except perhaps for the majority ($\uparrow$) spin in the valence region.

\subsection{Numerical stability}

\begin{figure}
  \begin{center}
    \includegraphics[width=\columnwidth]{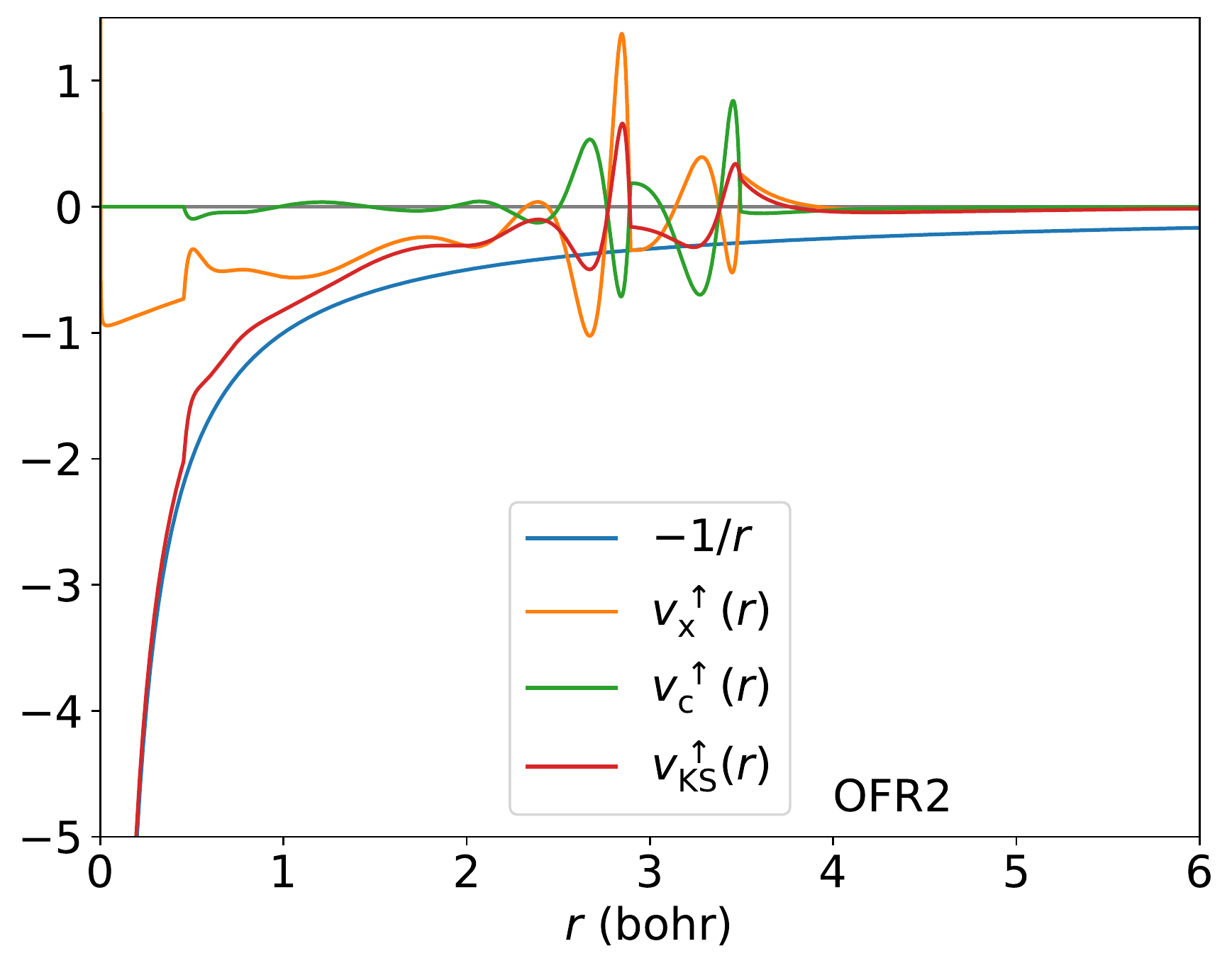}
  \end{center}
  \caption{ OFR2 Kohn-Sham potential calculated used Eq. (\ref{eq:vxc}) for the up-spin channel, evaluated on the exact density, $ n(r) = n_\uparrow(r) = e^{-2r}/\pi$ ($v\sxc^\downarrow=0$ identically for this system).
  A 6$^\text{th}$ order finite difference was used to calculate the requisite divergence and Laplacian terms.
  Oscillations are primarily due to inclusion of the density-Laplacian.
  \label{fig:ofr2_hatom_vks}}
\end{figure}

The LL-MGGA exchange-correlation potential is very sensitive to the dependence of $e\sxc$ on the density Laplacian.
Figure \ref{fig:ofr2_hatom_vks} demonstrates this for the hydrogen atom ($\alpha_\uparrow=0$) Kohn-Sham potential, using the exact density $n(r) = e^{-2r}/\pi$.
$v\sxc$ presents unusual oscillations that could be misinterpreted as shell structure.
Using this density,
\begin{align}
  \kf(r) &= (3\pi)^{1/3}e^{-2r/3} \\
  p(r) &= \kf^{-2} \\
  q(r) &= (1 - 1/r)\kf^{-2}.
\end{align}
Similar to the Cr atom in Fig. \ref{fig:cr_atom}, there is a region near $r = 1$ bohr that an LL-MGGA can mistakenly identify as slowly-varying, because $p\lesssim 1$, and $|q|\approx 0$.
This induces an artificial shell structure not seen in the semi-local part of the \rrscan Kohn-Sham potential \cite{furness2020}.
A sixth-order finite difference was used to evaluate $\nabla \cdot \left[\partial e\sxc/\partial (\nabla n_\sigma)\right]$ and $\nabla^2 \left[\partial e\sxc/\partial (\nabla^2 n_\sigma)\right]$.
The derivatives of $e\sxc$ with respect to $n$, $\nabla n$ and $\nabla^2 n$ were computed analytically.

\begin{figure}
  \centering
  \begin{center}
    \subfloat[\label{fig:js_rs_2_vxc_rpp}]{
      \includegraphics[width=0.48\textwidth]{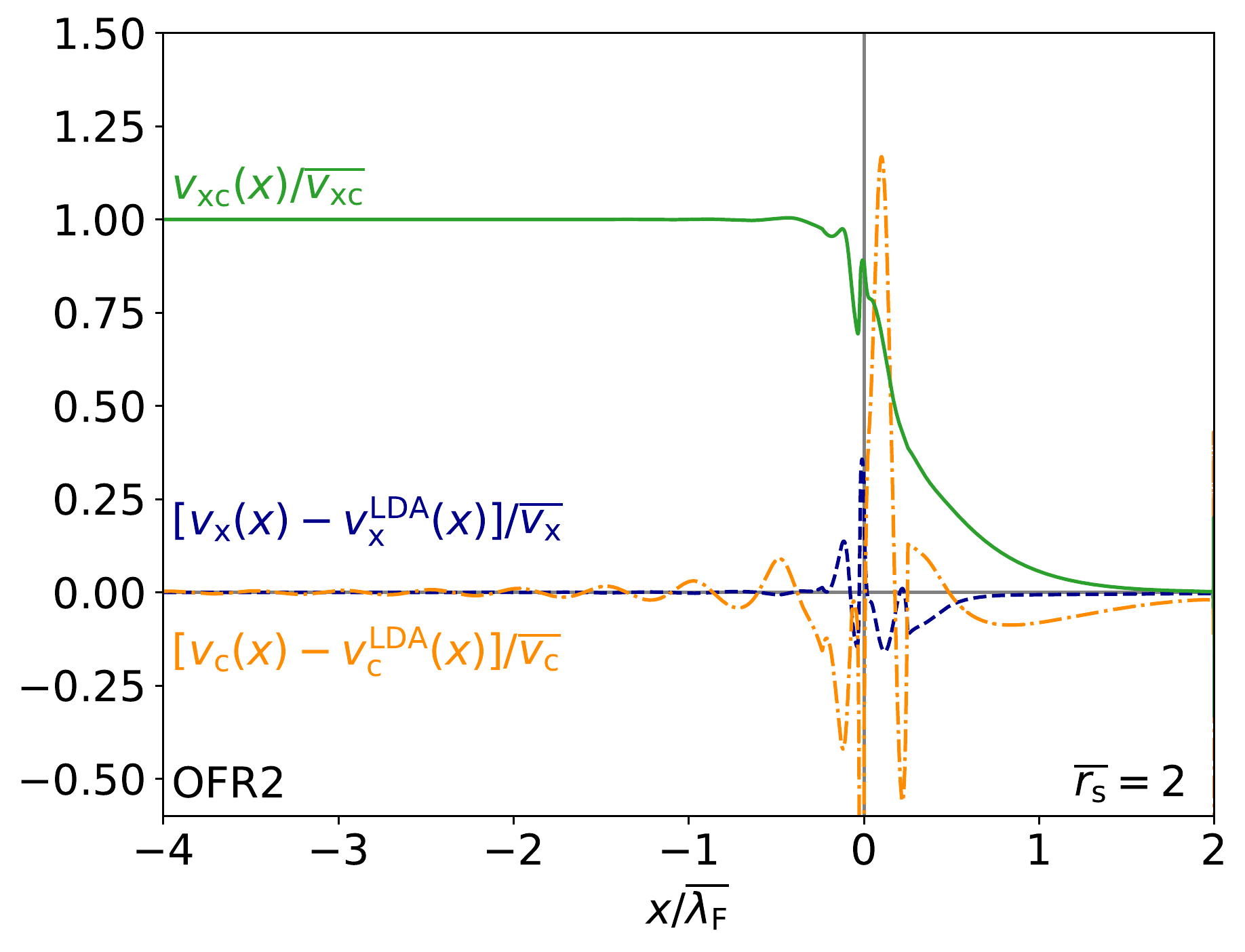}
    }
    \hfill
    \subfloat[\label{fig:js_rs_2_vxc_mrt}]{
      \includegraphics[width=0.48\textwidth]{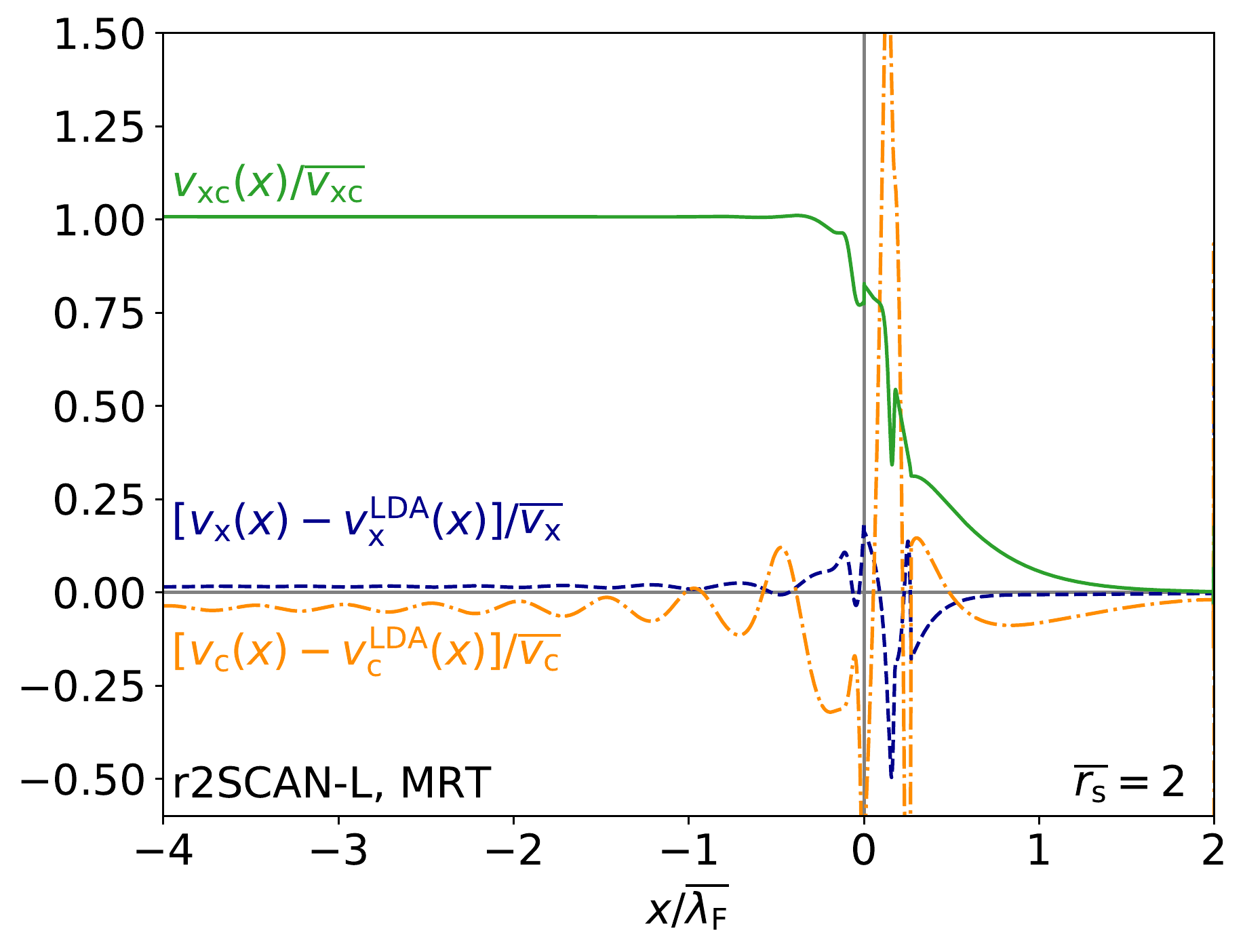}
    }
  \end{center}
  \caption{ The exchange and correlation potential in an $\overline{\rs}=2$ jellium surface, evaluated on the same LSDA densities used previously.
  The present OFR2 (RPP) (top, \ref{fig:js_rs_2_vxc_rpp}) and \rrscan{}-L (MRT) \cite{mejia2017} (bottom,\ref{fig:js_rs_2_vxc_mrt}) LL-MGGA potentials are shown.
  The same finite difference coefficients as in Fig. \ref{fig:ofr2_hatom_vks} were used to generate these plots.
  As before, the edge of the uniform positive background lies at $x=0$, and $x$ is scaled by the bulk Fermi wavelength, $\overline{\lambda}_\text{F}=2\pi/\overline{\kf}$.
  The potential is scaled by the corresponding LSDA potential evaluated at the bulk density.
  \label{fig:js_rs_2_vxc}}
\end{figure}

\begin{figure}
  \begin{center}
    \includegraphics[width=\columnwidth]{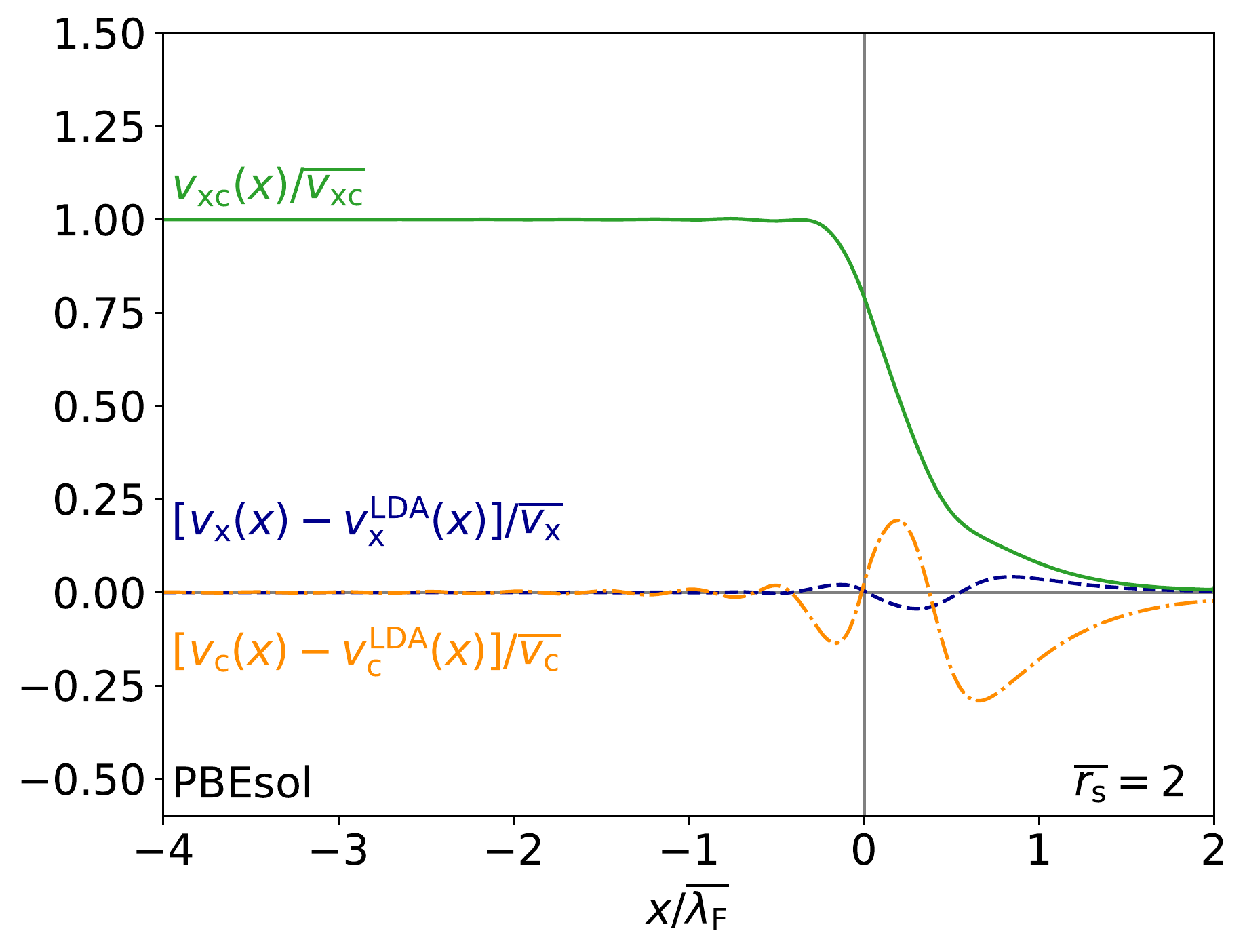}
  \end{center}
  \caption{Same as Fig. \ref{fig:js_rs_2_vxc}, but plotting the PBEsol exchange and correlation potentials evaluated on the LSDA density.
  \label{fig:js_rs_2_vxc_pbesol}}
\end{figure}

Similarly, Fig. \ref{fig:js_rs_2_vxc} plots the finite difference exchange and correlation potentials in a jellium surface with $\rs=2$, for OFR2 and and \rrscan{}-L.
As in the other calculations of the jellium surface, reference LSDA densities were used.
Both models manifest unphysical oscillations in the exchange and correlation potentials, which can be compared to the PBEsol potentials shown in Fig. \ref{fig:js_rs_2_vxc_pbesol} (using the same density).
PBEsol is expected to yield reasonable predictions of jellium surface properties by construction.
Despite the alarming appearance of Figs. \ref{fig:ofr2_hatom_vks} and \ref{fig:js_rs_2_vxc}, the method used by VASP to solve the generalized Kohn-Sham equations, summarized in Appendix \ref{app:wb_method}, is numerically efficient and stable.
It is clear, without plotting the associated electrostatic potential, that the oscillations in the LL-MGGA exchange-correlation potentials will be significant.

\subsection{Lattice constants \label{sec:latt_cons}}

All solid-state calculations were performed in the Vienna \textit{ab initio} Simulation Package (VASP) \cite{kresse1993,kresse1994,kresse1996,kresse1996a}, version 6.1.
We used a $\Gamma$-centered $\bm{k}$-point mesh of spacing 0.08 \AA{}$^{-1}$, with a plane-wave energy cutoff of 800 eV, except for a few cases, which we discuss below.
Energies were converged below $10^{-6}$ eV, and calculated using the Bl\"ochl tetrahedron method \cite{blochl1994}.
For reasons of numerical stability, ADDGRID was set to False.
Equilibrium structures were determined using the stabilized jellium equation of state (SJEOS) \cite{alchagirov2001,staroverov2004}.
12 single-point energy calculations in a range of $(1 \pm 0.1)V_\text{expt.}$, with $V_\text{expt.}$ the experimental (zero-point energy corrected) equilibrium volume were performed.
To fit hcp structures (hcp Co is discussed in Sec. \ref{sec:ferro}), we optimized the $c/a$ packing ratio at fixed volume, and found the optimal $c/a$ by fitting to a reduced SJEOS.
All input files can be found in the code repository.

Some of the standard VASP pseudopotentials cannot accommodate higher plane-wave energy cutoffs.
For example, ``PAW\_PBE Ba\_sv 06Sep2000'' (``PAW\_PBE Pd 04Jan2005'') can accommodate a maximum energy cutoff of about 600 eV (750 eV).
Both settings were used here instead of the 800 eV cutoff used for the other solids.
The LL-MGGAs exhibited a strong dependence on the number of bands used when the cutoff was exceeded, whereas the GGAs and T-MGGAs did not appear to be similarly affected.

Table \ref{tab:lc20} displays the relative error statistics in 20 cubic lattice constants (the LC20 set) \cite{sun2011} made by a variety of common, first-principles functionals: PBEsol \cite{perdew2008} (a benchmark GGA for this property), \rrscan \cite{furness2020}, \rrscan{}-L \cite{mejia2020} and OFR2.
Tables \ref{tab:lc20_full} and \ref{tab:lc20_b0_full} of Appendix \ref{app:lc20_data} present errors in the lattice constants and bulk moduli, respectively, for each solid in the LC20 set.

OFR2 exceeds the performance of \rrscan and \rrscan{}-L overall, for both metals and insulators in the set of lattice constants.
There are unusual cases where a LL-MGGA that is designed to mimic its parent T-MGGA, as \rrscan{}-L is, outperforms it: see the SCAN and SCAN-L binding energy of hexagonal BN and graphite out-of-plane lattice constant in Table VI of Ref. \cite{mejia2018}.
As OFR2 is not designed to mimic \rrscan{}, we find its superior performance for solid-state geometries less surprising.
However, \rrscan and PBEsol predict more accurate bulk moduli than do either of the orbital-free \rrscan meta-GGAs.

The lattice-constant results show the bias inherent in each meta-GGA's construction.
\rrscan{}-L does not have the correct uniform density limit and gradient expansion constraint that are critical to an accurate description of metallic condensed matter (those systems most like an electron gas with weak variations about a uniform density).
One might argue that the 10\% violation of the uniform density limit (see Eq. \ref{eq:tau_mrt_ge2}) is small even in the jellium surface exchange-correlation potential plot of Fig. \ref{fig:js_ked_b}.
However, it is clear that the loss of this limit is indeed important for accurate solid-state geometries.
The data used to fit \rrscan{}-L were biased toward finite systems (the 18 lightest neutral atoms were used to fit the PCopt model of $\tau$ \cite{mejia2017}).
OFR2 recovers the uniform density limit constraint of \rrscan{}, the second-order gradient expansion for correlation, and the fourth-order gradient expansion for exchange.
While the rare gas atoms were included in the training set of OFR2, this was done to prevent overfitting to the jellium norms, and does not ensure that OFR2 accurately describes finite systems.
This biases the construction of OFR2 toward solid-state properties.
Therefore, the \rrscan{}-L results show stronger performance for the lattice-constants of insulating solids than for those of the metals.
OFR2 is constructed in the spirit of PBEsol, and shows a large gain in performance over its parent functional \rrscan.

However an obvious question remains: Why do PBEsol and OFR2 describe the structures of insulators more accurately than PBE (a GGA with a slight bias towards molecules) and \rrscan{}-L?
Narrow-gap insulators (e.g., Si, Ge, GaAs), covalently bonded insulators (e.g. C and SiC), and ``strongly-correlated'' monoxides (e.g., MgO) have no classical turning surfaces in the Kohn-Sham potentials near equilibrium, whereas ``normally-correlated'' ionically-bound solids (e.g., LiF, LiCl, NaF, NaCl) do \cite{kaplan2021}.
The gradient expansions for the exchange and correlation energies are semiclassical in nature, and thus can only be valid inside a classical turning surface.
The lack of a turning surface permits these gradient expansions, which are preserved in PBEsol and OFR2 but not PBE and \rrscan{}-L, to have some validity for non-metallic solids.
There are caveats which we will discuss further in Sec. \ref{sec:alkalis}.

\begin{ruledtabular}
  \begin{table}
    \begin{center}
      \begin{tabular}{L{1.1cm}R{1.2cm}R{1.2cm}R{1.2cm}R{1.55cm}R{1.2cm}}
        (\AA{}) & PBEsol & SCAN & \rrscan & \rrscan{}-L & OFR2 \\
        \multicolumn{6}{c}{Metals} \\ \hline
        ME & -0.044 & 0.004 & 0.024 & 0.011 & -0.020 \\
        MAE & 0.044 & 0.021 & 0.033 & 0.044 & 0.021 \\
        \multicolumn{6}{c}{Insulators} \\ \hline
        ME & 0.024 & 0.004 & 0.017 & 0.016 & 0.005 \\
        MAE & 0.025 & 0.008 & 0.017 & 0.016 & 0.014 \\
        \multicolumn{6}{c}{Total} \\ \hline
        ME & -0.010 & 0.004 & 0.020 & 0.013 & -0.007 \\
        MAE & 0.035 & 0.015 & 0.025 & 0.030 & 0.018 \\
      \end{tabular}
    \end{center}
    \caption{Mean error (ME) and mean absolute error (MAE) statistics for 20 common cubic lattice constants (LC20) \cite{sun2011}, all in \AA{}.
    Subsets of metals and insulators are also shown.
    None of the OFR2 calculations failed to converge in the allotted number of self-consistency iterations (200 for each single-point calculation).
    Six (of the 240 total) \rrscan{}-L calculations failed to converge to $10^{-6}$ eV in 200 self-consistency steps.
    Troublesome convergence is a common issue for LL-MGGAs, and has been observed previously \cite{mejia2020}.
    Reference experimental equilibrium lattice constants (with zero-point corrections included) are taken from Ref. \cite{hao2012}.
     \label{tab:lc20}}
  \end{table}
\end{ruledtabular}

We derive a symmetric expression for the Laplacian contributions to the stress tensor in Appendix C.
The total exchange-correlation stress tensor $\Sigma\sxc^{ij}$, in a gauge appropriate for a code with periodic boundary conditions, is given by Eq. \ref{eq:xc_st_ibp}, reprinted here
\begin{align}
  \Sigma\sxc^{ij} &= \int \left[\left(e\sxc - v\sxc n \right)\delta_{ij}
    - \frac{1}{|\gn|}\dd{n}{r_i} \dd{n}{r_j} \dd{e\sxc}{|\gn|}
    \right.  \label{eq:xc_st_ibp_reprint} \\
  & \left. - 2 \dd{e\sxc}{\lan} \frac{\partial^2 n}{\partial r_i \partial r_j} \right] d^3 r.
  \nonumber
\end{align}
Here, $r_1=x$, $r_2=y$, and $r_3=z$, $e\sxc$ is the exchange-correlation energy density such that $E\sxc=\int e\sxc d^3r$, and $v\sxc$ is the exchange-correlation potential, Eq. \ref{eq:vxc}.
To use the stress tensor to minimize structures, we used a few additional computational parameters, keeping the others unchanged.
The magnitudes of forces were converged within $0.001$ eV/\AA{}.

By setting ISIF = 3, the ion positions, computational cell shape, and computational cell volume were permitted to relax; we verified that no change of symmetry occurred during the force minimization.
Generally, ISIF controls which degrees of freedom are permitted to relax, and if all elements or just the diagonal elements of the stress tensor are computed.
The minimization algorithm is controlled by the IBRION setting; we used the conjugate gradient algorithm, IBRION = 2.
First order Methfessel-Paxton smearing \cite{methfessel1989} (chosen by setting ISIGMA = 1) with width 0.2 eV was used for the metals (and Ge for PBEsol and \rrscan{}-L), Gaussian smearing of width 0.05 eV was used for the insulators.
ISIGMA selects a method for smearing electronic states near the Fermi level.
We refer the reader to the VASP manual \cite{vasp_manual} for other options.

The mean deviations in the LC20 lattice constants found by the equation of state fitting and by minimization of the stress tensor in VASP are presented in Tables \ref{tab:lc20_eos_st_md} and \ref{tab:lc20_eos_st_full}.
These tables also present results for PBEsol and \rrscan to benchmark how closely the lattice constants found from both methods agree.
The Laplacian-dependent stress tensor appears to agree to the same level of precision as the GGA and T-MGGA stress tensor.

\begin{table}
  \centering
  \begin{tabular}{l|rrrr} \hline \hline
     & PBEsol & \rrscan & \rrscan{}-L & OFR2 \\ \hline
    MD & $7.191 \times 10^{-4}$ & $7.499 \times 10^{-4}$ & $7.132 \times 10^{-3}$ & $3.598 \times 10^{-3}$ \\
    MAD & $2.013 \times 10^{-3}$ & $1.729 \times 10^{-3}$ & $8.073 \times 10^{-3}$ & $4.656 \times 10^{-3}$ \\ \hline
  \end{tabular}
  \caption{Mean deviation (MD) and mean absolute deviation (MAD) in the LC20 cubic lattice constants found by equation of state (EOS) fitting to the SJEOS and by minimization of the stress tensor (ST).
  From the PBEsol and \rrscan values, these lattice constants should agree to better than $10^{-2}$ \AA{} on average, which is satisfied.
  The deviations are $a_0^\text{EOS} - a_0^\text{ST}$.
  }
  \label{tab:lc20_eos_st_md}
\end{table}

\subsection{Transition metal magnetism \label{sec:ferro}}

As is well known by now \cite{ekholm2018,fu2018,mejia2019a}, some of the most sophisticated T-MGGAs predict correct structures for transition metals, but too large magnetic moments.
Previous works studied the simplest ferromagnetic materials: body-centered cubic (bcc) Fe, face-centered cubic (Ni), and hexagonal close-packed (hcp) Co.

Table \ref{tab:fm_mats} compares PBEsol, \rrscan \cite{furness2020}, \rrscan{}-L \cite{mejia2020}, and OFR2.
Consistent with Ref. \cite{mejia2019a}, OFR2 strikes a balance between the GGA and meta-GGA levels by providing more accurate geometries than PBEsol, and more accurate magnetic moments than \rrscan.
\rrscan{}-L and OFR2 are comparably accurate for these solids.

\begin{ruledtabular}
  \begin{table}[t]
    \begin{center}
      \begin{tabular}{p{1.2cm}lrrr}
        Solid (structure) & Functional & $a$ (\AA{}) & & $m\sus$ ($\mu_\text{B}$/atom) \\ \hline
        \multirow{5}{3cm}{Fe (bcc)} & PBEsol & 2.783 & & 2.094 \\
         & \rrscan & 2.864 & & 2.64 \\
         & \rrscan{}-L & 2.827 & & 2.20 \\
         & OFR2 & 2.791 & & 2.12 \\
         & Expt. & 2.855 & & 1.98 -- 2.13 \\ \hline
        \multirow{5}{3cm}{Ni (fcc)} & PBEsol & 3.465 & & 0.620 \\
         & \rrscan & 3.478 & & 0.74 \\
         & \rrscan{}-L & 3.500 & & 0.67 \\
         & OFR2 & 3.463 & & 0.66 \\
         & Expt. & 3.509 & & 0.52 -- 0.57 \\ \hline
         &  & $a$ (\AA{}) & $c/a$ & $m\sus$ ($\mu_\text{B}$/atom) \\ \hline
        \multirow{5}{3cm}{Co (hcp)} & PBEsol & 2.455 & 1.615 & 1.57 \\
         & \rrscan & 2.471 & 1.623 & 1.74 \\
         & \rrscan{}-L & 2.494 & 1.623 & 1.66 \\
         & OFR2 & 2.468 & 1.623 & 1.63 \\
         & Expt. & 2.503 & 1.621 & 1.52 -- 1.58 \\
      \end{tabular}
    \end{center}
    \caption{Comparison of structural and magnetic predictions for itinerant electron ferromagnets.
    Total energies for \rrscan and OFR2 are converged to 10$^{-6}$ eV.
    Total energies for \rrscan{}-L are converged to 10$^{-4}$ eV (the default for VASP); this is done for reasons of numerical stability.
    The experimental (expt.) equilibrium cubic lattice constants ($a$) are taken from Ref. \cite{hao2012}, and experimental zero-temperature extrapolated lattice constants for hcp Co are taken from Ref. \cite{ono1988}.
    The ranges of experimental magnetic moments ($m\sus$ in units of the Bohr magneton $\mu_\text{B}$ per atom) are taken from Ref. \cite{ekholm2018}.}
    \label{tab:fm_mats}
  \end{table}
\end{ruledtabular}

\subsection{Bandgaps \label{sec:insul}}

In a standard Kohn-Sham calculation, the exact exchange-correlation functional would lead to an underestimation of the fundamental (charge) bandgap equal to the ``exchange-correlation derivative discontinuity'' \cite{perdew1982}.
Even though GGAs like PBE may closely approximate the exact Kohn-Sham bandgap \cite{kaplan2021}, only functionals defined within a generalized Kohn-Sham (GKS) theory with nonzero derivative discontinuity can realistically estimate the observed fundamental bandgap \cite{perdew2017}.
For this reason, some T-MGGAs, which are orbital-dependent and thus defined within a GKS theory, can provide surprisingly reliable estimates of the bandgap \cite{aschebrock2019,neupane2021}.
Similarly, hybrid functionals reliably predict accurate bandgaps \cite{wing2021}, as single-determinant exchange is an explicit functional of the Kohn-Sham orbitals.

As LL-MGGAs are standard Kohn-Sham DFAs lacking a derivative discontinuity, we expect them to underestimate the fundamental bandgap.
This was shown in Ref. \cite{mejia2018} using SCAN-L.
Table \ref{tab:gaps} tabulates the bandgaps for a subset of the LC20 set of solids.
To compute the bandgap, the equilibrium lattice constants from Table \ref{tab:lc20_full} were used as input to a single-point total energy calculation.
From this, the Fermi energy was extracted, and a new density of states (DOS) grid was defined centered at the Fermi energy, evenly spaced in intervals of 0.01 eV.
The calculation was then repeated with the finer DOS grid.
A general-purpose functional should be able to reliably predict lattice parameters and bandgaps, thus we prefer to evaluate the bandgap using each DFA's relaxed structure.

\begin{ruledtabular}
  \begin{table}
    \begin{center}
      \begin{tabular}{lrrrrr}
        Solid & PBEsol & OFR2 & \rrscan{}-L & \rrscan & Expt. (eV) \\ \hline
        Ge & 0.00 & 0.22 & 0.06 & 0.31 & 0.74 \\
        Si & 0.48 & 0.70 & 0.83 & 0.79 & 1.17 \\
        GaAs & 0.42 & 0.73 & 0.65 & 0.94 & 1.52 \\
        SiC & 1.24 & 1.41 & 1.69 & 1.74 & 2.42 \\
        C & 4.03 & 4.06 & 4.23 & 4.34 & 5.48 \\
        MgO & 4.66 & 5.04 & 5.41 & 5.74 & 7.22 \\
        LiCl & 6.36 & 6.93 & 7.18 & 7.46 & 9.40 \\
        LiF & 9.03 & 9.57 & 10.01 & 10.59 & 13.60 \\ \hline
        ME & -1.92 & -1.61 & -1.44 & -1.20 & \\
        MAE & 1.92 & 1.61 & 1.44 & 1.20 & \\
      \end{tabular}
    \end{center}
    \caption{Comparison of bandgaps (eV), extracted from the DOS in VASP.
    GKS DFAs, like \rrscan, are expected to predict more realistic bandgaps than standard Kohn-Sham DFAs, like PBEsol, OFR2, and \rrscan{}-L.
    DFAs are listed in anticipated order of predicted bandgap accuracy.
    Experimental (expt.) values are taken from Ref. \cite{aschebrock2019}.
    Mean errors (MEs) and mean absolute errors (MAEs) are also reported.
     \label{tab:gaps}}
  \end{table}
\end{ruledtabular}

Interestingly, OFR2 and \rrscan{}-L show no consistent behavior with respect to gaps.
Both LL-MGGAs severely underestimate the fundamental gap, but often approximate the \rrscan bandgap well.
In Ref. \cite{mejia2018}, it was argued that the closeness of SCAN-L and SCAN bandgaps indicated that SCAN-L accurately approximated the SCAN optimized effective potential (OEP).
Recall that the OEP \cite{kummel2003} is a general procedure that transforms a non-local Kohn-Sham potential operator (such as that of a meta-GGA) into a local, multiplicative potential.
We lack a better explanation regarding the relative closeness of the \rrscan{}, \rrscan{}-L, and OFR2 bandgaps.
Moreover, we are unaware of OEP calculations of the \rrscan{} potential in real systems.
As was reported in Table V of Ref. \cite{mejia2018} for LiH computed using SCAN and SCAN-L, there are unusual cases where the orbital-free meta-GGA predicts a slightly larger bandgap than the parent T-MGGA: \rrscan{}-L appears to find a slightly larger gap for Si than \rrscan{}.

\subsection{Monovacancy in Platinum \label{sec:defect}}

Reference \cite{jana2018} found that SCAN predicts the formation of a monovacancy in Pt to be energetically favorable.
Here, we compute the equilibrium lattice constants and vacancy formation energies of Pt using SCAN, \rrscan, \rrscan{}-L, and OFR2.
The initial equilibrium lattice constants for face-centered cubic (fcc) Pt were found by fitting to the SJEOS, using the same computational parameters as before.
A $2\times 2\times 2$ supercell containing 32 atoms was constructed using that lattice constant, and the supercell was allowed to further relax (ISIF = 3, IBRION = 2), using first-order Methfessel-Paxton smearing of width 0.2 eV, and forces converged within 0.001 eV/\AA{}.
The total energy was determined from the relaxed supercell structure using the tetrahedron method (ISIGMA = -5).
An identical supercell, but with an ion nearest the center of the cell removed, was used to model the monovacancy, and the same procedure was repeated.
An $11\times 11 \times 11$ $\bm{k}$-point grid was used, as recommended in Ref. \cite{jana2018}.

Monovacancy formation (MVF) energies
\begin{equation}
  E_\text{MVF} = E(N-1) - \frac{N-1}{N}E(N),
\end{equation}
where $E(N)$ is the total energy of an $N$-atom supercell ($N=32$ here), are presented in Table \ref{tab:pt_mv}.
We found a small positive monovacancy formation energy for SCAN, unlike the negative value found in Ref. \cite{jana2018}.
A negative monovacancy formation energy implies that a solid is unstable.
We find it unlikely that SCAN predicts Pt to be unstable, as SCAN describes its other equilibrium properties with experimental accuracy.
OFR2 predicts a slightly larger monovacancy formation energy than PBE.
PBEsol predicts the most accurate Pt monovacancy formation energy, but still underestimates the lowest experimental value.

\begin{ruledtabular}
\begin{table}
  \centering
  \begin{tabular}{lrr}
    DFA & $a_0$ (SJEOS, \AA{}) & $E_\text{MVF}$ (eV) \\ \hline
    Expt. & 3.913 & 1.32--1.7 \\
    PBE & 3.971 & 0.676 \\
    PBEsol & 3.919 & 0.886 \\
    SCAN & 3.913 & 0.126 \\
    \rrscan & 3.943 & 0.593 \\
    \rrscan{}-L & 3.980 & 0.590 \\
    OFR2 & 3.928 & 0.684 \\
  \end{tabular}
  \caption{Monovacancy formation energy and equilibrium geometry of fcc Pt.
  The experimental, zero-point corrected lattice constant is taken from Ref. \cite{hao2012}, and the experimental monovacancy formation energy range is taken from Ref. \cite{jana2018}.
  Note that the SJEOS-determined lattice constant (second column) was later permitted to relax in the Pt supercell.
  For all DFAs shown, the supercell lattice constant after relaxation did not change to the stated precision, again verifying our implementation of the Laplacian-dependent stress tensor.
  }
  \label{tab:pt_mv}
\end{table}
\end{ruledtabular}

\subsection{Intermetallic formation energies \label{sec:intermet}}

We follow the methodology of Ref. \cite{isaacs2018} to probe whether \rrscan{}-L and OFR2 improve the \rrscan description of intermetallic formation energies.
All initial geometries were taken from the Open Quantum Materials Database (OQMD) \cite{saal2013,kirklin2015,oqmd_link}.
Following Ref. \cite{isaacs2018}, geometries were relaxed, with all ionic degrees of freedom permitted to change (ISIF = 3), and with first-order Methfessel-Paxton smearing of width 0.2 eV.
After relaxation, total energies were determined using the tetrahedron method at fixed geometry.
All ions were initialized with a (ferromagnetic) magnetic moment of 3.5 $\mu_\text{B}$.
The plane-wave cutoff was 600 eV, and the $\bm{k}$-grid was determined as follows: for a fixed density of $\bm{k}$-points $\kappa$ (\AA{}$^{-3}$), the spacing $\Delta k$ between adjacent $\bm{k}$-points along each axis (KSPACING tag) is
\begin{equation}
  \Delta k = \left(\frac{\prod_{i=1}^3 |\bm{b}_i|}{|\bm{a}_1 \cdot (\bm{a}_2\times \bm{a}_3)|}\frac{1}{\kappa} \right)^{1/3}, \label{eq:dk}
\end{equation}
where $\bm{a}_i$ and $\bm{b}_i$ are the direct and reciprocal lattice vectors, respectively, for the initial geometry.
As in Ref. \cite{isaacs2018}, we used $\kappa = 700$ $\bm{k}$-points/\AA{}$^{-3}$ and computed $\Delta k$ from Eq. \ref{eq:dk}.
For simplicity, we rounded $\Delta k$ and iteratively decreased its value (if needed) to ensure a uniformly-spaced grid with density of at least 700 $\bm{k}$-points/\AA{}$^{-3}$.
For VPt$_2$, we needed to manually determine a grid with an equal number of $\bm{k}$-points along each axis to ensure that VASP produced a $\bm{k}$-grid with the right symmetry.
Formation energies per atom $\Delta \varepsilon_\text{f}$ were computed from total energies per primitive unit cell $E$ as follows: for compound $Y = \prod_{i=1}^M (X_i)_{x_i}$ composed of $M$ elements $X_i$ with multiplicity $x_i$ as
\begin{equation}
  \Delta \varepsilon_\text{f} = \frac{1}{\sum_i x_i} \left[ E(Y) - \sum_{i=1}^M \frac{x_i }{N_i} E(X_i) \right]
\end{equation}
with $N_i$ the number of ions in the unit cell for the pure solid $X_i$.
We have assumed one formula unit per primitive cell for intermetallic compound $Y$.

Our results and those of Refs. \cite{isaacs2018,kingsbury2022} are presented in Table \ref{tab:intermet}.
None of the DFAs considered here accurately predict the formation energies of these solids, however \rrscan{}-L and OFR2 improve over SCAN and \rrscan.
Although scalar relativistic effects are included in the treatment of core electrons in the VASP pseudopotentials, relativistic corrections (e.g., spin-orbit coupling) for Hf, Os, and Pt may be needed here.
Moreover, these are uncommon alloys with little representation in the literature.
Other experimental references for the formation enthalpies could benefit further analysis.
A recent QMC calculation \cite{isaacs2022} found the enthalpy of formation for VPt$_2$ to be $-0.764 \pm 0.050$ eV/atom, in line with the SCAN values here, but much larger than the experimental and OFR2 values.
In that work, the spin-orbit effect was found to reduce the magnitude of the formation energy of VPt$_2$, by about 0.05 eV.
We therefore find it likely that the experimental reference values are unreliable.

\begin{ruledtabular}
\begin{table*}
  \centering
  \begin{tabular}{l|rrrr|rrrrrrr}
    \shortstack{$\Delta \varepsilon_\text{f}$ \\ (eV/atom)} & Expt. & \shortstack{PBE, \\ Ref. \cite{isaacs2018}}
      & \shortstack{SCAN, \\ Refs. \cite{isaacs2018,kingsbury2022} }
      & \shortstack{\rrscan, \\ Ref. \cite{kingsbury2022} }
      & LSDA & PBE & PBEsol & SCAN & \rrscan & \rrscan{}-L & OFR2 \\ \hline
    HfOs & $-0.482\pm 0.052$ & -0.707 & -0.874 & -0.846
      & -0.724 & -0.715 & -0.708 & -0.901 & -0.847 & -0.805 & -0.743 \\
    ScPt & $-1.086 \pm 0.056$ & -1.212 & -1.473 & -1.308
      & -1.233 & -1.214 & -1.204 & -1.461 & -1.301 & -1.243 & -1.193\\
    VPt$_2$ & $-0.386 \pm 0.026$ & -0.555 & -0.726 & -0.601
      & -0.562 & -0.548 & -0.566 & -0.712 & -0.592 & -0.524 & -0.570 \\
  \end{tabular}
  \caption{ Formation enthalpies $\Delta \varepsilon_\text{f}$, in eV/atom, of a few intermetallic elements.
  The DFT results are formation energies and neglect the $PV$ term in the enthalpy.
  The experimental formation enthalpy of HfOs is from Ref. \cite{mahdouk1998}; experimental values for ScPt and VPt$_2$ are taken from Ref. \cite{guo2001}.
  Reference PBE values are taken from Ref. \cite{isaacs2018}.
  Reference SCAN values are averages of those reported in Refs. \cite{isaacs2018} and \cite{kingsbury2022}.
  Reference \rrscan values are taken from Ref. \cite{kingsbury2022}.
  The LSDA uses the Perdew-Zunger parameterization \cite{perdew1981} of the uniform electron gas correlation energy.
  \label{tab:intermet}
  }
\end{table*}
\end{ruledtabular}

While PBE and SCAN overestimate the magnitudes of the intermetallic formation energies in comparison to the experimental values in Table \ref{tab:intermet}, these DFAs underestimate this magnitude for Cu-Au intermetallics \cite{levamaki2018}.
However, the Cu-Au formation energies have magnitudes of 0.1 eV/atom at most, and SCAN underestimates them only by about 0.03 eV/atom.
Even better agreement with experiment has been achieved by Ref. \cite{levamaki2018} in two different ways: (1) by using standard hybrid functionals, and (2) by using, for each element, a PBE GGA with its gradient coefficients for exchange and correlation tuned to the experimental lattice constant and bulk modulus for that element.
The latter approach is motivated by a physical picture in which the correction to LSDA comes mainly from the core-valence interaction, in agreement with the analysis of Ref. \cite{fuchs1998}.

The tests of intermetallic formation energies described here and in Refs. \cite{isaacs2018,kingsbury2022} test the ability of a DFA to predict the correct equilibrium structure, spin-densities, and total energies for a solid and its constituents (or benefit from a random cancellation of errors).
Thus it is hard to discern which aspect of this test a DFA fails.
The subject of density-driven and functional-driven errors \cite{sim2018} is a useful framework for decomposing the various errors in this kind of test.
However, we cannot apply this metric without having exact or nearly-exact spin-densities (and geometries).

Systems with a strong sensitivity to perturbations in the Kohn-Sham potential can exhibit density driven errors \cite{kim2013}.
Evaluating a semi-local DFA (GGA, meta-GGA) on the Hartree-Fock density can often eliminate density-driven errors in molecules, as has recently been shown for SCAN applied to liquid water \cite{dasgupta2021}.
It is unclear what an equivalent density-correction method would be for solid-state calculations, as such a method would need to produce a density with a realistic geometry.
A modern periodic Hartree-Fock calculation of face-centered cubic LiH \cite{paier2009} found an equilibrium lattice constant $a_0=4.105$ \AA{} and bulk modulus $B_0=32.3$ GPa, in significant error of the zero-point corrected experimental values $a_0=3.979$ \AA{} and $B_0=40.1$ GPa \cite{tran2016} (and less accurate than the PBE, PBEsol, and SCAN values reported in Ref. \cite{tran2016}).
We are unaware of periodic Hartree-Fock calculations for the equilibrium properties of metallic solids.

\subsection{Alkaline solids \label{sec:alkalis}}

As discussed in the Introduction, Ref. \cite{kovacs2019} demonstrated that SCAN less accurately describes the equilibrium properties of the alkali metals Li, Na, K, Rb, and Cs than PBE.
It is therefore worth investigating if a LL-MGGA remedies this behavior.

We note two interesting computational features of LL-MGGAs.
Reducing the plane-wave kinetic energy cutoff can stabilize the calculations of isolated atoms.
Therefore, the calculations of cohesive energies reported here use a cutoff of 600 eV for both the bulk system and isolated atoms.
The $\bm{k}$-point density was unchanged, and the energy convergence criteria were $10^{-6}$ eV for the bulk solid and $10^{-5}$ eV for the isolated atom.
The size of the computational cell for the isolated atom was $14 \times 14.1 \times 14.2$ \AA{}$^3$, and only the $\Gamma$ point was for $\bm{k}$-space integrations.
For atomic calculations, Gaussian smearing of the Fermi surface with width 0.1 eV were used.
Spin-symmetry was permitted to break, and the energy was minimized directly (ALGO=A, LSUBROT set to false).
ALGO controls the method used to minimize the total energy; ALGO = A selects a preconditioned conjugate gradient algorithm.
The Hamiltonian is diagonalized in the occupied and unoccupied subspaces using a perturbation-theory-like method \cite{kresse1996}; setting LSUBROT = False prevents further optimization of the density matrix via unitary transformations of the orbitals, as recommended for semilocal DFAs.
Convergence with a LL-MGGA is generally more challenging for atomic systems, at least within VASP at these higher computational settings.
Linear density mixing (AMIX=0.4, AMIX\_MAG=0.1, BMIX=BMIX\_MAG=0.0001) was found to be helpful.
Beyond this, the input parameters remained the same (ADDGRID set to false, etc.) as for the bulk solids.

The PBE pseudopotentials with $s$ semi-core states included in the valence pseudo-density (indicated with a suffix ``\_sv'') appear to be less transferrable to LL-MGGAs.
Convergence for the isolated Li, Na, and Ba atoms using $s$ semi-core pseudopotentials was slow due to charge sloshing.
Thus, following the suggestion of Mej\'ia-Rodr\'iguez and Trickey \cite{mejia2020}, in this section, we have used pseudopotentials without any suffix when possible.
For a few elements (K, Rb, Cs, Ca, Sr, and Ba), the $s$ semi-core pseudopotentials are the only ones available.
However, \rrscan{}-L and OFR2 failed to converge within 10$^{-5}$ eV only for the Ba atom, with 500 self-consistency steps permitted.
As both converged to about $1 \times 10^{-4}$ eV, we have not excluded Ba from the test set.

Both \rrscan{}-L and OFR2 found a double-minimum in the energy per volume curve for Rb.
We chose to exclude data for the second, deeper minimum, which occurred at a larger, unrealistic volume.

This section analyzes the ``LC23'' set, the LC20 set augmented with three alkali metals, K, Rb, and Cs.
Moreover, given the reduced computational parameters, this section is more likely to reflect real-world usage of the DFAs than the benchmark calculations reported previously.
Table \ref{tab:alkali_errors} reports error statistics in the equilibrium properties of the alkali metals.
Tables \ref{tab:lc23_a0}--\ref{tab:lc23_e0} of Appendix \ref{app:lc23_data} present the data for each individual solid in the set.

\begin{ruledtabular}
  \begin{table*}
    \centering
    \begin{tabular}{lrrrrrr}
      & PBE & PBEsol & SCAN & \rrscan & \rrscan{}-L & OFR2 \\ \hline
      $a_0$ ME (\AA{}) & 0.051 & -0.017 & 0.084 & 0.111 & -0.004 & 0.014 \\
      $a_0$ MAE (\AA{}) & 0.061 & 0.019 & 0.095 & 0.114 & 0.055 & 0.039 \\
      $B_0$ ME (GPa) & -0.105 & -0.056 & -0.164 & -0.329 & 2.481 & 0.008 \\
      $B_0$ MAE (GPa) & 0.446 & 0.340 & 0.467 & 0.360 & 3.639 & 0.760 \\
      $E_0$ ME (eV/atom) & -0.072 & -0.005 & -0.083 & -0.092 & -0.100 & -0.099 \\
      $E_0$ MAE (eV/atom) & 0.072 & 0.022 & 0.083 & 0.092 & 0.100 & 0.099 \\
    \end{tabular}
    \caption{ Error statistics in the equilibrium lattice constants $a_0$, bulk moduli $B_0$, and cohesive energies $E_0$ for the alkali metals Li, Na, K, Rb, and Cs.
    The PBE \cite{perdew1996} and PBEsol \cite{perdew2008} GGAs, SCAN \cite{sun2015} and \rrscan \cite{furness2020} T-MGGAs, and \rrscan{}-L \cite{mejia2020} and OFR2 LL-MGGAs are presented.
    \label{tab:alkali_errors}
    }
  \end{table*}
\end{ruledtabular}

From Table \ref{tab:alkali_errors}, OFR2 finds more accurate lattice constants $a_0$ and bulk moduli $B_0$ for the alkalis than SCAN, \rrscan, or \rrscan{}-L.
The average errors of the \rrscan{}-L bulk moduli are 5 or 10 times larger than those of the other DFAs in Table \ref{tab:alkali_errors}.
However, all meta-GGAs presented in Table \ref{tab:alkali_errors} yield similarly inaccurate cohesive energies $E_0$ for the alkalis.
PBEsol appears to be the best general choice for studies of alkali-containing solids, however OFR2 should yield similar accuracy for their structural properties.

Isolated atoms, which have negative chemical potentials and thus turning surfaces in the Kohn-Sham potential, are thus poorly described by the gradient expansions for exchange and correlation.
Therefore, PBEsol and OFR2, which likely predict realistic total energies for the solids in LC23, do not predict realistic atomic energies for those solids, and thus generally inaccurate cohesive energies, as shown in Table \ref{tab:lc23_e0} of App. \ref{app:lc23_data}.
Conversely, PBE and \rrscan{}-L benefit from error cancellation between the total energies of the solids and their atomic constituents, yielding generally more accurate cohesive energies.
This observation excludes the cohesive energies of insulators, where a cancellation of errors benefits PBEsol and OFR2, but not PBE and \rrscan{}-L.
Similar limitations do not apply to T-MGGAs like SCAN and \rrscan{}, except for the metallic systems emphasized here.

\subsection{Molecules \label{sec:ae6}}

Within the quantum chemistry community, the AE6 set of six molecular atomization energies \cite{lynch2003} is used to rapidly estimate the performance of a DFA on a much larger set of atomization energies.
Geometries were taken from the MGAE109 database \cite{peverati2011}.
Table \ref{tab:ae6_full} presents the results of the AE6 set for \rrscan, \rrscan{}-L, and OFR2.

These calculations were also performed in VASP.
Each atom or molecule was placed in an orthorhombic box of dimensions 10 \AA{} $\times$ 10.1 \AA{} $\times$ 10.2 \AA{} to sufficiently lower the lattice symmetry and reduce interactions with image cells.
A plane-wave energy cutoff of 1000 eV was used.
Beyond this, all other computational parameters used for the isolated atoms in Sec. \ref{sec:alkalis} were unchanged.

\begin{ruledtabular}
  \begin{table*}
    \centering
    \begin{tabular}{l|rrrrrr}
      Molecule & PBE & PBEsol & SCAN & \rrscan & \rrscan{}-L & OFR2 \\ \hline
      SiH$_4$ & 313.64 & 322.92 & 328.54 & 322.07 & 321.43 & 320.35 \\
      SiO & 195.93 & 204.09 & 191.06 & 186.81 & 188.03 & 186.46 \\
      S$_2$ & 115.68 & 129.62 & 108.68 & 110.36 & 110.51 & 112.26 \\
      C$_3$H$_4$ & 727.09 & 751.97 & 703.40 & 702.50 & 700.24 & 686.80 \\
      C$_2$H$_2$O$_2$ & 662.83 & 692.76 & 628.71 & 629.09 & 628.86 & 618.44 \\
      C$_4$H$_8$ & 1175.57 & 1221.27 & 1151.80 & 1147.71 & 1141.41 & 1126.86 \\ \hline
      ME LT03 & 14.57 & 36.55 & 1.48 & -0.79 & -2.14 & -8.69 \\
      MAE LT03 & 17.49 & 36.55 & 3.83 & 3.69 & 5.08 & 12.22 \\ \hline
      ME HK12 & 15.21 & 37.19 & 2.12 & -0.16 & -1.50 & -8.05 \\
      MAE HK12 & 18.86 & 37.75 & 3.80 & 3.65 & 3.93 & 11.06 \\
    \end{tabular}
    \caption{ Comparison of PBE \cite{perdew1996}, PBEsol \cite{perdew2008}, SCAN \cite{sun2015}, \rrscan \cite{furness2020}, \rrscan{}-L \cite{mejia2020}, and OFR2 atomization energies for the AE6 set \cite{lynch2003}.
    All values are in kcal/mol (1 eV $\approx 23.060548$ kcal/mol).
    We report mean errors (MEs) and mean absolute errors (MAEs) computed with respect to two sets of reference data: the original work of Ref. \cite{lynch2003} (LT03), and the more recent non-relativistic, frozen-core values from Table 4 of Ref. \cite{haunschild2012} (HK12).
    Given that the calculation in VASP is non-relativistic with a frozen-core pseudopotential, these latter reference values appear to be most appropriate.
    Absolute total energies have no physical meaning in a pseudopotential calculation, therefore we only report the energy differences here.
    \label{tab:ae6_full}}
  \end{table*}
\end{ruledtabular}

From Table \ref{tab:ae6_full}, we see that \rrscan{}-L broadly retains the accuracy of \rrscan for molecular systems.
OFR2, with a 11 kcal/mol mean absolute error (MAE) for AE6, appears to be the ``missing link'' DFA between the GGA level, with MAEs on the order of 20--40 kcal/mol, and the T-MGGA level, with MAEs less than 10 kcal/mol.
Convergence with OFR2 for finite systems is generally more challenging than with \rrscan{}-L.
Independent tests of OFR2 \cite{francisco2022} have confirmed our conclusions: \rrscan-L is faithful to the \rrscan description of molecules, whereas OFR2 is somewhat less accurate.

For an accurate description of solid state geometries and magnetic properties, we recommend OFR2.
To improve its description of cohesive energies, which lie between those of PBEsol and \rrscan{}-L in accuracy, one might perform a non-self-consistent evaluation of the \rrscan{} or \rrscan{}-L total energy using the (likely more accurate) relaxed OFR2 geometry and density for a solid as input.
For an accurate description of finite systems, we recommend \rrscan{}-L at the LL-MGGA level.
For greater accuracy and general-purpose calculations of finite or extended systems, we recommend \rrscan{}.

\section{Outlook: Machine learning and kinetic energy density \label{sec:HL}}

Machine learning has already made leaps and bounds in the construction of empirical DFAs.
The work of Ref. \cite{dick2021} suggests that the most sophisticated T-MGGAs have essentially reached a fundamental limit of accuracy for the meta-GGA level.
The work of Ref. \cite{kirkpatrick2021} built a local hybrid-level DFA that approximately satisfies fractional charge \cite{perdew1982} and spin \cite{cohen2008} exact constraints, heretofore seldom satisfied.

Doubtless, machine learning techniques will be applied to the three-dimensional kinetic energy density.
A machine-learned model is important for practical purposes, but excogitating the role of the parameters within the model is nigh impossible.
This section details a simple ``human-learned'' model (HL$M$) for the kinetic energy density, which can be instructive for future machine-learning work.
In particular, HL$M$ shows how heavy fitting can lead to wrong asymptotics and to numerical instability.

As in our RPP model of $\tau$ (but without consideration of the fourth-order gradient expansion), we will presume that the exact (spin-unpolarized) $\tau$ can be represented as an interpolation between exact limits,
\begin{align}
  \tau(n,p,q) &= \tu(n)\left[ F_\mathrm{W}(p) + z(p,q)\theta(z(p,q)) \right] \\
  z(p,q) &= F_\text{GE2}(p,q) - F_\mathrm{W}(p) \label{eq:z_for_theta} \\
  F_\mathrm{W}(p) &= \frac{5}{3}p \\
  F_\text{GE2}(p,q) &= 1 + \frac{20}{9} q + \frac{5}{27}p
\end{align}
We will model the function $\theta(z)$, which determines the mixing between Weizs\"acker and gradient expansion limits.
Moreover, $\theta(z)$ should permit extrapolation for arbitrary positive $z$, as suggested by Cancio and Redd \cite{cancio2017}.
Then for some of the appropriate norms considered here -- the neutral noble gas atoms Ne, Ar, Kr, and Xe, and the jellium surfaces of bulk densities $\overline{\rs} = 2,3,4,5$ -- we take a reference density and compute
\begin{equation}
  \theta(z) = \frac{\tau/\tau_\text{unif}(n) - F_\mathrm{W}(p)}{z(p,q)}.
  \label{eq:theta_def}
\end{equation}
Since the right-hand side of Eq. \ref{eq:theta_def} is not exactly a function of $z$, it is useful to bin the values of $\theta$ within a narrow range of $z$.

The form selected for $\theta$ enforces three constraints: the Weizs\"acker lower bound, the uniform density limit, and the second-order gradient expansion.
A machine can learn these constraints approximately by penalizing their violation, but cannot satisfy them by construction as a human-designed model can.
Because the ``exact'' $\theta(z)$ is complicated, we need an expression which has sufficient freedom for fitting.
Consider the $M$-parameter HL$M$ model
\begin{equation}
  \theta_M(z) =
    z^3\frac{1 + b_1 z + b_2 z^2}{1 + \sum_{i=1}^M c_i z^i}\Theta(z), \label{eq:hl_theta}
\end{equation}
where $\Theta(z\geq 0) = 1$ and $\Theta(z < 0) = 0$, and the $c_i$ are fit parameters.
To recover the uniform density limit requires $\theta_M(1)=1$; to recover the second-order gradient expansion of $\tau$ requires $\theta_M'(1)=0$.
Enforcing these constraints fixes the values of the $b_i$
\begin{align}
  b_1 &= 3 + \sum_{i=1}^M (5-i) c_i \label{eq:hl_b1} \\
  b_2 &= \sum_{i=1}^M c_i - b_1. \label{eq:hl_b2}
\end{align}
It appears that $\theta_M(z \gg 1) \sim b - a \log z$, for constants $a$ and $b$, however this model can approximately recover that behavior.
The minimum power of $z$ in the numerator is chosen to allow for sufficient smoothness of the exchange-correlation potential for $z \approx 0$.

\begin{figure}
  \centering
  \includegraphics[width=\columnwidth]{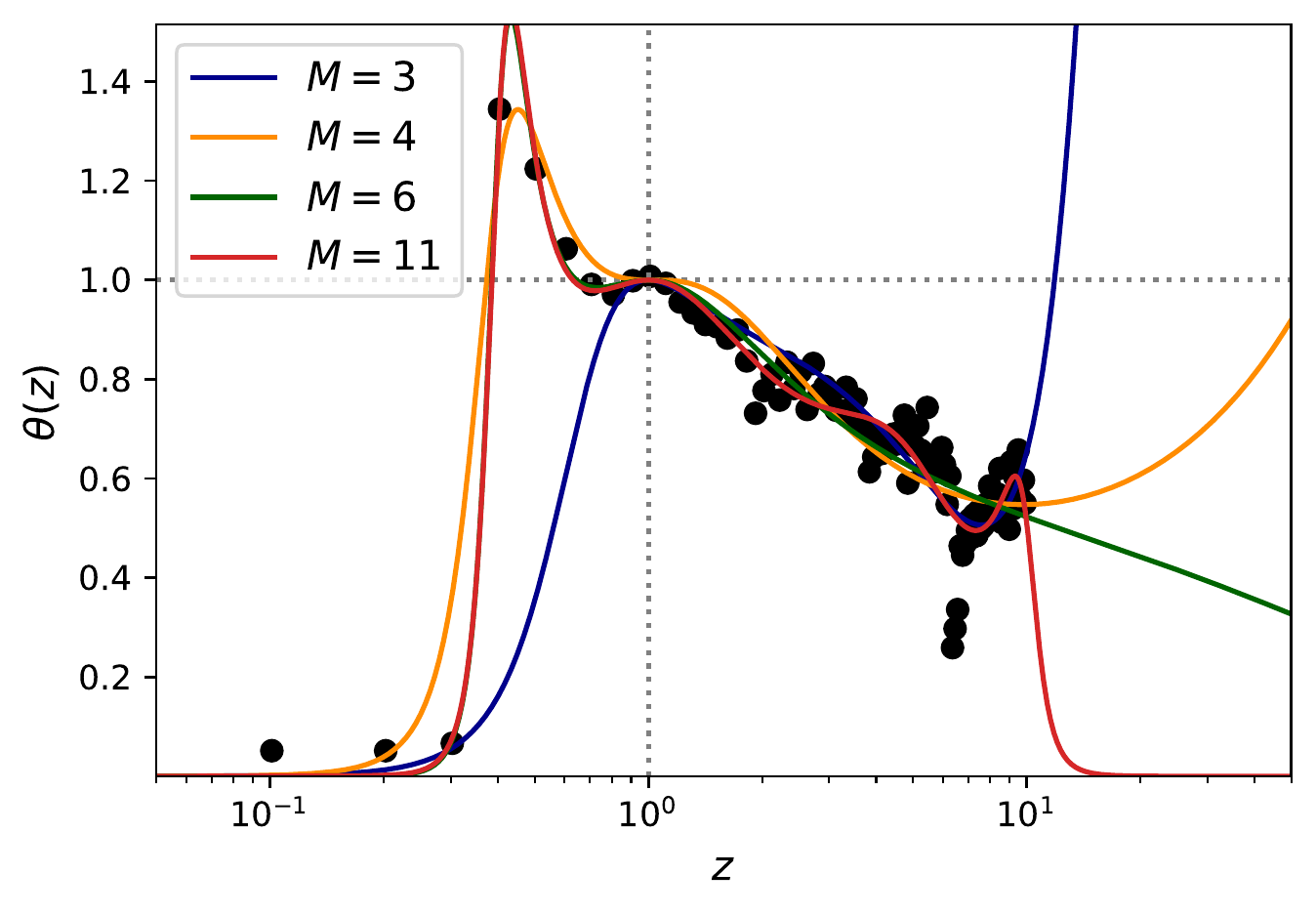}
  \caption{
  $M$-parameter mixing function $\theta_M(z)$ of Eq. \ref{eq:hl_theta} that determines the optimal mixing of Weizs\"acker and second-order gradient expansion kinetic energy densities.
  Acceptable (pole-free and non-negative $\theta(z\geq 0)$) parameter sets $M=3,4,6,\& 11$ are displayed.
  Solid points are the binned $\theta(z)$ data taken from the appropriate norms: the neutral noble gas atoms Ne, Ar, Kr, and Xe, and the jellium surfaces of bulk densities $\overline{\rs} = 2,3,4,5$.
  \label{fig:hl_ked}
  }
\end{figure}

We considered $2 \leq M \leq 20$; for $M \geq 5$, $\theta_M$ can be bounded as $z \to \infty$.
A non-linear least-squares fit was used to determine the $c_i$.
We discarded parameter sets for which the denominator of $\theta_M$ had positive polynomial roots or for which $\theta_M(z > 0) < 0$.
The possible acceptable parameters found were for $M=3,4,6,\& 11$, as shown in Fig. \ref{fig:hl_ked}.
Clearly, $M=3$ or 4 do not represent reliable extrapolations for $z \to \infty$.
$\theta_6$ appears to represent the most realistic, long-tailed extrapolation for $z\to \infty$, however $\theta_{11}$ more accurately captures the apparent oscillations in $\theta(z)$.

\begin{table}
  \centering
  \begin{tabular}{r|rrr|rrr} \hline \hline
    & \multicolumn{3}{c|}{\rrscan} & \multicolumn{3}{|c}{SCAN} \\
    $M$ & RGA & JS & JC & RGA & JS & JC \\ \hline
    3 & 0.73 & 9.20 & 11.48 & 0.95 & 6.71 & 10.39 \\
    4 & 0.91 & 2.82 & 1.15 & 1.01 & 2.87 & 3.74 \\
    6 & 0.53 & 3.60 & 2.61 & 0.55 & 1.51 & 2.11 \\
    11 & 0.48 & 3.73 & 2.72 & 0.49 & 1.53 & 1.88 \\ \hline
    Exact $\tau$ & 0.14 & 2.80 & 2.38 & 0.08 & 2.51 & 3.15 \\ \hline\hline
  \end{tabular}
  \caption{
  Orbital free \rrscan and SCAN appropriate norm performance using the highly-parameterized mixing function $\theta(F_\text{W}-F_\text{GE2})$ of Eq. \ref{eq:hl_theta}, compared to the orbital-dependent variants (bottom row).
  Increasing the number of parameters $M$ generally improves the fidelity of the approximate $\tau$, at the cost of more rapid oscillations.
  The mean absolute percentage errors of the rare gas atom (RGA) exchange-correlation energies, jellium surface (JS) exchange-correlation surface formation energies, and jellium cluster exchange-correlation surface formation energies are shown.
  \label{tab:hl_ked_norms}
  }
\end{table}

Thus we emphasize the need for human decision in highly-empirical DFA design.
Both $\theta_6$ and $\theta_{11}$ deliver similar performance for the appropriate norms, as shown in Table \ref{tab:hl_ked_norms}, however $\theta_6$ is much smoother and is thus likely more numerically stable.
It is purely for reasons of numeric stability that the HL$M$ models have been deferred to this section.
While we do not present plots of the \rrscan + HL6 or HL11 Kohn-Sham potential for the simple systems considered here, we have computed them and determined they are wholly unrealistic.

\section{Conclusions}

We developed a model Laplacian-level meta-GGA (LL-MGGA) OFR2 that is an orbital-free or ``deorbitalized'' variant of \rrscan \cite{furness2020}, in the tradition of Refs. \cite{mejia2017,mejia2018,mejia2020}, but recovering the fourth-order gradient expansion for exchange and the second-order gradient expansion for correlation.
Only $\ba$ has been modified, although the rest of \rrscan could be re-optimized in future work.
We extensively tested OFR2 against an existing deorbitalization of \rrscan, \rrscan{}-L \cite{mejia2020}, which breaks the uniform density limit of \rrscan.

OFR2 appears to improve upon \rrscan for the lattice constants of solids, matching or exceeding the accuracy of SCAN.
\rrscan{}-L and OFR2 more accurately describe transition-metal magnetism than \rrscan, which predicts substantially larger magnetic moments than found by experiment.
OFR2 better describes the structural properties of alkali metals than \rrscan and \rrscan{}-L, but not their cohesive energies.
We therefore recommend OFR2 for an orbital-free description of solids and liquids only, and particularly $sp$ or $sd$ metals.
For best accuracy in molecules and non-metallic condensed matter, we continue to recommend SCAN and \rrscan{}.

For an orbital-free description of molecules, we recommend \rrscan{}-L, which retains the accuracy of \rrscan for the AE6 set \cite{lynch2003} of atomization energies.
This conclusion was independently confirmed for a different set of molecules \cite{francisco2022}.
OFR2, which targets properties of metallic solids, bridges the gap between PBE GGA errors (MAE $\sim  19 $ kcal/mol) and \rrscan T-MGGA errors (MAE $\sim 4$ kcal/mol).

Like the SCAN \cite{sun2015} and TPSS \cite{tao2003} T-MGGAs, and unlike \rrscan, OFR2 recovers the fourth-order gradient expansion for the exchange energy.
Thus OFR2 has a correctly LSDA-like static linear density-response for the uniform electron gas, which, along with its correct description of slowly-varying densities and especially the weaker nonlocality of OFR2, should bolster its accuracy for metals.

Unlike chemistry, condensed matter physics must rely on experimental reference values whose uncertainties can be large or difficult to quantify.
The smallest experimental relative errors are probably those of lattice constants from X-ray diffraction.
Thus the high accuracy of OFR2 lattice constants for metals is encouraging.
Structural phase transitions are more challenging to DFAs than lattice constants are \cite{shahi2018}, but good results have been obtained \cite{shahi2018} for semiconductors from SCAN.
OFR2 might improve the critical pressures for transitions between metallic phases, especially for transition metals.

Obtaining highly-converged results with an LL-MGGA is generally more challenging than with other semi-local approximations.
Some PBE pseudopotentials also appear to be less transferrable to LL-MGGAs than $\tau$-meta-GGAs (T-MGGAs).
Mej\'ia-Rodr\'iguez and Trickey \cite{mejia2020} found that GW potentials were less transferrable to LL-MGGAs.
LL-MGGAs might have a particular niche for exploratory purposes: if benchmark-quality results are not desired, these can often match or surpass the accuracy of their T-MGGA counterparts.
Thus for computationally intensive tasks, such as mapping the phase diagram of transition metals, an LL-MGGA could be used to rapidly obtain a good starting guess for more sophisticated approximations.

The new OFR2 ``deorbitalizes'' the \rrscan{} meta-GGA while preserving and even enhancing the \rrscan exact constraints on the slowly-varying limit ($\ba \approx 1$, $p\ll 1$, $|q| \ll 1$).
Thus a comparison of OFR2 and \rrscan{} results for metals could reflect mainly the difference between the fully (if modestly) nonlocal argument $\tau(\br)$ and the semilocal argument $\nabla^2 n(\br)$ in the approximated exchange-correlation energy functional.
Weakening the nonlocality of \rrscan{} seems to improve (in comparison to experiment) the magnetic moments of the transition metals, the monovacancy formation energy of solid Pt, and the formation energies of intermetallics, producing results that are not very different (in the cases studied here) from those of the much less-sophisticated PBEsol \cite{perdew2008}.
However, for molecules and insulating materials, accuracy should improve from PBEsol to OFR2 to \rrscan.

\begin{acknowledgments}
  A.D.K. and J.P.P. acknowledge the support of the U.S. Department of Energy, Office of Science, Basic Energy Sciences, through Grant No. DE-SC0012575 to the Energy Frontier Research Center: Center for Complex Materials from First Principles.
  J.P.P. also acknowledges the support of the National Science Foundation under Grant No. DMR-1939528.
  A.D.K. thanks Temple University for a Presidential Fellowship.
  We thank C. Shahi for discussions on the monovacancy formation energy calculation, and J. Sun for discussions of solid-state phase diagrams.
\end{acknowledgments}

\section*{Code and data availability}

The Python 3 and Fortran code used to fit the orbital free \rrscan is made freely available at the code repository \cite{code_repo}.
Data files needed to run this code, general purpose Fortran subroutines, and VASP subroutines are included there as well.
All data is hosted publicly (without access restrictions) at Zenodo \cite{data_repo}.

%

\appendix

\section{Implementing the Laplacian in VASP \label{app:wb_method}}

White and Bird \cite{white1994} suggested a non-standard way to compute the exchange-correlation potential on a grid of $M$ finite points $\bR$ (minimum fast Fourier transform grid).
This robust method is used in many standard plane wave codes, including VASP, and was used in our VASP calculations.
We outline the method below.

Their analysis was tailored to the specific case of periodic boundary conditions, thus we define the reciprocal lattice vectors $\bG$.
Using Fourier series, we can write the density variables as
\begin{align}
  n(\br) &= \sum_{\bG} n(\bG) e^{i \bG \cdot \br} \\
  n(\bG) &= \frac{1}{M} \sum_{\bR} n(\bR) e^{-i \bG \cdot \bR} \\
  \nabla n(\br) &= i \sum_{\bG} \bG n(\bG) e^{i \bG \cdot \br} \\
    & = \frac{i}{M} \sum_{\bG,\bR} \bG n(\bR) e^{i \bG \cdot (\br - \bR)} \nonumber \\
  \nabla^2 n(\br) &= - \sum_{\bG} \bG^2 n(\bG) e^{i \bG \cdot \br} \\
    &= \frac{-1}{M} \sum_{\bG,\bR} \bG^2 n(\bR) e^{i \bG \cdot (\br - \bR)}. \nonumber
\end{align}
Now let the discrete $E\sxc$ within a cell volume $\Omega$ be
\begin{equation}
  \texc = \frac{\Omega}{M} \sum_{\bR} e\sxc(n(\bR),\nabla n(\bR),\nabla^2 n(\bR)),
\end{equation}
with $e\sxc = \varepsilon\sxc~ n(\bR)$. One can approximate the variations in \texc using
\begin{equation}
  \delta \texc = \frac{\Omega}{M} \sum_{\bR} \frac{d \texc}{d n(\bR)} \delta n(\bR) \equiv  \sum_{\bR} \widetilde v\sxc(\bR) \delta n(\bR),
\end{equation}
then the discrete potential $\widetilde v\sxc$ is represented as
\begin{align}
  \widetilde v\sxc(\bR) &= \frac{\partial e\sxc}{\partial n(\bR)} +
  \sum_{\bR'} \left\{ \frac{\partial e\sxc}{\partial \nabla n(\bR')}
   \cdot \frac{d (\nabla n(\bR'))}{d n(\bR)} \right. \\
  & \left. + \frac{\partial e\sxc}{\partial \nabla^2 n(\bR')}
  \frac{d (\nabla^2 n(\bR'))}{d n(\bR)} \right\}. \nonumber
\end{align}
It's now trivial to insert the Fourier series representations of the total derivatives on the RHS of the last equation.
Note that the density gradient \textit{vector} is never used in PBE-like GGAs, thus we can replace the derivatives with respect to $\nabla n$ by
\begin{equation}
  \frac{\partial}{\partial (\nabla n)} = \frac{\nabla n}{|\nabla n|}\frac{\partial}{\partial |\nabla n|}.
\end{equation}
The discrete potential then becomes
\begin{align}
  \widetilde v\sxc(\bR) &= \frac{\partial e\sxc}{\partial n(\bR)}
  + \frac{1}{M}\sum_{\bG,\bR'} \left\{ i \bG \cdot \frac{\nabla n(\bR')}{|\nabla n(\bR')|}
  \frac{\partial e\sxc}{\partial |\nabla n(\bR')|} \right. \nonumber \\
  & \left. - \bG^2 \frac{\partial e\sxc}{\partial \nabla^2 n(\bR')} \right\}
  e^{i \bG \cdot (\bR' - \bR)} .
\end{align}

Supplemental Tables S7, S8, and S9 of Ref. \cite{mejia2020} present lattice constants, bulk moduli, and cohesive energies for a variety of solids, computed with \rrscan{} and \rrscan{}-L.
As these tables include every solid in the LC23 set, we can roughly validate our implementation of \rrscan{}-L.
We use ``roughly'' here because not all computational parameters are available for that work.
Table \ref{tab:r2scan_l_valid} shows that the results of this work and Ref. \cite{mejia2020} agree to about 0.001 \AA{} (\rrscan{}) and 0.01 \AA{} (\rrscan{}-L) for the lattice constants; to about 0.3 GPa (\rrscan{}) and 1.4 GPa (\rrscan{}-L) for the bulk moduli; and to about 0.06 eV/atom (\rrscan{}) and 0.03 eV/atom (\rrscan{}-L) for the cohesive energies.
This is reasonable agreement.

\begin{ruledtabular}
  \begin{table}[t]
    \centering
    \begin{tabular}{llrrrrrrrr}
    &  & \multicolumn{2}{c}{Present work} & \multicolumn{2}{c}{Ref. \cite{mejia2020}} \\
    &  & \rrscan{} & \rrscan{}-L & \rrscan{} & \rrscan{}-L \\ \hline
    \multirow{4}{*}{$a_0$ (\AA{})} & ME & 0.037 & 0.010 & 0.037 & 0.019 \\
    & MAE & 0.049 & 0.040 & 0.048 & 0.045 \\
    & MD & 0.000 & -0.009 & & \\
    & MAD & 0.001 & 0.013 & & \\ \hline
    \multirow{4}{*}{$B_0$ (GPa)} & ME & 0.843 & -3.284 & 0.692 & -3.731 \\
    & MAE & 3.522 & 7.074 & 3.512 & 6.510 \\
    & MD & 0.151 & 0.447 & & \\
    & MAD & 0.258 & 1.403 & & \\ \hline
    \multirow{4}{1cm}{$E_0$ (eV/atom)} & ME & 0.032 & -0.134 & -0.022 & -0.162 \\
    & MAE & 0.109 & 0.150 & 0.102 & 0.172 \\
    & MD & 0.053 & 0.028 & & \\
    & MAD & 0.057 & 0.032 & & \\
    \end{tabular}
    \caption{Comparison of the \rrscan{} and \rrscan{}-L LC23 equilibrium lattice constants $a_0$ (\AA{}), bulk moduli $B_0$ (GPa), and cohesive energies $E_0$ (eV/atom) from this work and Ref. \cite{mejia2020}.
    Mean deviations (MDs) and mean absolute deviations (MADs) between \rrscan{}(-L) in this work and Ref. \cite{mejia2020} are also included.
    }
    \label{tab:r2scan_l_valid}
  \end{table}
\end{ruledtabular}

\clearpage

\onecolumngrid
\section{Restoring the fourth-order gradient expansion for exchange to \rrscan \label{app:r2ge4}}

This section builds upon the derivation of \rrscan presented in Ref. \cite{furness2022}.
By construction, \rrscan recovers the exact second-order gradient expansion for exchange, but not the fourth-order terms.
It will be shown in a forthcoming work that \rrscan severely overestimates the magnitude of the fourth-order gradient expansion coefficients.
The exact exchange enhancement factor has a fourth-order gradient expansion in $p$ and $q$ \cite{svendsen1996}
\begin{equation}
  F\sx = 1 + \frac{10}{81}p + \frac{146}{2025}q^2 - \frac{73}{405}p q
    + \mathcal{O}(|\nabla n|^6). \label{eq:fxge4}
\end{equation}
Note that the coefficient of $p q$ is known within some uncertainty, as is the coefficient of $p^2$.
We take the best estimates from Ref. \cite{svendsen1996}.

However, an orbital-free \rrscan can be made to recover the right fourth-order gradient expansion for exchange.
This is accomplished by using different fourth-order terms in the gradient expansion of the approximate $\tau(p,q)$ than those that appear in the gradient expansion of the exact $\tau$ \cite{brack1976}.
To maintain the second-order gradient expansion constraint of \rrscan, we retain the correct \textit{second-order} gradient expansion of $\tau$,
\begin{align}
  \frac{\tau(p,q)}{\tau_0} &= 1 + \frac{5}{27}p + \frac{20}{9}q
  + b_{qq}q^2 + b_{pq} p q  + b_{pp}p^2 + \mathcal{O}(|\nabla n|^6),
\end{align}
with fourth-order coefficients $b_{qq}$, $b_{pq}$, and $b_{pp}$ to be determined below.
The iso-orbital indicator used in \rrscan is the numerically-stable
\begin{equation}
  \ba = \frac{\tau - \tau_\text{W}}{\tau_0 + \eta\tau_\text{W}}
\end{equation}
where $\eta = 0.001$ \cite{furness2020}.
It can be seen that the gradient expansion of the approximate $\ba(p,q)$ is
\begin{align}
  \ba(p,q) &= 1 - \frac{5(8 + 9\eta)}{27} p + \frac{20}{9}q + b_{qq}q^2
    + \left( b_{pq} - \frac{100\eta}{27} \right)p q
    + \left(b_{pp} + \frac{200\eta}{81} + \frac{25\eta^2}{9} \right)p^2
    + \mathcal{O}(|\nabla n|^6). \label{eq:ba_app_ge}
\end{align}
Note that the gradient expansion \cite{brack1976} of $\ba$ using the exact $\tau$ cannot be expressed in terms of a polynomial in $p$ and $q$.

We turn our attention to the enhancement factor $F\sx^\text{\rrscan}$,
\begin{equation}
  F\sx^\text{\rrscan} = \{ h\sx^1(p) + f\sx(\ba)[h\sx^0 - h\sx^1] \} g\sx(p).
\end{equation}
In \rrscan, $g\sx(p)$ is a non-analytic smooth function, with Taylor series $1 + \mathcal{O}(|\nabla n|^\infty)$.
Therefore, $g\sx(p)$ does not contribute to the gradient expansion of the enhancement factor beyond order zero.
Note that $h\sx^0 = 1 + k_0$, where $k_0 = 0.174$.
As is done in Ref. \cite{furness2022} to construct the model r$^4$SCAN functional, we seek a Taylor expansion of $F\sx$ in $p$ and $\ba-1$, which approximately define the slowly-varying limit,
\begin{align}
  F\sx^\text{\rrscan} =& 1  + h\sx'(0)p + \frac{h\sx''(0)}{2}p^2
     + \left[f\sx'(1)(\ba - 1) + \frac{f\sx''(1)}{2}(\ba - 1)^2 \right]
     \left[1 + k_0 - 1 - h\sx'(0)p - \frac{h\sx''(0)}{2}p^2\right] \\
    & + \mathcal{O}(|\nabla n|^6) .\nonumber
\end{align}
Here, $h\sx'(0)=dh\sx^1/dp(0)$, etc.
Now, $(\ba - 1)$ contains terms of both second- and fourth-order, whereas $(\ba - 1)^2$ and $(\ba-1)p$ contain terms of fourth- and sixth-order,
\begin{align}
  (\ba - 1)^2 &= \frac{400}{81}q^2 - \frac{200(8+9\eta)}{243}p q
    + \frac{25(8+9\eta)^2}{729}p^2 + \mathcal{O}(|\nabla n|^6) \label{eq:bam12} \\
  (\ba - 1)p &= \frac{20}{9}p q -\frac{5(8+9\eta)}{27}p^2 + \mathcal{O}(|\nabla n|^6). \label{eq:bam1p}
\end{align}
The Taylor series of the enhancement factor can be simplified as
\begin{align}
  F&\sx^\text{\rrscan} = 1 + h\sx'(0)p + k_0 f\sx'(1)(\ba - 1)
    + \frac{h\sx''(0)}{2}p^2 - h\sx'(0) f\sx'(1)(\ba - 1)p
    + \frac{k_0}{2} f\sx''(1)(\ba - 1)^2 + \mathcal{O}(|\nabla n|^6).
\end{align}
After inserting Eq. \ref{eq:ba_app_ge} for the gradient expansion of the approximate $\ba$, Eq. \ref{eq:bam12} for $(\ba-1)^2$, and Eq. \ref{eq:bam1p} for $(\ba-1)p$, we find the OFR2 enhancement factor,
\begin{align}
  F\sx^\text{OFR2}& = 1 + \left[h\sx'(0) - \frac{5(8 + 9\eta)}{27}k_0 f\sx'(1)\right]p
    + \frac{20}{9}k_0 f\sx'(1) q
    + \left[\frac{200}{81} f\sx''(1) + f\sx'(1) b_{qq} \right]k_0 q^2 \\
    & -\left[\left( \frac{100\eta}{27} -b_{pq} \right)k_0 f\sx'(1)
      +  \frac{20}{9} h\sx'(0) f\sx'(1)
      + \frac{100(8+9\eta)}{243}k_0 f\sx''(1) \right]pq \nonumber \\
    & + \left[\frac{h\sx''(0)}{2} + \left(b_{pp} + \frac{200\eta}{81}
      + \frac{25\eta^2}{9} \right) k_0 f\sx'(1) + \frac{5(8+9\eta)}{27}h\sx'(0) f\sx'(1)
      + \frac{25(8+9\eta)^2}{1458}k_0 f\sx''(1) \right]p^2 + \mathcal{O}(|\nabla n|^6). \nonumber
\end{align}

As was shown in Ref. \cite{furness2022}, the divergence theorem may be used to eliminate the term linear in $q$ in favor of a term linear in $p$ plus a gauge function.
Suppose an enhancement factor can separated as $F\sx = \widetilde F\sx + n^{-4/3} \nabla \cdot \bm{G}\sx$.
Under integration over a volume $\Omega$ with bounding surface $\text{bdy}\, \Omega$, the exchange energy is
\begin{equation}
  E\sx[n] = \int_\Omega F\sx \epsilon\sx^\text{LDA} d^3r
  = A\sx \int_\Omega F\sx n^{4/3} d^3 r
  = A\sx \int_\Omega \widetilde F\sx n^{4/3} d^3r
    + A\sx \int_{\text{bdy}\, \Omega} \bm{G}\sx \cdot d\bm{S}.
\end{equation}
Provided that the integral of $\bm{G}\sx$ vanishes at the bounding surface, $F\sx$ and the ``integrated-by-parts'' $\widetilde F\sx$ will yield the same exchange energy and potential, but different exchange energy densities.
Note that $A\sx = -3(3\pi^2)^{1/3}/(4\pi)$.
As is easily seen,
\begin{equation}
  q n^{4/3} = \frac{p}{3}n^{4/3} +
  \nabla \cdot \left[\frac{\nabla n}{4(3\pi^2)^{2/3}n^{1/3}} \right],
\end{equation}
therefore the overall gauge function is $n^{-4/3}\nabla \cdot [n^{-1/3}\nabla n ]/[4(3\pi^2)^{2/3}]$.
Then the integrated-by-parts enhancement factor is
\begin{align}
  \widetilde F\sx^\text{OFR2} & = 1 + \left[h\sx'(0) - \frac{5(4 + 9\eta)}{27}k_0 f\sx'(1)\right]p
    + \left[\frac{200}{81} f\sx''(1) + f\sx'(1) b_{qq} \right]k_0 q^2 \label{eq:fxibp} \\
    & -\left[\left( \frac{100\eta}{27} -b_{pq} \right)k_0 f\sx'(1)
      + h\sx'(0) f\sx'(1) \frac{20}{9}
      + \frac{100(8+9\eta)}{243}k_0 f\sx''(1) \right]pq \nonumber \\
    & + \left[\frac{h\sx''(0)}{2} + \left(b_{pp} + \frac{200\eta}{81}
      + \frac{25\eta^2}{9} \right) k_0 f\sx'(1) + \frac{5(8+9\eta)}{27}h\sx'(0) f\sx'(1)
      + \frac{25(8+9\eta)^2}{1458}k_0 f\sx''(1) \right]p^2 + \mathcal{O}(|\nabla n|^6). \nonumber
\end{align}

Now equate the terms in Eq. \ref{eq:fxibp} with the terms of matching order in Eq. \ref{eq:fxge4} to constrain $\widetilde F\sx^\text{OFR2}$ to have the correct fourth-order gradient expansion,
\begin{align}
  h\sx'(0) - \frac{5(4 + 9\eta)}{27}k_0 f\sx'(1) &= \frac{10}{81} \label{eq:r2_ge2_cons} \\
  \left[\frac{200}{81} f\sx''(1) + f\sx'(1) b_{qq} \right]k_0 &= \frac{146}{2025} \\
  \left( \frac{100\eta}{27} -b_{pq} \right)k_0 f\sx'(1)
    +  \frac{20}{9} h\sx'(0) f\sx'(1) + \frac{100(8+9\eta)}{243}k_0 f\sx''(1) &= \frac{73}{405} \\
  \frac{h\sx''(0)}{2} + \left(b_{pp} + \frac{200\eta}{81}
    + \frac{25\eta^2}{9} \right) k_0 f\sx'(1) + \frac{5(8+9\eta)}{27}h\sx'(0) f\sx'(1)
    + \frac{25(8+9\eta)^2}{1458}k_0 f\sx''(1) &= 0.
\end{align}
By construction, in \rrscan, $h\sx'(0)$ is constrained to satisfy Eq. \ref{eq:r2_ge2_cons}.
Therefore we need only solve for the $b_i$,
\begin{align}
  b_{qq} &= \left[ \frac{146}{2025 k_0} - \frac{200}{81} f\sx''(1) \right]\frac{1}{f\sx'(1)}
    \approx 1.8010191875490722 \\
  b_{pq} &= \frac{1}{k_0 f\sx'(1)}\left[h\sx'(0) f\sx'(1) \frac{20}{9}
  + \frac{100(8+9\eta)}{243}k_0 f\sx''(1) - \frac{73}{405} \right] + \frac{100\eta}{27}
    \approx -1.850497151349339\\
  b_{pp} &= -\frac{1}{k_0 f\sx'(1)} \left[ \frac{h\sx''(0)}{2}
    + \frac{5(8+9\eta)}{27}h\sx'(0) f\sx'(1) + \frac{25(8+9\eta)^2}{1458}k_0 f\sx''(1) \right]
    - \frac{200\eta}{81} - \frac{25\eta^2}{9} \approx 0.974002499350257.
\end{align}
In \rrscan \cite{furness2020}, the interpolation function $f\sx$ is a piecewise function, but is a polynomial for $0 \leq \ba \leq 2.5$,
\begin{align}
  f\sx(0 \leq \ba \leq 2.5) &= \sum_{i=0}^7 c_{\text{x}i} \ba^i \\
  f\sx'(1) &= \sum_{i=1}^7 i c_{\text{x}i} \approx -0.9353000875519996 \\
  f\sx''(1) &= \sum_{i=2}^7 i(i-1) c_{\text{x}i} \approx 0.8500359204920018,
\end{align}
with the coefficients $c_{\text{x}i}$ taken from rSCAN \cite{bartok2019}.
The $h\sx^1$ function is unique to \rrscan,
\begin{align}
  h\sx^1(p) &= 1 + k_1 - k_1[1 + x(p)/k_1]^{-1} \\
  x(p) &= \left\{ \frac{5(4 + 9\eta)}{27}k_0 f\sx'(1) \exp[-p^2/d_{p2}^4]
    + \frac{10}{81} \right\}p,
\end{align}
therefore
\begin{align}
  h\sx'(0) &= \frac{5(4 + 9\eta)}{27}k_0 f\sx'(1) + \frac{10}{81} \approx 0.0026357640358089796  \\
  h\sx''(0) &= -\frac{2h\sx'(0)^2}{k_1} \approx -0.00021376160161427815.
\end{align}

It should be noted that the fourth-order terms in $\tau(p,q)$ are positive semi-definite, as they can be written in the form
\[
  b_{qq}q^2 + b_{pq} p q + b_{pp} p^2 = \left(\sqrt{b_{qq}}q +\frac{b_{pq}}{2\sqrt{b_{qq}}}p \right)^2
    + \left(b_{pp} - \frac{b_{pq}^2}{4b_{qq}}\right)p^2,
\]
and $b_{pp}-b_{pq}^2/(4b_{qq})>0$.

\onecolumngrid

\section{Laplacian-dependent stress tensor}

For practical calculations, the exchange-correlation stress tensor, $\Sigma^{ij}\sxc$, defined as \cite{dalcorso1994}
\begin{equation}
  \Sigma^{ij}\sxc = \int_\Omega n(\br) r_j \dd{v\sxc}{r_i} d^3 r, \label{eq:sxc_int}
\end{equation}
is greatly useful.
Here, the system volume is $\Omega$.
We take $r_1 = x$, $r_2 = y$, and $r_3 = z$.
Thus the exchange-correlation stress density,
\begin{equation}
  \sigma^{ij}\sxc = n(\br) r_j \dd{v\sxc}{r_i}, \label{eq:xc_st_dens}
\end{equation}
is only defined up to a certain gauge, like the exchange-correlation energy density $e\sxc$.
The gauge can be chosen up to the curl of a tensor, as the divergence of this tensor must yield the force on the system due to the exchange-correlation potential \cite{bartolotti1980}.
An overall choice of sign corresponds to consideration of internal or external stresses (for example, VASP appears to use the opposite sign convention as Eq. \ref{eq:xc_st_dens}).
Moreover, the stress tensor and its density should be symmetric.

While Eq. \ref{eq:sxc_int} is well-defined in a finite system, the term linear in $r_j$ makes this intractable in an extended system.
Following Ref. \cite{dalcorso1994}, we therefore take the system volume $\Omega$ to be \textit{finite}, and seek an expression for $\sigma^{ij}\sxc$ that is independent of the boundary conditions.
The latter expression will be well-defined as the thermodynamic average in an extended system.
Consider that
\begin{equation}
  \sigma^{ij}\sxc = \dd{}{r_i}\left( n r_j v\sxc \right) - v\sxc n \delta_{ij} - v\sxc r_j \dd{n}{r_i},
\end{equation}
where $\delta_{ij}=1$ if $i=j$ and 0 if $i\neq j$ is the Kronecker delta.
In a finite system, the integral of the total derivative will vanish, as it can be evaluated on a bounding surface at infinity.
Thus we will collect all terms that involve total derivatives and use those as a choice of gauge.

Suppose that an exchange-correlation functional depends upon $n, |\gn|,$ and $\lan$, and further that $e\sxc$ and $v\sxc$ are the exchange-correlation energy density and potential, respectively,
\begin{align}
  E\sxc &= \int e\sxc(n, |\gn|, \lan) d^3 r \\
  v\sxc &= \dd{e\sxc}{ n} - \dd{}{r_k} \left[ \dd{e\sxc}{(\partial_k n)}\right] + \dd{}{r_k}\dd{}{r_k} \left( \dd{e\sxc}{\lan}\right).
\end{align}
We use the Einstein or summation convention, wherein repeated indices imply summation,
\[
  \dd{}{r_k} \dd{e\sxc}{(\partial_k n)} \equiv \sum_{k=1}^3 \dd{}{r_k} \dd{e\sxc}{(\partial_k n)},
\]
and the shorthand $\partial_k n \equiv \partial n / \partial r_k$.
Then
\begin{align}
  \sigma^{ij}\sxc = - v\sxc n \delta_{ij} - \left[ \dd{e\sxc}{ n}
    - \dd{}{r_k} \left[ \dd{e\sxc}{(\partial_k n)}\right]
    + \dd{}{r_k}\dd{}{r_k} \left( \dd{e\sxc}{\lan}\right) \right] r_j \dd{n}{r_i}
    + \dd{}{r_i}\left(n r_j v\sxc \right).
\end{align}
We can express the gradient of $e\sxc$ as
\begin{equation}
  \dd{e\sxc}{r_i} = \dd{e\sxc}{n}\dd{n}{r_i}
    + \dd{e\sxc}{(\partial_k n)} \frac{\partial^2 n}{\partial r_k \partial r_i}
    + \dd{e\sxc}{\lan} \frac{\partial^3 n}{\partial r_k \partial r_k \partial r_i},
\end{equation}
and thus replace
\begin{align}
  \sigma^{ij}\sxc &= - v\sxc n \delta_{ij} - r_j \dd{e\sxc}{r_i}
    + r_j \dd{e\sxc}{(\partial_k n)} \frac{\partial^2 n}{\partial r_k \partial r_i}
    + r_j \dd{e\sxc}{\lan} \frac{\partial^3 n}{\partial r_k \partial r_k \partial r_i}
    + r_j \dd{n}{r_i} \dd{}{r_k} \left[ \dd{e\sxc}{(\partial_k n)}\right] \nonumber \\
  & - r_j \dd{n}{r_i} \dd{}{r_k}\dd{}{r_k} \left( \dd{e\sxc}{\lan}\right)
    + \dd{}{r_i}\left(n r_j v\sxc \right) \nonumber\\
  &= \left(e\sxc - v\sxc n \right)\delta_{ij}
    + r_j \dd{e\sxc}{(\partial_k n)} \frac{\partial^2 n}{\partial r_k \partial r_i}
    + r_j \dd{e\sxc}{\lan} \frac{\partial^3 n}{\partial r_k \partial r_k \partial r_i}
    + r_j \dd{n}{r_i} \dd{}{r_k} \left[ \dd{e\sxc}{(\partial_k n)}\right] \nonumber \\
  & - r_j \dd{n}{r_i} \dd{}{r_k}\dd{}{r_k} \left( \dd{e\sxc}{\lan}\right)
    + \dd{}{r_i}\left(n r_j v\sxc - r_j e\sxc \right). \label{eq:sxc_1}
\end{align}
Rearranging the term
\begin{align}
  r_j \dd{e\sxc}{(\partial_k n)} \frac{\partial^2 n}{\partial r_k \partial r_i}
  &= \dd{}{r_k} \left[r_j \dd{n}{r_i} \dd{e\sxc}{(\partial_k n)}  \right]
    - \delta_{ik} \dd{n}{r_i} \dd{e\sxc}{(\partial_k n)}
    - r_j \dd{n}{r_i} \dd{}{r_k} \left[\dd{e\sxc}{(\partial_k n)} \right]
\end{align}
shows that it partly cancels with another term in Eq. \ref{eq:sxc_1},
\begin{align}
  \sigma^{ij}\sxc &= \left(e\sxc - v\sxc n \right)\delta_{ij}
    -  \dd{n}{r_i} \dd{e\sxc}{(\partial_j n)}
    + r_j \dd{e\sxc}{\lan} \frac{\partial^3 n}{\partial r_k \partial r_k \partial r_i}
    \nonumber \\
  & - r_j \dd{n}{r_i} \dd{}{r_k}\dd{}{r_k} \left( \dd{e\sxc}{\lan}\right)
    + \dd{}{r_i}\left(n r_j v\sxc - r_j e\sxc \right)
    + \dd{}{r_k} \left[r_j \dd{e\sxc}{(\partial_k n)} \dd{n}{r_i} \right]. \label{eq:sxc_2}
\end{align}
Now, assuming that $\partial n/ \partial r_k$ has equal mixed partials,
\[
  \frac{\partial^3 n}{\partial r_k \partial r_k \partial r_i} =
    \frac{\partial^3 n}{\partial r_k \partial r_i \partial r_k},
\]
we rearrange
\begin{align}
  r_j \dd{e\sxc}{\lan} \frac{\partial^3 n}{\partial r_k \partial r_k \partial r_i}
  &= \dd{}{r_k} \left[r_j \dd{e\sxc}{\lan} \frac{\partial^2 n}{\partial r_k \partial r_i} \right]
    - \delta_{jk} \dd{e\sxc}{\lan} \frac{\partial^2 n}{\partial r_k \partial r_i}
    - r_j\frac{\partial^2 n}{\partial r_k \partial r_i}\dd{}{r_k} \left(\dd{e\sxc}{\lan}\right) \nonumber \\
  &= \dd{}{r_k} \left[r_j \dd{e\sxc}{\lan} \frac{\partial^2 n}{\partial r_k \partial r_i} \right]
    - \dd{e\sxc}{\lan} \frac{\partial^2 n}{\partial r_i \partial r_j}
    - \dd{}{r_k}\left[ r_j \dd{n}{r_i} \dd{}{r_k} \left( \dd{e\sxc}{\lan} \right) \right]
    \nonumber \\
  & + \delta_{jk} \dd{n}{r_i} \dd{}{r_k}\left( \dd{e\sxc}{\lan} \right)
    + r_j \dd{n}{r_i} \dd{}{r_k} \dd{}{r_k}\left( \dd{e\sxc}{\lan} \right) \nonumber \\
  &=  \dd{}{r_k} \left[r_j \dd{e\sxc}{\lan} \frac{\partial^2 n}{\partial r_k \partial r_i}
    - r_j \dd{n}{r_i} \dd{}{r_k} \left( \dd{e\sxc}{\lan} \right) \right]
    + \dd{}{r_j}\left[\dd{n}{r_i}\dd{e\sxc}{\lan} \right]
    - 2 \dd{e\sxc}{\lan} \frac{\partial^2 n}{\partial r_i \partial r_j} \nonumber \\
  & + r_j \dd{n}{r_i} \dd{}{r_k} \dd{}{r_k}\left( \dd{e\sxc}{\lan} \right).
\end{align}
Inserting this latter equality into Eq. \ref{eq:sxc_2} shows further cancellation
\begin{align}
  \sigma^{ij}\sxc &= \left(e\sxc - v\sxc n \right)\delta_{ij}
    - \dd{e\sxc}{(\partial_j n)} \dd{n}{r_i}
    - 2 \dd{e\sxc}{\lan} \frac{\partial^2 n}{\partial r_i \partial r_j}
    + \dd{}{r_j}\left[\dd{n}{r_i}\dd{e\sxc}{\lan} \right]
    \nonumber \\
  & + \dd{}{r_i}\left(n r_j v\sxc - r_j e\sxc \right)
    + \dd{}{r_k} \left[r_j \dd{n}{r_i} \dd{e\sxc}{(\partial_k n)}  \right]
    + \dd{}{r_k} \left[r_j \dd{e\sxc}{\lan} \frac{\partial^2 n}{\partial r_k \partial r_i}
      - r_j \dd{n}{r_i} \dd{}{r_k} \left( \dd{e\sxc}{\lan} \right) \right].
\end{align}
Let
\begin{align}
  \sigma^{ij}\sxc &= \widetilde{\sigma}\sxc^{ij} + \mathcal{G}\sxc^{ij} \\
  \widetilde{\sigma}\sxc^{ij} &= \left(e\sxc - v\sxc n \right)\delta_{ij}
    -  \dd{n}{r_i} \dd{e\sxc}{(\partial_j n)}
    - 2 \dd{e\sxc}{\lan} \frac{\partial^2 n}{\partial r_i \partial r_j} \\
  \mathcal{G}\sxc^{ij} &= \dd{}{r_j}\left[\dd{n}{r_i}\dd{e\sxc}{\lan} \right]
    + \dd{}{r_i}\left(n r_j v\sxc - r_j e\sxc \right)
    + \dd{}{r_k} \left[r_j \dd{e\sxc}{(\partial_k n)} \dd{n}{r_i} \right] \nonumber \\
  & + \dd{}{r_k} \left[r_j \dd{e\sxc}{\lan} \frac{\partial^2 n}{\partial r_k \partial r_i}
    - r_j \dd{n}{r_i} \dd{}{r_k} \left( \dd{e\sxc}{\lan} \right) \right].
\end{align}
The total stress due to the volume integral of $\sigma^{ij}\sxc$ and its integrated-by-parts counterpart $\widetilde{\sigma}\sxc^{ij}$ will be the same provided
\begin{equation}
  \int_\Omega \mathcal{G}\sxc^{ij} d^3 r = 0,
\end{equation}
again in a \textit{finite} system.
Looking term by term, this requires that the factors multiplying $r_i$ in $\mathcal{G}\sxc^{ij}$ vanish faster than $1/r$.
As the density decays exponentially as $r \to \infty$ \cite{almbladh1985}, we can safely assume that the integral of $\mathcal{G}\sxc^{ij}$ vanishes in a finite system.

As a final note of simplification, modern DFAs tend not to depend upon the direction of the density gradient, only its magnitude,
\begin{equation}
  \dd{e\sxc}{(\partial_j n)} =
    \dd{e\sxc}{|\gn|} \dd{}{(\partial_j n)}\left[(\partial_k n)( \partial_k n) \right]^{1/2}
    =
    \frac{1}{|\gn|} \dd{n}{r_j} \dd{e\sxc}{|\gn|} ,
\end{equation}
and thus the stress tensor density $\widetilde{\sigma}\sxc^{ij}$ appropriate for extended systems is
\begin{equation}
  \widetilde{\sigma}\sxc^{ij} = \left(e\sxc - v\sxc n \right)\delta_{ij}
    - \frac{1}{|\gn|}\dd{n}{r_i} \dd{n}{r_j} \dd{e\sxc}{|\gn|}
    - 2 \dd{e\sxc}{\lan} \frac{\partial^2 n}{\partial r_i \partial r_j},
    \label{eq:xc_st_ibp}
\end{equation}
and the stress tensor is $\Sigma\sxc^{ij} = \int \widetilde{\sigma}\sxc^{ij} d^3 r$.

\clearpage
\section{Full LC20 data \label{app:lc20_data}}

\begin{table}[h]
  \begin{center}
    \begin{tabular}{p{2cm}p{1.5cm}rrrrr}\hline\hline
      \multirow{2}{2cm}{Solid \\ (structure)} & \multirow{2}{1.5cm}{Reference \\ (\AA{})} & PBEsol & SCAN & \rrscan & \rrscan{}-L & OFR2 \\ \\ \hline
      Li (bcc) & 3.451 & -0.018 & -0.022 & 0.024 & -0.039 & -0.012 \\
      Na (bcc) & 4.207 & -0.036 & -0.012 & 0.007 & -0.039 & -0.056 \\
      Ca (fcc) & 5.555 & -0.095 & -0.003 & 0.023 & -0.044 & -0.046 \\
      Sr (fcc) & 6.042 & -0.129 & 0.041 & 0.061 & 0.015 & -0.023 \\
      Ba (bcc) & 5.004 & -0.110 & 0.046 & 0.073 & 0.069 & -0.006 \\
      Al (fcc) & 4.019 & -0.004 & -0.014 & -0.032 & -0.046 & -0.029 \\
      Cu (fcc) & 3.595 & -0.026 & -0.029 & -0.013 & 0.017 & -0.028 \\
      Rh (fcc) & 3.793 & -0.013 & -0.006 & 0.012 & 0.037 & -0.006 \\
      Pd (fcc) & 3.876 & -0.003 & 0.018 & 0.037 & 0.062 & 0.006 \\
      Ag (fcc) & 4.063 & -0.011 & 0.021 & 0.044 & 0.076 & 0.002 \\
      C (ds) & 3.555 & 0.001 & -0.000 & 0.007 & 0.014 & 0.023 \\
      SiC (zb) & 4.348 & 0.011 & 0.004 & 0.007 & 0.008 & 0.022 \\
      Si (ds) & 5.422 & 0.014 & 0.006 & 0.018 & 0.001 & 0.009 \\
      Ge (ds) & 5.644 & 0.031 & 0.022 & 0.035 & 0.057 & 0.014 \\
      GaAs (zb) & 5.641 & 0.023 & 0.019 & 0.028 & 0.048 & 0.003 \\
      LiF (rs) & 3.974 & 0.035 & -0.005 & 0.010 & 0.004 & 0.002 \\
      LiCl (rs) & 5.072 & -0.008 & 0.009 & 0.016 & -0.002 & -0.021 \\
      NaF (rs) & 4.57 & 0.066 & -0.015 & 0.011 & 0.016 & 0.020 \\
      NaCl (rs) & 5.565 & 0.041 & -0.002 & 0.026 & 0.005 & -0.022 \\
      MgO (rs) & 4.188 & 0.023 & -0.002 & 0.008 & 0.004 & 0.003 \\ \hline
      ME & (metals) & -0.044 & 0.004 & 0.024 & 0.011 & -0.020\\
      MAE & (metals) & 0.044 & 0.021 & 0.033 & 0.044 & 0.021 \\ \hline
      ME & (insulators) & 0.024 & 0.004 & 0.017 & 0.016 & 0.005 \\
      MAE & (insulators) & 0.025 & 0.008 & 0.017 & 0.016 & 0.014 \\ \hline
      ME & (total) & -0.010 & 0.004 & 0.020 & 0.013 & -0.007 \\
      MAE & (total) & 0.035 & 0.015 & 0.025 & 0.030 & 0.018 \\
      \hline\hline
    \end{tabular}
  \end{center}
  \caption{Relative errors ($a_0^{\text{approx}}-a_0^{\text{ref.}}$) for the LC20 test set \cite{sun2011} of 20 cubic lattice constants, all in \AA{}.
  Reference experimental lattice constants (with zero-point vibration effects removed) are taken from Ref. \cite{hao2012}.
  We include mean absolute (MAE) and mean errors (ME).
  The structures considered are face-centered cubic (fcc), body-centered cubic (bcc), cubic diamond structure (ds), rock-salt (rs), and zinc-blende (zb).
  OFR2 exceeds the accuracy of the parent meta-GGA \rrscan overall and for the metallic and insulating subsets of LC20.
  \label{tab:lc20_full}
  }
\end{table}

\begin{table}[h]
  \begin{center}
    \begin{tabular}{p{1.5cm}rrrrrr}\hline\hline
      \multirow{2}{2cm}{Solid \\ (structure)} & \multirow{2}{1.5cm}{Reference \\ (GPa)} & PBEsol & SCAN & \rrscan & \rrscan{}-L & OFR2 \\ \\ \hline
      Li (bcc) & 13.1 & 0.619 & -1.471 & -4.659 & -4.143 & -1.461 \\
      Na (bcc) & 7.9 & 0.021 & 0.683 & 0.254 & 1.546 & -0.480 \\
      Ca (fcc) & 15.9 & 2.084 & 2.141 & 1.959 & 3.237 & 3.190 \\
      Sr (fcc) & 12.0 & 0.397 & -0.739 & -0.627 & 0.019 & 0.269 \\
      Ba (bcc) & 10.6 & -1.161 & -2.062 & -2.051 & -1.001 & -1.265 \\
      Al (fcc) & 77.1 & 4.995 & 1.611 & 15.956 & 20.322 & 14.243 \\
      Cu (fcc) & 144.3 & 20.498 & 24.233 & 15.450 & -0.281 & 24.019 \\
      Rh (fcc) & 277.1 & 19.283 & 15.178 & 4.888 & -20.918 & 14.439 \\
      Pd (fcc) & 187.2 & 17.506 & 8.133 & -0.978 & -19.524 & 11.245 \\
      Ag (fcc) & 105.7 & 12.824 & 4.225 & -2.764 & -12.636 & 6.744 \\
      C (ds) & 454.7 & -5.144 & 3.611 & -5.483 & -21.214 & -28.634 \\
      SiC (zb) & 229.1 & -8.101 & -3.061 & -2.166 & -9.657 & -11.991 \\
      Si (ds) & 101.3 & -7.744 & -1.713 & -4.034 & -5.194 & -6.490 \\
      Ge (ds) & 79.4 & -11.809 & -8.053 & -8.147 & -17.672 & -8.620 \\
      GaAs (zb) & 76.7 & -7.721 & -4.294 & -4.104 & -30.596 & -3.777 \\
      LiF (rs) & 76.3 & -2.860 & 7.068 & 3.965 & 4.592 & 5.766 \\
      LiCl (rs) & 38.7 & -3.517 & 1.040 & -0.413 & -3.648 & -3.061 \\
      NaF (rs) & 53.1 & -4.571 & 7.039 & 2.988 & 2.640 & 3.033 \\
      NaCl (rs) & 27.6 & -1.714 & 0.763 & -0.103 & 0.791 & 2.324 \\
      MgO (rs) & 169.8 & -9.361 & 2.552 & 0.801 & 0.774 & -0.966 \\ \hline
      ME & (metals) & 7.707 & 5.193 & 2.743 & -3.338 & 7.094\\
      MAE & (metals) & 7.939 & 6.048 & 4.959 & 8.363 & 7.735 \\ \hline
      ME & (insulators) & -6.254 & 0.495 & -1.669 & -7.918 & -5.241 \\
      MAE & (insulators) & 6.254 & 3.919 & 3.220 & 9.678 & 7.466 \\ \hline
      ME & (total) & 0.726 & 2.844 & 0.537 & -5.628 & 0.926 \\
      MAE & (total) & 7.096 & 4.983 & 4.090 & 9.020 & 7.601 \\
      \hline\hline
    \end{tabular}
  \end{center}
  \caption{Relative errors ($B_0^{\text{approx}}-B_0^{\text{ref.}}$) for the LC20 test set \cite{sun2011} of bulk moduli for 20 cubic solids, all in GPa (1 eV/\AA{}$^3$ $\approx 160.2176634$ GPa).
  Reference experimental bulk moduli (with zero-point vibration effects removed) are taken from Ref. \cite{tran2016}.
  It should be noted that the \rrscan and \rrscan{}-L values presented here and in Ref. \cite{mejia2020} agree to within a few GPa for each solid, generally.
  In a few cases, like Ge and GaAs for \rrscan{}-L or NaCl for \rrscan and \rrscan{}-L, agreement is quite poor.
  We attribute this to the different pseudopotentials used: Ref. \cite{mejia2020} used ``no-suffix'' pseudopotentials, whereas we used the recommended pseudopotentials from VASP.
  In these cases, the Ge\_d (which treats $d$-semicore states as valence states), Ga\_d, and Na\_pv (which treats $p$-semicore states as valence states) pseudopotentials might give very different behaviors than their no-suffix counterparts (which treat fewer electrons as valence electrons).
  \label{tab:lc20_b0_full}
  }
\end{table}

\begin{table}
  \centering
  \begin{tabular}{l|rrrr} \hline \hline
    Solid (struc) & PBEsol & r$^2$SCAN & r$^2$SCAN-L & OFR2 \\ \hline
    Li (bcc) & $3.9698 \times 10^{-3}$ & $9.1842 \times 10^{-3}$ & $1.5553 \times 10^{-2}$ & $-9.6238 \times 10^{-3}$ \\
    Na (bcc) & $9.8549 \times 10^{-4}$ & $7.0530 \times 10^{-4}$ & $4.1485 \times 10^{-3}$ & $1.1015 \times 10^{-2}$ \\
    Ca (fcc) & $2.5486 \times 10^{-3}$ & $2.9326 \times 10^{-3}$ & $1.5698 \times 10^{-2}$ & $1.6111 \times 10^{-2}$ \\
    Sr (fcc) & $-1.2412 \times 10^{-2}$ & $1.2689 \times 10^{-3}$ & $2.5362 \times 10^{-2}$ & $6.6702 \times 10^{-3}$ \\
    Ba (bcc) & $2.5548 \times 10^{-4}$ & $1.0728 \times 10^{-3}$ & $5.9928 \times 10^{-2}$ & $1.4873 \times 10^{-2}$ \\
    Al (fcc) & $6.1313 \times 10^{-6}$ & $-7.9749 \times 10^{-4}$ & $2.6118 \times 10^{-3}$ & $3.0966 \times 10^{-3}$ \\
    Cu (fcc) & $3.0698 \times 10^{-4}$ & $9.6047 \times 10^{-4}$ & $3.5296 \times 10^{-3}$ & $1.4137 \times 10^{-3}$ \\
    Rh (fcc) & $2.9099 \times 10^{-4}$ & $3.6044 \times 10^{-5}$ & $3.8564 \times 10^{-4}$ & $3.7767 \times 10^{-4}$ \\
    Pd (fcc) & $-3.3150 \times 10^{-4}$ & $-6.9784 \times 10^{-4}$ & $-7.5960 \times 10^{-4}$ & $-3.4265 \times 10^{-4}$ \\
    Ag (fcc) & $5.6017 \times 10^{-4}$ & $1.2080 \times 10^{-4}$ & $1.2583 \times 10^{-4}$ & $5.0807 \times 10^{-4}$ \\
    C (ds) & $7.3743 \times 10^{-4}$ & $9.5259 \times 10^{-4}$ & $8.4424 \times 10^{-4}$ & $2.6446 \times 10^{-3}$ \\
    SiC (zb) & $6.5009 \times 10^{-4}$ & $6.9223 \times 10^{-4}$ & $1.5169 \times 10^{-3}$ & $2.2265 \times 10^{-3}$ \\
    Si (ds) & $1.5047 \times 10^{-4}$ & $1.8607 \times 10^{-4}$ & $-9.5398 \times 10^{-4}$ & $3.3177 \times 10^{-3}$ \\
    Ge (ds) & $4.8177 \times 10^{-4}$ & $1.7996 \times 10^{-3}$ & $1.9719 \times 10^{-3}$ & $3.7134 \times 10^{-3}$ \\
    GaAs (zb) & $-1.9404 \times 10^{-4}$ & $-3.2999 \times 10^{-4}$ & $1.0211 \times 10^{-2}$ & $3.0868 \times 10^{-3}$ \\
    LiF (rs) & $5.7602 \times 10^{-3}$ & $2.0001 \times 10^{-3}$ & $-2.7121 \times 10^{-3}$ & $7.1041 \times 10^{-4}$ \\
    LiCl (rs) & $1.6706 \times 10^{-3}$ & $-1.0409 \times 10^{-3}$ & $-4.9830 \times 10^{-3}$ & $-6.1942 \times 10^{-4}$ \\
    NaF (rs) & $6.0002 \times 10^{-3}$ & $1.7042 \times 10^{-3}$ & $3.5240 \times 10^{-3}$ & $8.9884 \times 10^{-3}$ \\
    NaCl (rs) & $1.6417 \times 10^{-3}$ & $-6.9238 \times 10^{-3}$ & $6.5536 \times 10^{-3}$ & $1.8502 \times 10^{-3}$ \\
    MgO (rs) & $1.3037 \times 10^{-3}$ & $1.1726 \times 10^{-3}$ & $7.7154 \times 10^{-5}$ & $1.9402 \times 10^{-3}$ \\ \hline
    MD & $7.1911 \times 10^{-4}$ & $7.4993 \times 10^{-4}$ & $7.1316 \times 10^{-3}$ & $3.5979 \times 10^{-3}$ \\
    MAD & $2.0129 \times 10^{-3}$ & $1.7289 \times 10^{-3}$ & $8.0725 \times 10^{-3}$ & $4.6564 \times 10^{-3}$ \\ \hline\hline
  \end{tabular}
  \caption{Comparison of the LC20 cubic lattice-constant differences found by fitting (EOS) to the SJEOS and by minimization of the stress tensor (ST) using Eq. \ref{eq:xc_st_ibp}.
  The deviations are $a_0^\text{EOS} - a_0^\text{ST}$; mean deviations (MDs) and mean absolute deviations (MADs) are also presented, in \AA{}.
  }
  \label{tab:lc20_eos_st_full}
\end{table}

\clearpage
\section{Full LC23 data \label{app:lc23_data}}

\begin{table}[h]
  \centering
  \begin{tabular}{l|rrrrrrr}\hline\hline
    Solid (structure) & \multirow{2}{1.5cm}{Reference \\ (\AA{})} & PBE & PBEsol & SCAN & \rrscan & \rrscan-L & OFR2 \\ \\ \hline
    Li (bcc) & 3.451 & -0.012 & -0.008 & 0.018 & 0.029 & -0.021 & 0.010 \\
    Na (bcc) & 4.207 & -0.014 & -0.038 & -0.026 & -0.007 & -0.083 & -0.057 \\
    K (bcc) & 5.211 & 0.072 & 0.004 & 0.111 & 0.139 & -0.042 & -0.006 \\
    Rb (bcc) & 5.58 & 0.088 & -0.012 & 0.132 & 0.166 & 0.025 & 0.054 \\
    Cs (bcc) & 6.043 & 0.119 & -0.032 & 0.186 & 0.228 & 0.103 & 0.069 \\
    Ca (fcc) & 5.555 & -0.024 & -0.095 & -0.005 & 0.024 & -0.049 & -0.046 \\
    Sr (fcc) & 6.042 & -0.020 & -0.129 & 0.042 & 0.062 & 0.007 & -0.018 \\
    Ba (bcc) & 5.004 & 0.026 & -0.110 & 0.045 & 0.073 & 0.055 & 0.000 \\
    Al (fcc) & 4.019 & 0.021 & -0.004 & -0.014 & -0.032 & -0.046 & -0.029 \\
    Cu (fcc) & 3.595 & 0.040 & -0.026 & -0.027 & -0.013 & 0.014 & -0.028 \\
    Rh (fcc) & 3.793 & 0.031 & -0.018 & -0.014 & 0.011 & 0.031 & -0.010 \\
    Pd (fcc) & 3.876 & 0.064 & -0.003 & 0.018 & 0.037 & 0.062 & 0.005 \\
    Ag (fcc) & 4.063 & 0.084 & -0.011 & 0.021 & 0.044 & 0.076 & 0.002 \\
    C (ds) & 3.555 & 0.018 & 0.002 & 0.001 & 0.008 & 0.015 & 0.024 \\
    SiC (zb) & 4.348 & 0.032 & 0.011 & 0.004 & 0.007 & 0.008 & 0.023 \\
    Si (ds) & 5.422 & 0.047 & 0.014 & 0.005 & 0.018 & 0.004 & 0.005 \\
    Ge (ds) & 5.644 & 0.138 & 0.057 & 0.040 & 0.037 & 0.061 & 0.039 \\
    GaAs (zb) & 5.641 & 0.121 & 0.043 & 0.024 & 0.031 & 0.056 & 0.024 \\
    LiF (rs) & 3.974 & 0.099 & 0.042 & 0.005 & 0.022 & 0.039 & 0.043 \\
    LiCl (rs) & 5.072 & 0.081 & -0.002 & 0.021 & 0.039 & 0.006 & -0.003 \\
    NaF (rs) & 4.57 & 0.062 & -0.014 & -0.091 & -0.067 & -0.056 & -0.042 \\
    NaCl (rs) & 5.565 & 0.090 & -0.005 & -0.047 & -0.019 & -0.047 & -0.058 \\
    MgO (rs) & 4.188 & 0.060 & 0.023 & -0.002 & 0.008 & 0.009 & 0.006 \\ \hline
    ME & (metals) & 0.037 & -0.037 & 0.037 & 0.058 & 0.010 & -0.004\\
    MAE & (metals) & 0.047 & 0.038 & 0.051 & 0.066 & 0.047 & 0.026 \\ \hline
    ME & (alkalis) & 0.051 & -0.017 & 0.084 & 0.111 & -0.004 & 0.014 \\
    MAE & (alkalis) & 0.061 & 0.019 & 0.095 & 0.114 & 0.055 & 0.039 \\ \hline
    ME & (insulators) & 0.075 & 0.017 & -0.004 & 0.008 & 0.009 & 0.006 \\
    MAE & (insulators) & 0.075 & 0.021 & 0.024 & 0.026 & 0.030 & 0.027 \\ \hline
    ME & (total) & 0.053 & -0.013 & 0.019 & 0.037 & 0.010 & 0.000 \\
    MAE & (total) & 0.059 & 0.031 & 0.039 & 0.049 & 0.040 & 0.026 \\ \hline
    ME & (LC20) & 0.047 & -0.014 & 0.001 & 0.016 & 0.007 & -0.005 \\
    MAE & (LC20) & 0.054 & 0.033 & 0.024 & 0.029 & 0.037 & 0.024 \\
  \hline\hline
  \end{tabular}
  \caption{
  Relative errors in the equilibrium lattice constants $a_0$ (in \AA{}) for the LC23 set (LC20 augmented with K, Rb, and Cs).
  The PBE \cite{perdew1996} and PBEsol \cite{perdew2008} GGAs, SCAN \cite{sun2015} and \rrscan \cite{furness2020} T-MGGAs, and \rrscan{}-L \cite{mejia2020} and OFR2 LL-MGGAs are presented.
  Reference experimental lattice constants (with zero-point vibration effects removed) are taken from Ref. \cite{hao2012}, except for Rb, which is taken from \cite{tran2016}.
  LC20 error statistics are also reported to demonstrate the level of convergence with respect to the benchmark results presented in Table \ref{tab:lc20_full}.
  \label{tab:lc23_a0}
  }
\end{table}

\begin{table}[h]
  \centering
  \begin{tabular}{l|rrrrrrr}\hline\hline
    Solid (structure) & \multirow{2}{1.5cm}{Reference \\ (GPa)} & PBE & PBEsol & SCAN & \rrscan & \rrscan-L & OFR2 \\ \\ \hline
    Li (bcc) & 13.1 & 0.839 & 0.583 & 0.596 & 0.013 & 1.239 & 0.045 \\
    Na (bcc) & 7.9 & 0.014 & 0.125 & 0.163 & 0.065 & -2.894 & -0.671 \\
    K (bcc) & 3.8 & -0.207 & -0.077 & -0.349 & -0.360 & 11.257 & 1.370 \\
    Rb (bcc) & 3.6 & -0.821 & -0.648 & -0.905 & -0.963 & 2.295 & -1.210 \\
    Cs (bcc) & 2.3 & -0.348 & -0.265 & -0.324 & -0.400 & 0.509 & 0.506 \\
    Ca (fcc) & 15.9 & 1.327 & 2.084 & 2.100 & 1.879 & 3.302 & 3.089 \\
    Sr (fcc) & 12.0 & -0.689 & 0.399 & -0.745 & -0.615 & -0.108 & 0.038 \\
    Ba (bcc) & 10.6 & -1.761 & -1.162 & -2.070 & -2.055 & -3.387 & -1.543 \\
    Al (fcc) & 77.1 & 0.260 & 4.965 & 1.574 & 15.934 & 13.496 & 11.678 \\
    Cu (fcc) & 144.3 & -6.910 & 20.643 & 17.327 & 16.028 & 3.719 & 23.695 \\
    Rh (fcc) & 277.1 & -18.422 & 21.063 & 17.606 & 4.758 & -20.703 & 14.192 \\
    Pd (fcc) & 187.2 & -18.081 & 17.501 & 8.447 & -0.768 & -18.533 & 11.649 \\
    Ag (fcc) & 105.7 & -16.360 & 12.857 & 3.767 & -2.716 & -14.355 & 6.616 \\
    C (ds) & 454.7 & -19.906 & -3.552 & 3.901 & -3.551 & -19.449 & -24.790 \\
    SiC (zb) & 229.1 & -16.873 & -8.254 & -2.853 & -2.359 & -10.102 & -13.234 \\
    Si (ds) & 101.3 & -12.494 & -7.742 & -1.521 & -4.008 & -6.276 & -6.069 \\
    Ge (ds) & 79.4 & -20.223 & -11.949 & -7.579 & -6.319 & -11.531 & -8.730 \\
    GaAs (zb) & 76.7 & -14.665 & -6.497 & -1.881 & -2.929 & -8.244 & -3.676 \\
    LiF (rs) & 76.3 & -8.886 & -3.567 & 3.680 & 2.138 & -0.878 & -7.408 \\
    LiCl (rs) & 38.7 & -6.865 & -3.591 & -2.399 & -3.768 & -1.930 & -1.460 \\
    NaF (rs) & 53.1 & -5.934 & -0.959 & 9.802 & 6.859 & 5.564 & 5.555 \\
    NaCl (rs) & 27.6 & -3.345 & -0.746 & 2.736 & 1.551 & 2.196 & 3.204 \\
    MgO (rs) & 169.8 & -17.938 & -9.140 & 2.450 & 0.967 & -0.729 & -0.859 \\ \hline
    ME & (metals) & -4.704 & 6.005 & 3.630 & 2.369 & -1.859 & 5.343\\
    MAE & (metals) & 5.080 & 6.336 & 4.306 & 3.581 & 7.369 & 5.869 \\ \hline
    ME & (alkalis) & -0.105 & -0.056 & -0.164 & -0.329 & 2.481 & 0.008 \\
    MAE & (alkalis) & 0.446 & 0.340 & 0.467 & 0.360 & 3.639 & 0.760 \\ \hline
    ME & (insulators) & -12.713 & -5.600 & 0.634 & -1.142 & -5.138 & -5.747 \\
    MAE & (insulators) & 12.713 & 5.600 & 3.880 & 3.445 & 6.690 & 7.498 \\ \hline
    ME & (total) & -8.186 & 0.960 & 2.327 & 0.843 & -3.284 & 0.521 \\
    MAE & (total) & 8.399 & 6.016 & 4.121 & 3.522 & 7.074 & 6.578 \\ \hline
    ME & (LC20) & -9.346 & 1.153 & 2.755 & 1.055 & -4.480 & 0.566 \\
    MAE & (LC20) & 9.590 & 6.869 & 4.660 & 3.964 & 7.432 & 7.410 \\
  \hline\hline
  \end{tabular}
  \caption{
  Relative errors in the equilibrium bulk moduli $B_0$ (in GPa) for the LC23 set (LC20 augmented with K, Rb, and Cs).
  The PBE \cite{perdew1996} and PBEsol \cite{perdew2008} GGAs, SCAN \cite{sun2015} and \rrscan \cite{furness2020} T-MGGAs, and \rrscan{}-L \cite{mejia2020} and OFR2 LL-MGGAs are presented.
  Reference experimental bulk moduli (with zero-point vibration effects removed) are taken from Ref. \cite{tran2016}.
  LC20 error statistics are also reported to demonstrate the level of convergence with respect to the benchmark results presented in Table \ref{tab:lc20_b0_full}.
  \label{tab:lc23_b0}
  }
\end{table}

\begin{table}[h]
  \centering
  \begin{tabular}{l|rrrrrrr} \hline\hline
    Solid (structure) & \multirow{2}{1.5cm}{Reference \\ (eV/atom)} & PBE & PBEsol & SCAN & \rrscan & \rrscan-L & OFR2 \\ \\ \hline
    Li (bcc) & 1.67 & -0.065 & 0.005 & -0.105 & -0.096 & -0.060 & -0.102 \\
    Na (bcc) & 1.12 & -0.033 & 0.038 & -0.018 & -0.031 & -0.056 & -0.050 \\
    K (bcc) & 0.94 & -0.073 & -0.011 & -0.074 & -0.089 & -0.100 & -0.090 \\
    Rb (bcc) & 0.86 & -0.088 & -0.025 & -0.097 & -0.111 & -0.131 & -0.101 \\
    Cs (bcc) & 0.81 & -0.099 & -0.032 & -0.121 & -0.131 & -0.154 & -0.149 \\
    Ca (fcc) & 1.87 & 0.032 & 0.233 & 0.206 & 0.201 & 0.181 & 0.174 \\
    Sr (fcc) & 1.73 & -0.122 & 0.077 & 0.078 & 0.060 & 0.001 & 0.078 \\
    Ba (bcc) & 1.91 & -0.035 & 0.203 & 0.117 & 0.077 & -0.006 & 0.079 \\
    Al (fcc) & 3.43 & 0.080 & 0.432 & 0.170 & 0.172 & -0.006 & 0.016 \\
    Cu (fcc) & 3.51 & -0.025 & 0.522 & 0.375 & 0.350 & -0.018 & 0.385 \\
    Rh (fcc) & 5.78 & -0.021 & 0.933 & 0.072 & 0.052 & -0.335 & 0.462 \\
    Pd (fcc) & 3.93 & -0.189 & 0.541 & 0.437 & 0.236 & -0.244 & 0.363 \\
    Ag (fcc) & 2.96 & -0.441 & 0.118 & -0.075 & -0.082 & -0.450 & -0.037 \\
    C (ds) & 7.55 & 0.264 & 0.763 & -0.051 & -0.090 & -0.196 & -0.186 \\
    SiC (zb) & 6.48 & -0.012 & 0.411 & -0.037 & 0.046 & -0.203 & -0.218 \\
    Si (ds) & 4.68 & -0.100 & 0.246 & 0.029 & 0.190 & -0.084 & -0.092 \\
    Ge (ds) & 3.89 & -0.180 & 0.211 & 0.246 & 0.133 & -0.314 & 0.042 \\
    GaAs (zb) & 3.34 & -0.158 & 0.233 & 0.029 & -0.016 & -0.284 & 0.013 \\
    LiF (rs) & 4.46 & -0.023 & 0.085 & -0.066 & -0.065 & -0.171 & -0.271 \\
    LiCl (rs) & 3.59 & -0.189 & -0.056 & -0.102 & -0.121 & -0.179 & -0.246 \\
    NaF (rs) & 3.97 & 0.027 & 0.128 & 0.041 & 0.044 & -0.074 & -0.169 \\
    NaCl (rs) & 3.34 & -0.181 & -0.071 & -0.041 & -0.056 & -0.136 & -0.205 \\
    MgO (rs) & 5.2 & -0.196 & 0.134 & 0.062 & 0.055 & -0.060 & -0.182 \\ \hline
    ME & (metals) & -0.083 & 0.233 & 0.074 & 0.047 & -0.106 & 0.079\\
    MAE & (metals) & 0.100 & 0.244 & 0.150 & 0.130 & 0.134 & 0.160 \\ \hline
    ME & (alkalis) & -0.072 & -0.005 & -0.083 & -0.092 & -0.100 & -0.099 \\
    MAE & (alkalis) & 0.072 & 0.022 & 0.083 & 0.092 & 0.100 & 0.099 \\ \hline
    ME & (insulators) & -0.075 & 0.208 & 0.011 & 0.012 & -0.170 & -0.152 \\
    MAE & (insulators) & 0.133 & 0.234 & 0.070 & 0.082 & 0.170 & 0.163 \\ \hline
    ME & (total) & -0.079 & 0.222 & 0.047 & 0.032 & -0.134 & -0.021 \\
    MAE & (total) & 0.115 & 0.239 & 0.115 & 0.109 & 0.150 & 0.161 \\ \hline
    ME & (LC20) & -0.078 & 0.259 & 0.068 & 0.053 & -0.135 & -0.007 \\
    MAE & (LC20) & 0.119 & 0.272 & 0.118 & 0.109 & 0.153 & 0.169 \\
  \hline\hline
  \end{tabular}
  \caption{
  Relative errors in the equilibrium cohesive energies $E_0$ (in eV/atom) for the LC23 set (LC20 augmented with K, Rb, and Cs).
  The PBE \cite{perdew1996} and PBEsol \cite{perdew2008} GGAs, SCAN \cite{sun2015} and \rrscan \cite{furness2020} T-MGGAs, and \rrscan{}-L \cite{mejia2020} and OFR2 LL-MGGAs are presented.
  Reference experimental cohesive energies (with zero-point vibration effects removed) are taken from Ref. \cite{tran2016}.
  LC20 error statistics are also reported.
  \label{tab:lc23_e0}
  }
\end{table}

\end{document}